\documentstyle[prd,aps,epsf]{revtex}

\def\mps{m_{\rm PS}}
\def\mv{m_{\rm V}}

\newcommand{\STRUT}{\rule{0in}{2.5ex}}

\def\simgt{\,\rlap{\lower 3.5 pt\hbox{$\mathchar \sim$}}\raise 1pt \hbox {$>$}\,}
\def\simlt{\,\rlap{\lower 3.5 pt\hbox{$\mathchar \sim$}}\raise 1pt \hbox {$<$}\,}

\begin{document}

\title{
%
%
\begin{flushright}
\normalsize
UTCCP-P-104          \\
\end{flushright}
%
%
Light Hadron Spectroscopy with Two Flavors of Dynamical 
Quarks on the Lattice}  

\author{A.~Ali Khan$^1$\thanks{address till 31 August, 2000},
        S.~Aoki$^2$,
        G.~Boyd$^1$,
        R.~Burkhalter$^{1,2}$,
        S.~Ejiri$^1$\thanks{present address :
            Department of Physics, University of Wales, 
            Swansea SA2 8PP, U.K.},
        M.~Fukugita$^3$,
        S.~Hashimoto$^4$,
        N.~Ishizuka$^{1,2}$,
        Y.~Iwasaki$^{1,2}$,
        K.~Kanaya$^{2}$,
        T.~Kaneko$^4$,
        Y.~Kuramashi$^4$,
        T.~Manke$^1$\thanks{present address :
            Physics Department, Columbia University, 
            New York, NY 10027, U. S. A.},
        K.~Nagai$^1$\thanks{present address :
            CERN, Theory Division, CH--1211 Geneva 23, Switzerland},
        M.~Okawa$^4$,
        H.P.~Shanahan$^1$\thanks{present address :
            Department of Biochemistry and Molecular
            Biology, University College London, London, England, U.K.},
        A.~Ukawa$^{1,2}$ and
        T.~Yoshi\'e$^{1,2}$ \\[2mm]
        (CP-PACS Collaboration)
}

\address{
$^1$Center for Computational Physics,
University of Tsukuba, Tsukuba, Ibaraki 305--8577, Japan \\
$^2$Institute of Physics, University of
Tsukuba, Tsukuba, Ibaraki 305--8571, Japan \\
$^3$Institute for Cosmic Ray Research,
University of Tokyo, Kashiwa 277--8582, Japan \\
$^4$High Energy Accelerator Research Organization
(KEK), Tsukuba, Ibaraki 305--0801, Japan
}

\date{May 17, 2001}

\maketitle

\begin{abstract}

We present results of a numerical calculation of lattice QCD with two
degenerate flavors of dynamical quarks, identified with up and down quarks,
and with a strange quark treated in the quenched approximation. The
lattice action and simulation parameters are chosen with a view to carrying
out an extrapolation to the continuum limit as well as chiral
extrapolations in dynamical up and down quark masses. Gauge configurations
are generated with a renormalization-group improved gauge action and a mean
field improved clover quark action at three values of $\beta=6/g^2$,
corresponding to lattice spacings of $a \approx 0.22$, 0.16 and 0.11~fm,
and four sea quark masses corresponding to $\mps/\mv\approx 0.8$, 0.75, 0.7
and 0.6. The sizes of lattice are chosen to be $12^3\times 24$, $16^3\times
32$ and $24^3\times 48$ so that the physical spatial size is kept constant
at $La\approx 2.5$~fm. Hadron masses, light quark masses and meson decay
constants are measured at five valence quark masses corresponding to
$\mps/\mv\approx 0.8$, 0.75, 0.7, 0.6 and 0.5. We also carry out
complementary quenched simulations with the same improved actions. The
quenched spectrum from this analysis agrees well in the continuum limit
with the one of our earlier work using the standard action, quantitatively
confirming the systematic deviation of the quenched spectrum from
experiment. We find the two-flavor full QCD meson masses in the continuum
limit to be much closer to experimental meson masses than those from
quenched QCD. When using the $K$ meson mass to fix the strange quark mass,
the difference between quenched QCD and experiment of $2.6^{+0.3}_{-0.9}$\%
for the $K^*$ meson mass and of $4.1^{+0.5}_{-1.6}$\% for the $\phi$ meson
mass is reduced to $0.7^{+1.1}_{-1.7}$\% and $1.3^{+1.8}_{-2.5}$\% in full
QCD, where the errors include estimates of systematic errors of the
continuum extrapolation as well as statistical errors. Analyses of the $J$
parameter yield a similar trend in that the quenched estimate in the
continuum limit $J=0.375^{+0.039}_{-0.009}$ increases to
$J=0.440^{+0.061}_{-0.031}$ in two-flavor full QCD, approaching the
experimental value $J\approx 0.48$. We take these results as manifestations
of sea quark effects in two-flavor full QCD. For baryon masses full QCD
values for strange baryons such as $\Xi$ and $\Omega$ are in agreement with
experiment, while they differ increasingly with decreasing strange quark
content, resulting in a nucleon mass higher than experiment by 10\% and a
$\Delta$ mass by 13\%. The pattern suggests finite size effects as a
possible origin for this deviation. For light quark masses in the continuum
limit we obtain $m_{ud}^{\overline{\rm MS}}(2\,{\rm
GeV})=3.44^{+0.14}_{-0.22}$~MeV and $m_s^{\overline{\rm MS}}(2\,{\rm
GeV})=88^{+4}_{-6}$~MeV ($K$--input) and $m_s^{\overline{\rm MS}}(2\,{\rm
GeV})=90^{+5}_{-11}$~MeV ($\phi$--input), which are reduced by about 25\%
compared to the values in quenched QCD. We also present results for decay
constants where large scaling violations obstruct a continuum
extrapolation. Need for a non-perturbative estimate of renormalization
factors is discussed.

\end{abstract}


\newpage


\section{Introduction}
\label{sec:intro}

The mass spectrum of hadrons represents a fundamental manifestation of the
long-distance dynamics of quarks and gluons governed by QCD.
Non-perturbative calculations through numerical simulations on a space-time
lattice~\cite{wilson} provide a method to obtain this quantity from the QCD
Lagrangian without approximations. Such calculations also lead to a
determination of the light quark masses~\cite{weinberg}, which are
fundamental constants of Nature and yet not directly measurable in
experiments. These reasons underlie the large number of attempts toward
the hadron spectrum carried out since the pioneering studies of
Ref.~\cite{HamberParisiWeingarten}.

Most of these calculations employed the quenched approximation of ignoring
sea quarks, since dynamical quark simulations incorporating their effects
place quite severe demands on computational resources. Significant advance
has been made over the years within this approximation. In particular,
Weingarten and collaborators~\cite{GF11} made a pioneering attempt toward a
precision calculation of the spectrum in the continuum limit through
control of all systematic errors other than quenching within a single set
of simulations.

This approach was pushed further in Ref.~\cite{quench.CPPACS} where the
precision of the calculation reached the level of a few percent for hadron
masses. Scrutinized with this accuracy, the quenched hadron spectrum shows
a clear and systematic deviation from experiment; when one uses $\pi, \rho$
and $K$ meson masses as input to fix the physical scale and light quark
masses, the $K^*-K$ hyperfine splitting is too small by about 10\% compared
to the experimental value, the octet baryon masses are systematically
lower, and the decuplet baryon mass splitting is smaller than experiment by
about 30\%.

Clearly further progress in lattice calculations of the hadron mass
spectrum requires a departure from the quenched approximation. In fact
simulations of full QCD with dynamical quarks have a long history
\cite{Langevin,LosAlamosFull,HEMCGC-KS,ColumbiaFull,LosAlamosFullII,FukugitaAoki,MILC-fullKS,HEMCGC-Wilson,SCRI-full,MILC-fullrecent},
leading up to the recent efforts of 
Refs.~\cite{SESAM-spectrum,UKQCD-IMPfull,QCDSF}.
In contrast to quenched simulations, however, no attempt to control all of
the systematic errors within a single set of simulations has been made so
far. Except for the work of the MILC Collaboration~\cite{MILC-fullrecent},
employing the Kogut-Susskind quark action, previous calculations have been
restricted to a few quark masses within a small range and/or a single value
of the lattice spacing. Furthermore, until recently, statistics have been
rather limited due to the limitation of available computing power.

In the present work, we wish to advance an attempt toward simulations of
full QCD which includes extrapolations to the chiral limit of light quark
masses and the continuum limit of zero lattice spacing. This is an endeavor
demanding considerable computing resources, which we hope to meet with the
use of the CP-PACS parallel computer with a peak speed of 614~GFLOPS
developed at the University of Tsukuba~\cite{CPPACS-machine,Nakazawa}. We
explore, as a first step toward a realistic simulation of QCD, the case of
dynamical up and down quarks, which are assumed degenerate, treating the
strange quark in the quenched approximation. Preliminary results of the
present work have been reported previously~\cite{CPPACS-preliminary}.

A crucial computational issue in this attempt is how one copes with the
large amount of computation necessary in full QCD, and still covers a range
of lattice spacings required for the continuum extrapolation. We deal with
this problem with the use of improved lattice actions, which are designed
to reduce scaling violations, and hence should allow a continuum
extrapolation from coarse lattice spacings.

In Ref.~\cite{comparative} we have carried out a preparatory study on the
efficiency of improved actions in full QCD. Based on the results from this
study we employ a renormalization group improved action~\cite{RGIA} for the
gauge field and a mean field improved Sheikholeslami-Wohlert clover
action~\cite{clover} for the quark field. With these actions, hadron masses
show reasonable scaling behavior and the static quark potential good
rotational symmetry, at a coarse lattice spacing of $a\approx 0.2$~fm, as
compared to the range $a\simlt 0.1$~fm needed for the standard plaquette
and Wilson quark actions. This leads us to make a continuum extrapolation
from the range of lattice spacings $a\approx 0.2$--0.1~fm.

Previous studies of finite size effects (see, {\it e.g.,}
Refs.~\cite{GF11,FukugitaAoki,MILC-fullKS}) indicate that physical lattice
sizes larger than $La \approx 2.5$--3.0~fm are required to avoid
size-dependent errors in hadron masses. Compromising on a lattice of
physical size $La \approx 2.5$~fm leads to a $12^3\times 24$ lattice at
$a\approx 0.2$~fm, and $24^3\times 48$ at $a\approx 0.1$~fm. Estimates of
CPU time obtained in our preparatory study~\cite{comparative} show that
simulations on such a set of lattices are feasible with the full use of the
CP-PACS computer.

Since we employ the quenched approximation for strange quark, the
calculation of the strange spectrum requires the introduction of a valence
strange quark which only appears in hadron propagators. We generalize this
treatment and analyze hadron masses as functions of valence and sea quark
masses regarded as independent variables. The benefit of this approach is
that it gives us better control over the whole spectrum (strange and
non-strange) and its cross-over from quenched to full QCD when the mass of
the underlying sea quark is decreased.

There are a number of physics issues we wish to explore in our full QCD
simulation. An important question is whether effects of dynamical quarks
can be seen in the light hadron spectrum. In particular we wish to examine
if and to what extent the deviation of the quenched spectrum from
experiment established in our extensive study with the standard plaquette
and Wilson quark actions~\cite{quench.CPPACS} can be explained
as effects of sea quarks. Answering this question requires a
detailed comparison with hadron masses in quenched QCD for which we use
results of Ref.~\cite{quench.CPPACS}. We also carry out a set of new
quenched simulations with the same RG-improved gluon action and the clover
quark action as employed in the simulation of full QCD in order to make a
point-to-point comparison of full and quenched QCD at the same range of
lattice spacings.

Another question concerns light quark masses. Quenched calculations of
light quark masses have made considerable progress in recent
years~\cite{FNALquark,JLQCDquark,UKQCDquark,QCDSFquark,quench.CPPACS}. It
has become clear~\cite{quench.CPPACS} that the quenched estimate for the
strange quark mass extrapolated to the continuum limit suffers from a large
systematic uncertainty of order 20\% depending on the choice of hadron mass
for input, {\it e.g.,} $K$ meson mass or $\phi$ meson mass. This is a
reflection of the systematic deviation of the quenched spectrum from
experiment. It is an important issue to examine how dynamical quarks
affect light quark masses and resolve the systematic uncertainty of strange
quark mass. A recent attempt at a full QCD determination of light quark
masses~\cite{SESAMquark} was restricted to a single lattice spacing. We
extracted light quark masses through analyses of hadron mass data obtained
in the spectrum calculation. The main findings of our light quark mass
calculation have been presented in Ref.~\cite{QuarkLetter}. We give here a
more detailed report of the analysis and results.

Full QCD configurations generated in this work can be used to calculate a
large variety of physical quantities and examined for sea quark effects. We
have already pursued calculations of several quantities. Among these, the
flavor singlet meson spectrum and its relation with topology and $U(1)$
anomaly is of particular interest from the theoretical viewpoint, and
preliminary results have been published in
Ref.~\cite{CPPACS-topology}. Other calculations concern the prediction of
hadronic matrix elements important for phenomenological analyses of the
Standard Model. Results have been published for heavy quark quantities such
as $B$ and $D$ meson decay
constants~\cite{CPPACS-fB-clover,CPPACS-fB-NRQCD} as well as bottomonium
spectra~\cite{CPPACS-onia}. A report of the analysis of the light
pseudoscalar and vector meson decay constants is included in this article.

The outline of this paper is as follows. We first describe details of the
lattice action, the choice of simulation parameters and the algorithm for
configuration generation in Sec.~\ref{sec:simulations}. Measurements of
hadron masses, the static quark potential and a discussion of
autocorrelations are presented in Sec.~\ref{sec:measurements}. In
Sec.~\ref{sec:chiral} we discuss the procedure of chiral extrapolation.
Section~\ref{sec:spectrum} contains the main results for the full QCD light
hadron spectrum. In Sec.~\ref{sec:quenched} we then turn to a presentation
of quenched QCD simulations with improved actions. This sets the stage for
a discussion of sea quark effects which is contained in
Sec.~\ref{sec:effects}. Calculations of light quark masses are presented in
Sec.~\ref{sec:quarkmass}. Section~\ref{sec:decay} contains a discussion of
decay constants. Finally, we present our conclusions in
Sec.~\ref{sec:conclusions}.


\section{Simulation}
\label{sec:simulations}

\subsection{Choice of improved lattice action}
\label{sec:action}

Based on our preparatory study in Ref.~\cite{comparative} we choose
improved gauge and quark actions for full QCD configuration generation.
The improved gluon action has the form
\begin{equation}
S_g = \frac{\beta}{6}  \left\{\;c_0\sum_{x,\mu <\nu} W_{\mu\nu}^{1\times 1}(x) 
+c_1\sum_{x,\mu ,\nu} W_{\mu\nu}^{1\times 2}(x)\right\}.
\label{eq:6LinkAct} 
\end{equation}
The coefficient $c_1 = -0.331$ of the $1\times
2$ Wilson loop $W_{\mu\nu}^{1\times 2}$ is fixed by an 
approximate renormalization group analysis~\cite{RGIA},  
and $c_0 = 1-8c_1=3.648$ of the $1\times 1$ Wilson loop by the normalization 
condition, which defines the bare coupling $\beta=6/g^2$.
From the point of view of Symanzik improvement the leading scaling
violation of this action is $O(a^2)$, the same as for the standard action.  

For the quark part we use the clover quark action~\cite{clover} defined by  
\begin{eqnarray}
S_q & = & \sum_{x,y}\overline{q}_x D_{x,y}q_y,\\
D_{x,y} & = & \delta_{xy}
- \kappa \sum_\mu \left\{(1-\gamma_\mu)U_{x,\mu} \delta_{x+\hat\mu,y}
      + (1+\gamma_\mu)U_{x,\mu}^{\dag} \delta_{x,y+\hat\mu} \right\}
- \, \delta_{xy} c_{\rm SW} \kappa \sum_{\mu < \nu}
         \sigma_{\mu\nu} F_{\mu\nu},
\label{eq:clover}
\end{eqnarray}
where $\kappa$ is the usual hopping parameter and $F_{\mu\nu}$ the standard 
lattice discretization of the field strength. 

For the clover coefficient $c_{\rm SW}$
we adopt a mean field improved choice defined by
\begin{equation}
c_{\rm SW} = \left ( W^{1\times 1} \right )^{-3/4} 
= \left (1-0.8412 \beta^{-1} \right )^{-3/4},
\label{eq:csw} 
\end{equation}
where for the plaquette $W^{1\times 1}$ the value calculated in one-loop
perturbation theory~\cite{RGIA} is substituted. This choice is based on our
observation in Ref.~\cite{comparative} that the one-loop calculation
reproduces the measured values well. Indeed, an inspection of
Table~\ref{tab:wloop} in Appendix~\ref{app:Z} shows that $W^{1\times 1}$ in
the simulations agree with one-loop values with a difference of at most
8\%. The agreement for $c_{\rm SW}$ is not fortuitous; the one-loop value
for the RG gauge action (\ref{eq:6LinkAct}), which was
calculated~\cite{aoki} after the present work was started, equals $c_{\rm
SW}=1+0.678/\beta$, which differs from our choice $c_{\rm
SW}=1+0.631/\beta+\ldots$ only by a few per cent. We do not employ the
measured plaquette for the clover coefficient, as prescribed by the usual
mean field approximation, which would have required a time-consuming
self-consistent tuning. The leading scaling violation with our choice of
$c_{\rm SW}$ is $O(g^2a)$.

\subsection{Simulation parameters}
\label{sec:params}

The target of this work is a calculation of the two-flavor QCD light hadron
spectrum in the continuum limit and at physical quark masses. For this
purpose we carry out simulations at three lattice spacings in the range
$a\approx 0.2$--0.1~fm for continuum extrapolation, and at four sea quark
masses corresponding to $m_\pi/m_\rho\approx 0.8$--0.6 for chiral
extrapolation for each lattice spacing. The simulation parameters are
given in Table~\ref{tab:overview}.

We employ three lattices of size $12^3\times 24$, $16^3\times 32$ and
$24^3\times 48$ for our runs. The coupling constants $\beta=1.8, 1.95$ and
$2.1$ are chosen so that the physical lattice size remains approximately
constant at $La\approx 2.5$~fm. The resulting lattice spacings determined
from the $\rho$ meson mass equal $a = 0.2150(22)$, $0.1555(17)$ and
$0.1076(13)$~fm or $a^{-1} = 0.9177(92)$, $1.269(14)$ and $1.834(22)$~GeV.

We have also performed an initial run at $\beta=2.2$ on a $24^3\times 48$
lattice for which the lattice spacing turned out to be $a=0.087$~fm. The
physical lattice size $La=2.08$~fm is significantly smaller than the other
three lattices. In order to avoid different magnitude of possible finite
size effects, we do not include data from this run when we make
extrapolations to the continuum limit. They will be included in figures and
tables for completeness, however.

We carry out hadron mass analyses distinguishing the sea and valence quark
hopping parameters $\kappa_{\rm sea}$ and $\kappa_{\rm val}$. At each value
of $\beta$, configurations are generated at four sea quark hopping
parameters $\kappa_{\rm sea}$ such that the mass ratio of pseudoscalar to
vector mesons made of sea quarks takes $\mps/\mv \approx 0.8$, 0.75, 0.7
and 0.6. At each sea quark mass, hadron propagators are measured for five
valence hopping parameters $\kappa_{\rm val}$ with approximate ratios of
$\mps/\mv \approx 0.8$, 0.75, 0.7, 0.6 and 0.5. The four heavier
$\kappa_{\rm val}$ coincide with those chosen for sea quarks.

\begin{figure*}[p]
\vspace{3mm}
\centerline{
\epsfxsize=13cm \epsfbox{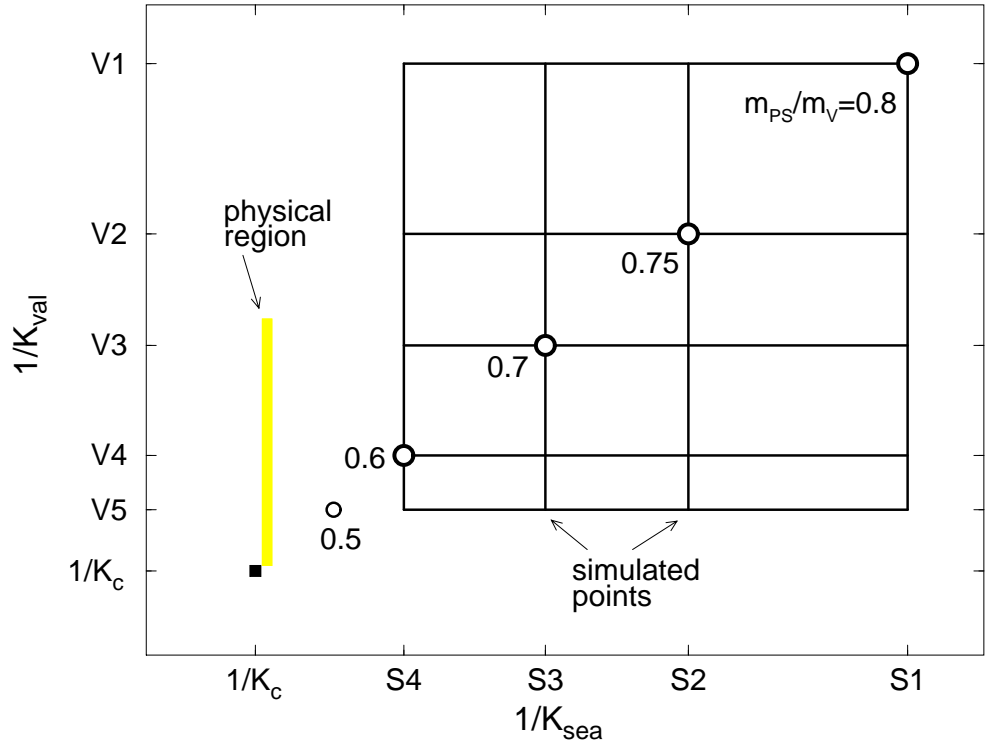}
}
\caption{Schematic plot for the choice of sea and valence hopping
parameters. For circles at the points $\kappa_{\rm val}=\kappa_{\rm sea}$ the
corresponding pseudoscalar to vector meson mass ratio is
indicated.}
\label{fig:KseaKval}
\vspace{-4mm}
\end{figure*}

A schematic representation of our choice on the $(1/\kappa_{\rm sea},
1/\kappa_{\rm val})$ plane is shown in Fig.~\ref{fig:KseaKval}. The
physical point is characterized by $1/\kappa_{\rm sea} = 1/\kappa_{\rm val}
= 1/\kappa_{ud}$ for degenerate up and down quarks, and $1/\kappa_{\rm sea}
= 1/\kappa_{ud}$ and $1/\kappa_{\rm val} = 1/\kappa_{\rm strange}$ for
strange quark, {\it i.e.}, lying in the shaded region on the left hand side
of the diagram. The additional points with $1/\kappa_{\rm val} = V5$ in the
bottom part of the diagram are not directly needed in exploring the
physical region. As we will see in Sec.~\ref{sec:chiral}, they help in
the description of hadron masses as a combined function of sea {\it and}
valence quark masses and are therefore indirectly useful for the
extrapolation to physical points. Including them also keeps the possibility
open for a future extension of the present work towards the chiral limit by
adding the fifth sea quark and completing the grid of
Fig.~\ref{fig:KseaKval}.

Our choice of hopping parameters enables us to obtain the full strange and
non-strange hadron spectrum in a sea of degenerate up and down quarks. If
we denote with $S$ a valence quark with $\kappa_{\rm val} = \kappa_{\rm
sea}$ and with $V$ a valence quark with $\kappa_{\rm val} \neq \kappa_{\rm
sea}$, we obtain mesons of the form $SS$, $SV$ and $VV$ and baryons of the
form $SSS$, $SSV$, $SVV$ and $VVV$.

\subsection{Configuration generation}
\label{sec:generation}

Configurations are generated for two flavors of degenerate quarks with the
Hybrid Monte Carlo (HMC) algorithm. In Table~\ref{tab:param} we give
details of the parameters and statistics of the runs. At the main coupling
constants $\beta=1.8$--2.1, runs are made with a length of 4000--7000 HMC
unit-trajectories per sea quark mass. The additional runs at $\beta=2.2$
are stopped at 2000 HMC trajectories per sea quark mass for the reason
described in Sec.\ref{sec:params}.

To speed up the calculation we have implemented several improvements in our
code. For the inversion of the quark matrix during the HMC update we use
the even/odd preconditioned BiCGStab algorithm~\cite{BiCGStab}. Test runs
confirmed that the performance of this algorithm is better than that of the
MR algorithm and that the advantage increases toward lighter quark
masses~\cite{Frommer}. In a test run at $m_\pi/m_\rho\approx 0.7$ we
observed a 43\% gain in computer time for the same accuracy of inversion
compared to the MR algorithm.

As a stopping condition for the inversion of the equation 
$D(\kappa )G=B$ during the fermionic force evaluation we use the criterion
\begin{equation}
||DG-B||^2  \leq {\em stop},
\end{equation}
with values of ${\em stop}$ given in Table~\ref{tab:param} where we also
give the number of iterations necessary for the inversion. For the
evaluation of the Hamiltonian we use a stricter stopping condition which is
smaller by a factor of $10^8$ than the one used for the force
evaluation. With these stopping conditions, the Hamiltonian is evaluated
with a relative error of less than $10^{-10}$. We have also checked that
the reversibility over trajectories of unit length is satisfied to a
relative level better than $10^{-7}$ for the gluon link variables.

Another improvement concerns the scheme for the integration of molecular
dynamics equations. For our runs we have used the following three schemes.

a) the standard leap-frog integration scheme: The operator to evolve gauge
fields and conjugate momenta by a step $\Delta\tau$ in fictitious time can
be written in the form
\begin{equation}
T_P \left (\frac{1}{2}\Delta\tau \right ) T_Q(\Delta\tau) 
T_P \left (\frac{1}{2}\Delta\tau \right ),
\end{equation}
where the operator $T_P(\Delta\tau) = \exp(\Delta\tau \sum_i
p_i\partial_i)$ moves the gauge field $U$ by a step $\Delta\tau$, whereas
the operator
$T_Q(\Delta\tau)=\exp(-\Delta\tau\sum_i\partial_iS(U,\Phi)\partial/\partial
p_i)$ moves the conjugate momenta $p$ by a step $\Delta\tau$. The leap-frog
integrator has an error of $O(\Delta\tau^3)$ for a single step and of
$O(\Delta\tau^2)$ for a unit-trajectory.

b) an improved scheme: The discretization error of the leap-frog
integration scheme can be reduced by using an improved scheme. The simplest
improvement has the form,
\begin{equation}
T_P \left (\frac{b}{2}\Delta\tau \right ) 
T_Q \left (\frac{\Delta\tau}{2} \right )
T_P \left ((1-b)\Delta\tau \right )
T_Q \left (\frac{\Delta\tau}{2} \right ) 
T_P \left (\frac{b}{2}\Delta\tau \right ). 
\label{eq:molecimp}
\end{equation}
This scheme has errors of the same order as the standard leap-frog scheme
but the main contribution to the error is removed by the choice $b =
(3-\sqrt{3})/3$~\cite{SexWein}. Test runs have shown that $\Delta\tau$ can
be taken a factor 3 larger than for leap-frog without losing the acceptance
rate for the heaviest sea quark. This leads to a gain of about 30\% in
computer time. The gain, however, decreases toward lighter quark masses,
and the computer time required for the improved scheme at the lightest
quark mass is roughly the same as for the standard leap-frog scheme (see
parameters of the run at $\beta=1.8$ and $\kappa=0.1464$ in
Table~\ref{tab:param}).

c) Sexton-Weingarten scheme~\cite{SexWein}: In this scheme the evolution
with the gauge field force $\sum_i\partial_iS_g(U)$ is made with an $n$
times smaller time step than that with the fermionic force
$\sum_i\partial_iS_f(U,\Phi)$ according to
\begin{equation}
\left [ T_1 \left (\frac{\Delta\tau}{2n} \right ) \right ]^n 
T_2(\Delta\tau)
\left [ T_1 \left (\frac{\Delta\tau}{2n} \right ) \right ]^n, 
\label{eq:sw1}
\end{equation}
where
\begin{eqnarray}
T_1 \left (\Delta\tau \right ) 
& = & T_P \left (\frac{1}{2}\Delta\tau \right )
  \exp \left (-\Delta\tau\sum_i\partial_iS_g(U)\partial/\partial p_i \right ) 
        T_P \left (\frac{1}{2}\Delta\tau \right ), \label{eq:sw2} \\
T_2 \left (\Delta\tau \right ) 
& = & \exp \left (-\Delta\tau\sum_i\partial_iS_f(U,\Phi)\partial/\partial p_i
             \right ).
\end{eqnarray}
We have implemented a scheme for which both Eq.~(\ref{eq:sw1}) and
Eq.~(\ref{eq:sw2}) are improved as in Eq.~(\ref{eq:molecimp}). For $n=2$
the time step $\Delta\tau$ can be chosen 10\% larger than in scheme b)
while maintaining a similar acceptance. However, this improvement is offset
by an increase of a factor 4 in the number of operations for the gauge
field force. This leads to an increase of 30\% in the total number of
operations at $\beta=1.8$, $\kappa=0.1445$. Hence the performance of scheme
c) is similar to the leap-frog scheme, as can be seen in
Table~\ref{tab:param}.

The scheme employed for each run is listed in Table~\ref{tab:param}. After
some trials on the smaller lattices ($12^3$ and $16^3$) we found the scheme
b) to be most practical, and we used it for all the runs on the larger
$24^3$ lattices. The step size $\Delta\tau$ for molecular dynamics has been
chosen so that the acceptance ratio turns out to be 70--80\%.

Light hadron propagators are measured simultaneously with the configuration
generation with a separation of 5 HMC trajectories. The number of
measurements is given in Table~\ref{tab:param}. We stored configurations
with a separation of 10 HMC trajectories (at $\beta=1.8$ and 1.95) or 5 HMC
trajectories (at $\beta=2.1$ and 2.2) on tapes for later measurement of
other observables such as the topological charge and flavor singlet meson
mass~\cite{CPPACS-topology}, quarkonium spectra~\cite{CPPACS-onia} and the
$B$ meson decay constant~\cite{CPPACS-fB-clover,CPPACS-fB-NRQCD}.

In the last column of Table~\ref{tab:param}, we list the number of
configurations removed by hand because of the occurrence of exceptional
propagators. We did not encounter exceptional configurations in full QCD
where $\kappa_{\rm val}=\kappa_{\rm sea}$. However, strange behavior of
propagators did occur for the lightest valence quark mass for some
configurations. We have removed all the propagators obtained on such
configurations in order to allow a jack-knife error analysis.

Our criterion for removal of a configuration is a deviation of hadron
propagator by more than 10 standard deviations from the ensemble average
for at least one channel and at least one timeslice. The fraction of
removed configurations drops from 1.2\% at $\beta=1.8$ to 0.1\% at
$\beta=2.1$. No configurations needed to be removed at $\beta=2.2$.

\subsection{Coding and runs on the CP-PACS computer}

We have spent much effort in optimizing the double precision codes for
configuration generation on the CP-PACS computer as described in
Ref.~\cite{CP-PACS-perform}. Actual runs took advantage of the
partitioning capability of the CP-PACS, using 64 PU (processing units), 256
PU and 512 PU for the lattice size $12^3\times 24$, $16^3\times 32$ and
$24^3\times 48$, and executing runs at different values of $\kappa_{\rm
sea}$ at the same time. For some of the runs at smaller quark masses, which
need longer execution times, we made two or more independent parallel runs
which are combined for the purposes of measurements.

The CPU time needed per trajectory is listed in Table~\ref{tab:param}.
Converted to the number of days with the full use of the CP-PACS computer,
the configuration generation took 10 days for $\beta=1.8$ on a $12^3\times
24$ lattice, 40 days at $\beta=1.95$ on a $16^3\times 32$ lattice, 186 days
at $\beta=2.1$ on a $24^3\times 48$ lattice and 82 days on the same size
lattice at $\beta=2.2$. Adding 3+12+46+23 days for measurements of
observables and 1+3+6+3 days for I/O loss, the entire CPU time spent for
the simulations equals 415 days of the full operation of the CP-PACS
computer.


\section{Measurements}
\label{sec:measurements}

\subsection{Hadron masses}
\label{sec:massmeas}

We employ meson operators defined by
\begin{equation}
M_A^{fg}(n)=\overline{f}_n\Gamma_A g_n,  \label{eq:mesonop}
\end{equation}
where $f$ and $g$ are quark fields with flavor indices $f$ and $g$, and
$\Gamma_A$ represents one of the 16 spin matrices $\Gamma_A=I$,
$\gamma_5$, $i\gamma_{\mu}\gamma_5$, $\gamma_{\mu}$ and
$i[\gamma_{\mu},\gamma_{\nu}]/2$ of the Dirac algebra. Using these operators, 
meson propagators are calculated as
\begin{equation}
\langle M_A(n)M_A(0)\rangle. 
\end{equation}
For the operator of octet baryons with spin $J=1/2$ we use the definition
\begin{equation}
O^{fgh}_\alpha(n)=\epsilon^{abc}(f_n^{Ta}C\gamma_5g_n^b)h^c_{n\alpha}, 
\label{eq:octetop}
\end{equation}
where $a,b,c$ are color indices, $C=\gamma_4\gamma_2$ is the charge
conjugation matrix and $\alpha=1,2$ represents the $z$-component of the
spin $J_z=\pm 1/2$. To distinguish $\Sigma$ and $\Lambda$-like octet
baryons we antisymmetrize flavor indices, written symbolically as
\begin{eqnarray}
\Sigma&=&-\frac{[fh]g+[gh]f}{\sqrt{2}}, \\
\Lambda&=&\frac{ [fh]g-[gh]f-2[fg]h}{\sqrt{6}},
\end{eqnarray}
where $[fg]=fg-gf$.

The operator of decuplet baryons with spin $J=3/2$ is given by
\begin{equation}
D^{fgh}_{\mu,\alpha}(n) = 
\epsilon^{abc}(f_n^{Ta}C\gamma_\mu g_n^b)h^c_{n\alpha}. 
\label{eq:decupletop}
\end{equation}
Writing out the spin structure $(\mu,\alpha)$ explicitly, we employ
operators for the four $z$-components of the
spin $J_z=\pm 3/2$,~$\pm 1/2$ defined as
\begin{eqnarray}
 D_{3/2}  = && \epsilon^{abc}(f^{Ta}C\Gamma_+g^b)h^c_1, \\
 D_{1/2}  = && \epsilon^{abc}[(f^{Ta}C\Gamma_0g^b)h^c_1 
             - (f^{Ta}C\Gamma_+g^b)h^c_2]/3, \\
 D_{-1/2} = && \epsilon^{abc}[(f^{Ta}C\Gamma_0g^b)h^c_2 
             - (f^{Ta}C\Gamma_-g^b)h^c_1]/3, \\
 D_{-3/2} = && \epsilon^{abc}(f^{Ta}C\Gamma_-g^b)h^c_2,
\end{eqnarray}
where $\Gamma_{\pm} =(\gamma_1 \mp i\gamma_2)/2$ and $\Gamma_0 = \gamma_3$.

Using operators defined as above, we calculate 8 baryon propagators given by
\begin{eqnarray}
&&\langle \Sigma_\alpha(n)\overline{\Sigma}_\alpha(0)\rangle, 
\quad \alpha=1,2, \\
&&\langle \Lambda_\alpha(n)\overline{\Lambda}_\alpha(0)\rangle, 
\quad \alpha=1,2, \\
&&\langle D_S(n)\overline{D}_S(0)\rangle, \quad S=3/2,1/2,-1/2,-3/2,
\end{eqnarray}
together with 8 antibaryon propagators similarly defined.

We average zero momentum hadron propagators over three polarization states
for vector mesons, two spin states for octet baryons and four spin states
for decuplet baryons. We also average the propagators for the particles
with the ones for the corresponding antiparticles.

For each configuration quark propagators are calculated with a point source
and a smeared source. For the smeared source we fix the gauge configuration
to the Coulomb gauge and use an exponential smearing function $\psi(r) =
A\exp(-Br)$ for $r>0$ with $\psi(0)=1$. We chose $A$ and $B$ based on
experiences from previous quenched measurements of the pion wave
function~\cite{JLQCD-smearing} and from our preparatory full QCD
study~\cite{comparative} and readjusted them by hand so that hadron
effective masses reach a plateau as soon as possible on average. The values
of $A$ and $B$ are given in Table~\ref{tab:smearing}.

\begin{figure*}[p]
\vspace{3mm}
\centerline{
\epsfxsize=11cm \epsfbox{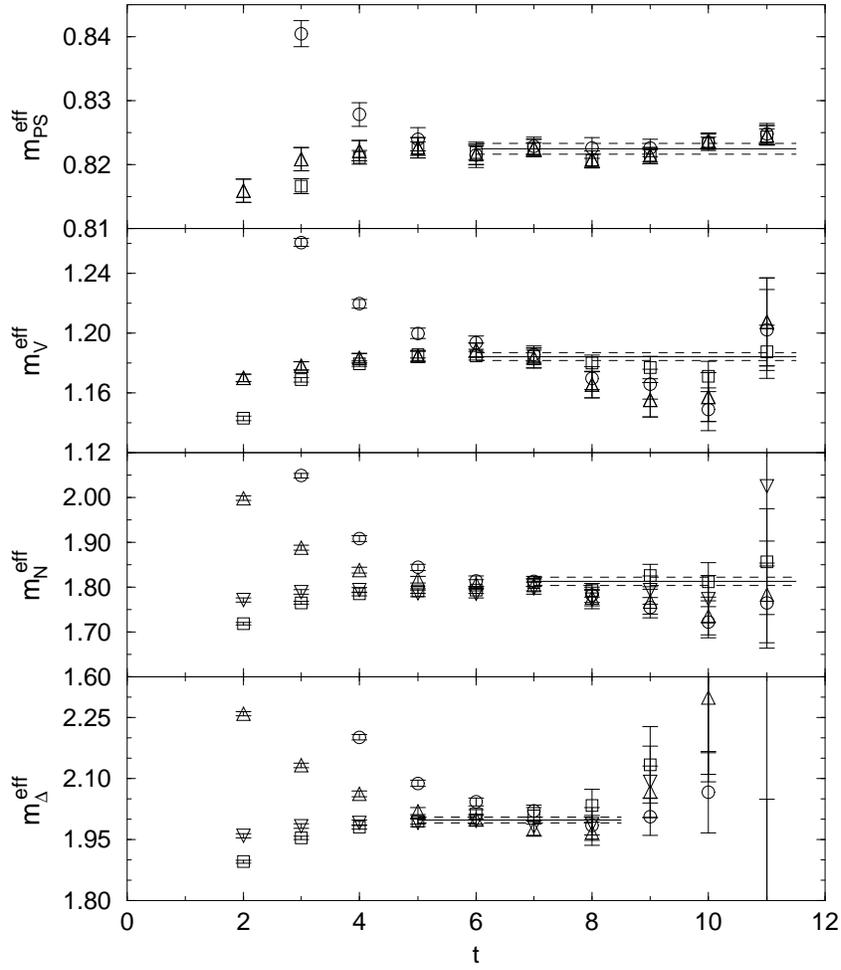}
}
\vspace{5mm}
\caption{Effective mass plots for pseudoscalar, vector, nucleon and
$\Delta$ channels with degenerate valence hopping parameters $\kappa_{\rm
val}=\kappa_{\rm sea}=0.1445$ at $\beta=1.8$. Circles represent effective
masses obtained when all quark propagators are calculated with point
sources. For squares all quark propagators have smeared sources and
triangles are for mixed combinations of sources. Solid lines denote the
results from correlated mass fits to smeared source hadron
propagators. Dashed lines show the one standard deviation error band
determined by jack-knife analysis with a bin size of 10 configurations.}
\label{fig:effmassB18}
\vspace{-4mm}
\end{figure*}

\begin{figure*}[p]
\vspace{3mm}
\centerline{
\epsfxsize=11cm \epsfbox{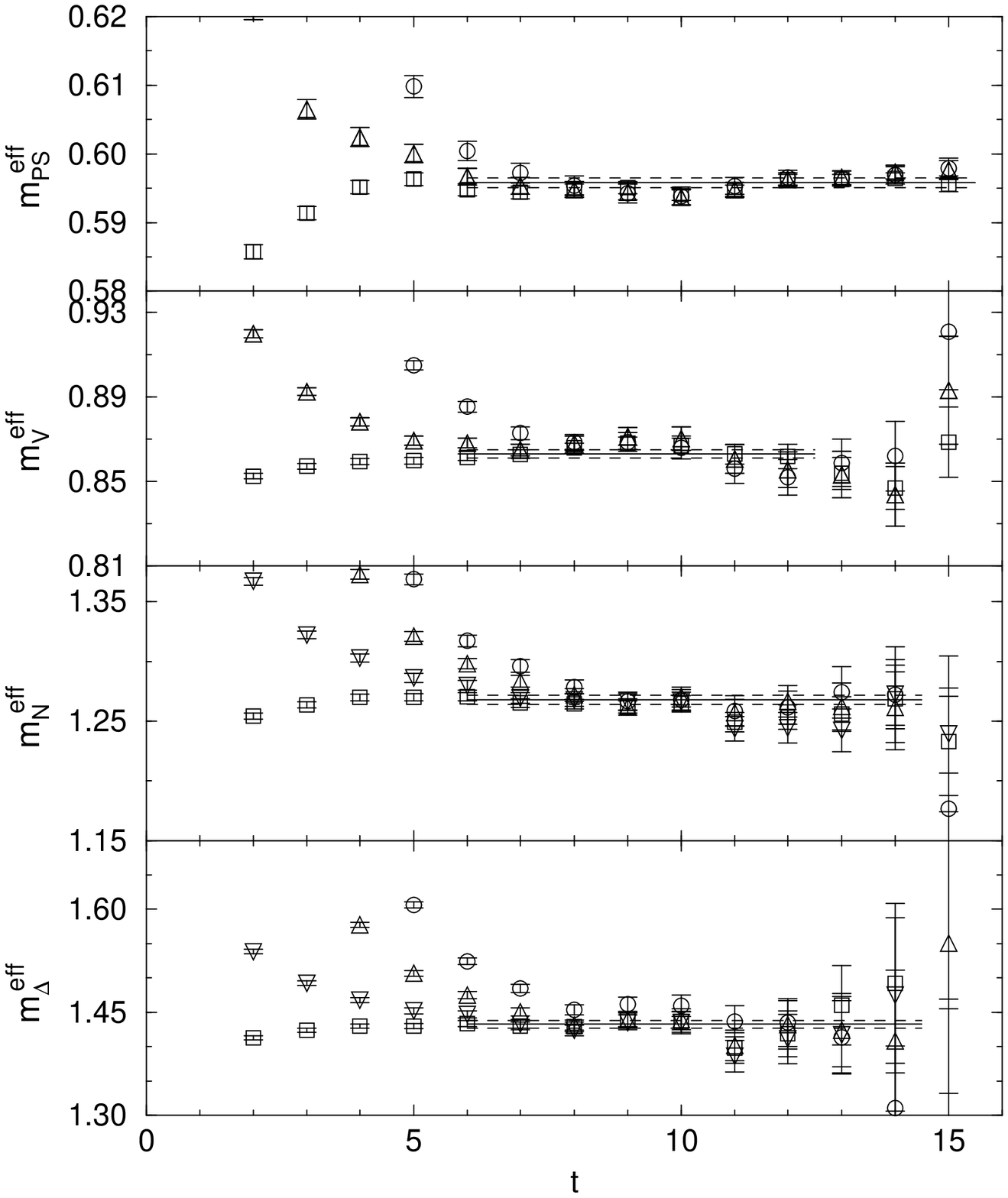}
}
\vspace{5mm}
\caption{Effective mass plots for pseudoscalar, vector, nucleon and
$\Delta$ channels with degenerate valence hopping parameters
$\kappa_{\rm val}=\kappa_{\rm sea}=0.1400$ at $\beta=1.95$. 
Symbols have the same meaning as in Fig.~\ref{fig:effmassB18}.}
\label{fig:effmassB195}
\vspace{-4mm}
\end{figure*}

\begin{figure*}[p]
\vspace{3mm}
\centerline{
\epsfxsize=11cm \epsfbox{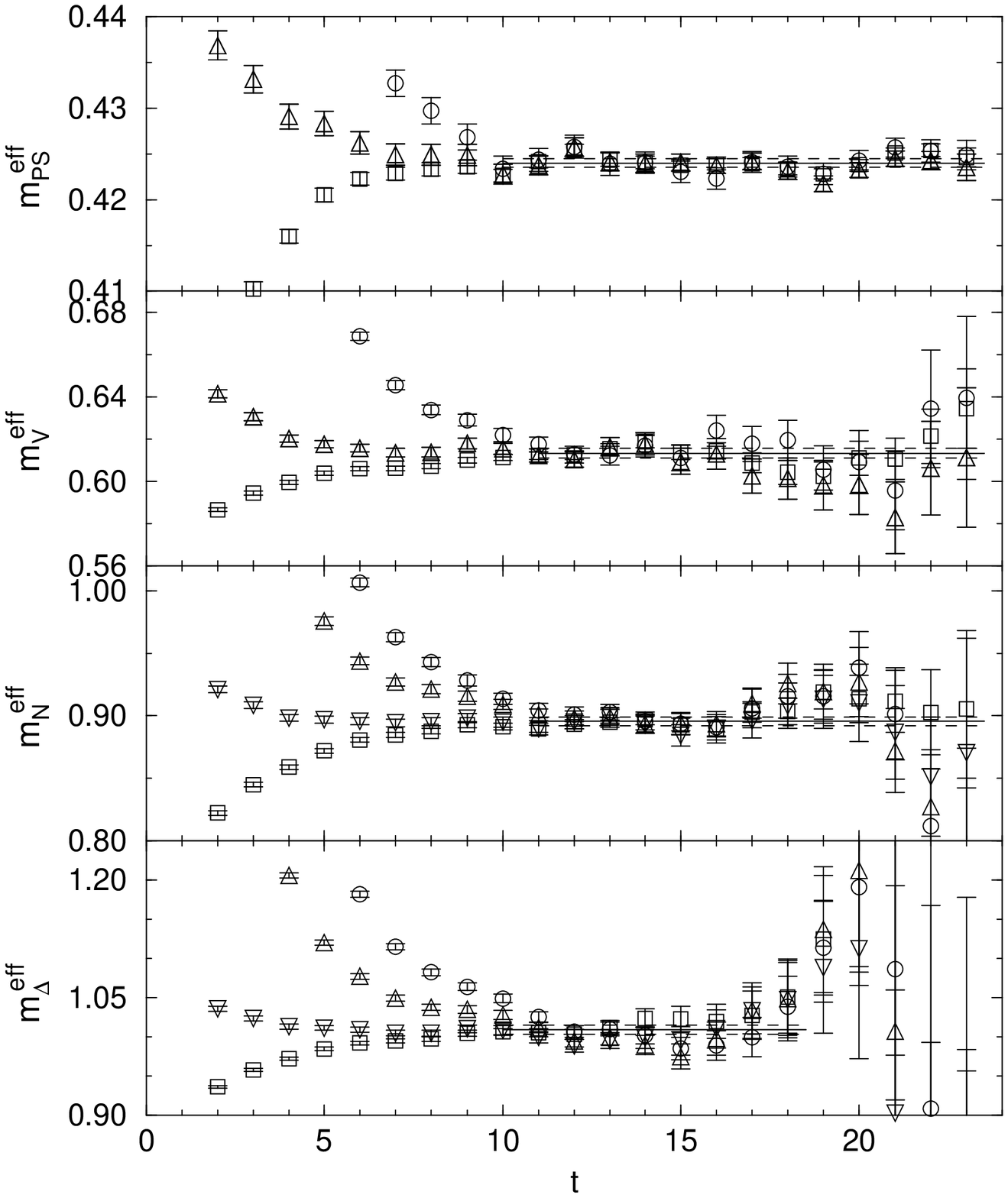}
}
\vspace{5mm}
\caption{Effective mass plots for pseudoscalar, vector, nucleon and
$\Delta$ channels with degenerate valence hopping parameters
$\kappa_{\rm val}=\kappa_{\rm sea}=0.1374$ at $\beta=2.1$. 
Symbols have the same meaning as in Fig.~\ref{fig:effmassB18}.}
\label{fig:effmassB21}
\vspace{-4mm}
\end{figure*}

In Figs.~\ref{fig:effmassB18}--\ref{fig:effmassB21} we show examples of
effective mass plots for hadron propagators with degenerate valence quarks
equal to the sea quark. Effective masses from hadron propagators where all
the quark propagators have been calculated with smeared sources have the
smallest statistical errors and exhibit good plateaux starting at smaller
values of $t$ than those containing point sources. We therefore use smeared
propagators for hadron mass fits.

Fit ranges $[t_{\rm min},t_{\rm max}]$ are determined by inspecting
effective mass plots. As a general guideline, we choose $t_{\rm min}$ to be
in the same range for given particle and gauge coupling. The approach to a
plateau changes with the quark mass and this allows for a variation of
$t_{\rm min}$. To be confident that contributions of excited states die out
at $t_{\rm min}$ we also consult effective masses from propagators with
point and mixed sources. The upper end of the fit range, $t_{\rm max}$, is
chosen to extend as far as the effective mass exhibits a plateau and the
signal is not lost in the noise.

Hadron masses are derived from correlated fits to propagators with
correlations among different time slices taken into account. We assume a
single hyperbolic cosine for mesons and a single exponential for
baryons. With a statistics of 4000--7000 HMC trajectories (corresponding to
80--140 binned configurations, see Sec.~\ref{sec:autocor}) for hadron
propagators at $\beta=1.8$, 1.95 and 2.1, the covariance matrix is
determined well. Typically, the errors of eigenvalues of the covariance
matrix are around 15\%, and fits have a $\chi^2/N_{\rm DF}$ around 1 and at
most 3. For $\beta=2.2$, however, where fewer configurations are available,
eigenvalues of the covariance matrix have typical errors of 30\%, and the
correlated fits are less stable. For all the cases we also made uncorrelated
fits and checked that masses are consistent within error bars.

Errors in hadron masses and in $\chi^2/N_{\rm DF}$ are estimated with the
jack-knife procedure with a bin size of 10 configurations or 50 HMC
trajectories. A discussion of the choice of this bin size will follow in
Sec.~\ref{sec:autocor}.

Resulting hadron masses are collected in Appendix~\ref{app:masses}. There and
in the following, lower case symbols are used for observables in lattice
units, for which the lattice spacing $a$ is not explicitly written.

\subsection{Quark mass}
\label{sec:quarkmass-def}

Another quantity which can be obtained from meson correlation functions is
the quark mass based on the axial vector Ward identity
(AWI)~\cite{Bochicchio,Itoh}. It is defined from matrix elements of the
pseudoscalar density $P$ and the fourth component of the axial vector
current $A_4$ by the expression
\begin{equation}
m^{\rm AWI} 
=  \frac{\langle 0 | \nabla_4 A^{\rm imp}_4 | {\rm PS} \rangle}
        {2\langle  0 | P | {\rm PS} \rangle}, 
\label{eq:awi-mass}
\end{equation}
where we employ the improved axial vector current $A^{\rm
imp}_4=A_4+c_A\tilde{\partial}_{4}P$ with $c_A$ calculated in one-loop
perturbation theory and $\tilde{\partial}_{\mu}$ representing the symmetric 
lattice derivative (see Appendix~\ref{app:Z}). 

In practice we extract the AWI quark mass from single-exponential fits to
meson correlators. For the analysis of pseudoscalar masses we assume the form
\begin{equation}
\langle P(t) P(0) \rangle = C_P 
\left [ \exp(-\mps t) + \exp(-\mps (L_t-t) ) \right ],
\end{equation}
which has already been described above. Keeping the value of $\mps$ obtained
from this fit, we make an additional fit to the correlator
\begin{equation}
\langle A^{\rm imp}_4(t) P(0) \rangle = C_A 
\left [ \exp(-\mps t) - \exp(-\mps (L_t-t) \right ],
\end{equation}
where $C_A$ is the only fit parameter. The AWI bare quark mass before
renormalization is then obtained through 
\begin{equation}
m^{\rm AWI} =  \frac{\mps C_A}{2 C_P}.
\end{equation}
Results for $m^{\rm AWI}$ are given in Appendix~\ref{app:masses}.

\subsection{Static quark potential}
\label{sec:potmeas}

\begin{figure*}[p]
\vspace*{1cm}
\centerline{
\epsfxsize=8cm \epsfbox{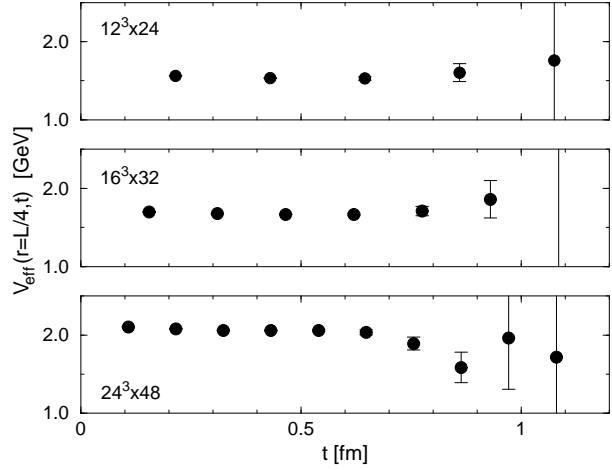}
}
\caption{Effective mass plots of potential data at $r = L/4$ for sea quark
mass corresponding to $m_{\rm PS}/m_{\rm V} \approx 0.7$. The scale is
fixed from $\rho$ meson mass at the physical point.}
\label{fig:pot-em}
\end{figure*}

We measure the temporal Wilson loops applying the smearing procedure of
Ref.~\cite{Wsmearing}. The number of smearing steps is fixed to 2, 4 and 10
on $12^3 \times 24$, $16^3 \times 32$, and $24^3 \times 48$ lattices,
respectively, which we find sufficient to ensure a good overlap of Wilson
loops onto the ground state. The static quark potential $V(r)$ is
determined from a correlated fit of the form
\begin{equation}
W(r,t) = C(r) \exp \left( - V(r) t \right).
\label{eq:wilfit}
\end{equation}
As shown in Fig.~\ref{fig:pot-em} noise dominates the signal when the
temporal size of $W(r,t)$ exceeds $t \approx$~0.9~fm. We therefore take fit
ranges, listed in Table~\ref{tab:pot-range}, which approximately correspond
to $t \approx$~0.45--0.90~fm at $\beta=1.8$, 1.95 and 2.1. At $\beta=2.2$,
we use the same fit ranges as those taken at $\beta=2.1$.

\begin{figure*}[p]
\centerline{
\epsfxsize=8cm \epsfbox{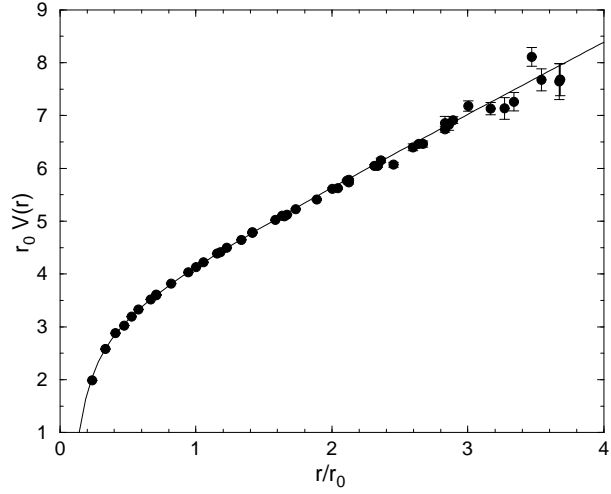}
}
\caption{Static quark potential on $24^3 \times 48$ lattice at $\kappa_{\rm
sea} =$~0.1374. Both vertical and horizontal lines are normalized by the
Sommer scale $r_0$. The solid line represents the fit curve of
Eq.~(\ref{eq:potfit}).}
\label{fig:VvsR}
\end{figure*}

A typical result for $V(r)$ is plotted in Fig.~\ref{fig:VvsR}. Since we do
not observe signs of string breaking, we parametrize $V(r)$ in the form,
\begin{equation}
V(r) = V_0 - \frac{\alpha}{r} + \sigma r.
\label{eq:potfit}
\end{equation}
The lattice correction to the Coulomb term calculated from one lattice
gluon exchange diagram~\cite{1LGXchg} is not included since breaking of
rotational symmetry is found to be small with the improved actions we
employ~\cite{comparative}.

The Sommer scale $r_0$ is defined through~\cite{SommerScale} 
\begin{equation}
r_0^2 \left . \frac{dV(r)}{dr} \right |_{r=r_0} = 1.65 .
\end{equation}
Using the fit parameters in Eq.~(\ref{eq:potfit}), 
$r_0$ can be obtained from
\begin{equation}
r_0 = \sqrt{\frac{1.65 - \alpha}{\sigma}}.
\end{equation}

We fit potential data to Eq.~(\ref{eq:potfit}) and determine $r_0$ for
several fitting ranges lying in the interval $[R_{\rm min}, R_{\rm
max}]$. Values of $R_{\rm min}$ and $R_{\rm max}$ are listed in
Table~\ref{tab:pot-range}. We take the average of fit results as central
values for $V_0, \alpha, \sigma$ and $r_0$, and use the standard deviation
as an estimate of the systematic error. Results of $\sigma$ and $r_0$ are
summarized in Table~\ref{tab:sig}.

\subsection{Autocorrelations}
\label{sec:autocor}

The autocorrelation function of a time series of a variable $f$ 
is defined as 
\begin{equation}
\rho_f(t) = \frac{C_f(t)}{C_f(0)}, 
\end{equation}
where the unnormalized autocorrelation function is given by 
\begin{equation}
C_f(t) = \langle f_s f_{s+t} \rangle - \langle f_s \rangle^2.
\end{equation}
The quantity relevant for the determination of the statistical error of $f$ 
is the integrated autocorrelation time $\tau^{\rm int}_f$, defined as
\begin{equation}
\tau^{\rm int}_f = \frac{1}{2} \sum_{t=-\infty}^{\infty} \rho_f(t) =
\frac{1}{2} + \sum_{t=1}^{\infty} \rho_f(t).
\label{eq:tauint}
\end{equation}
The naive error estimate is smaller than the true error by a factor of
$\sqrt{2\tau^{\rm int}_f}$. In numerical estimations of $\tau^{\rm
int}_f$, the sum in Eq.~(\ref{eq:tauint}) has to be cut off. It has been
found to be practical~\cite{sokal} to calculate the sum self-consistently
up to $t\approx (4$--$10) \tau^{\rm int}_f$. A convenient quantity for this
purpose is the cumulative autocorrelation time
\begin{equation}
\tau^{\rm cum}_f(t) = \frac{1}{2} + \sum_{s=1}^{t} \rho_f(s),
\label{eq:taucum}
\end{equation}
which should run into a plateau for $\tau^{\rm cum}_f(t) \approx
t/4$--$t/10$.

We calculate autocorrelation times for three different quantities:
\begin{enumerate}
\item The gauge action $c_0 W^{1\times1} + c_1 W^{1\times2}$.
Measurements are made after every HMC trajectory.
\item
The number of iterations $N_{\rm inv}$ for the inversion of the Dirac
matrix during the HMC update. Since this quantity is governed by the ratio
of the largest to the smallest eigenvalue of the Dirac matrix, it is
expected to be the quantity which takes the longest simulation time to
decorrelate. Measurements are made during every HMC trajectory.
\item
The effective pion mass $m_{\pi,{\rm eff}}$ measured at the onset of
a plateau. Measurements are made only after every $5^{\rm th}$ HMC trajectory.
\end{enumerate}

\begin{figure*}[p]
\centerline{
              \epsfxsize=8cm \epsfbox{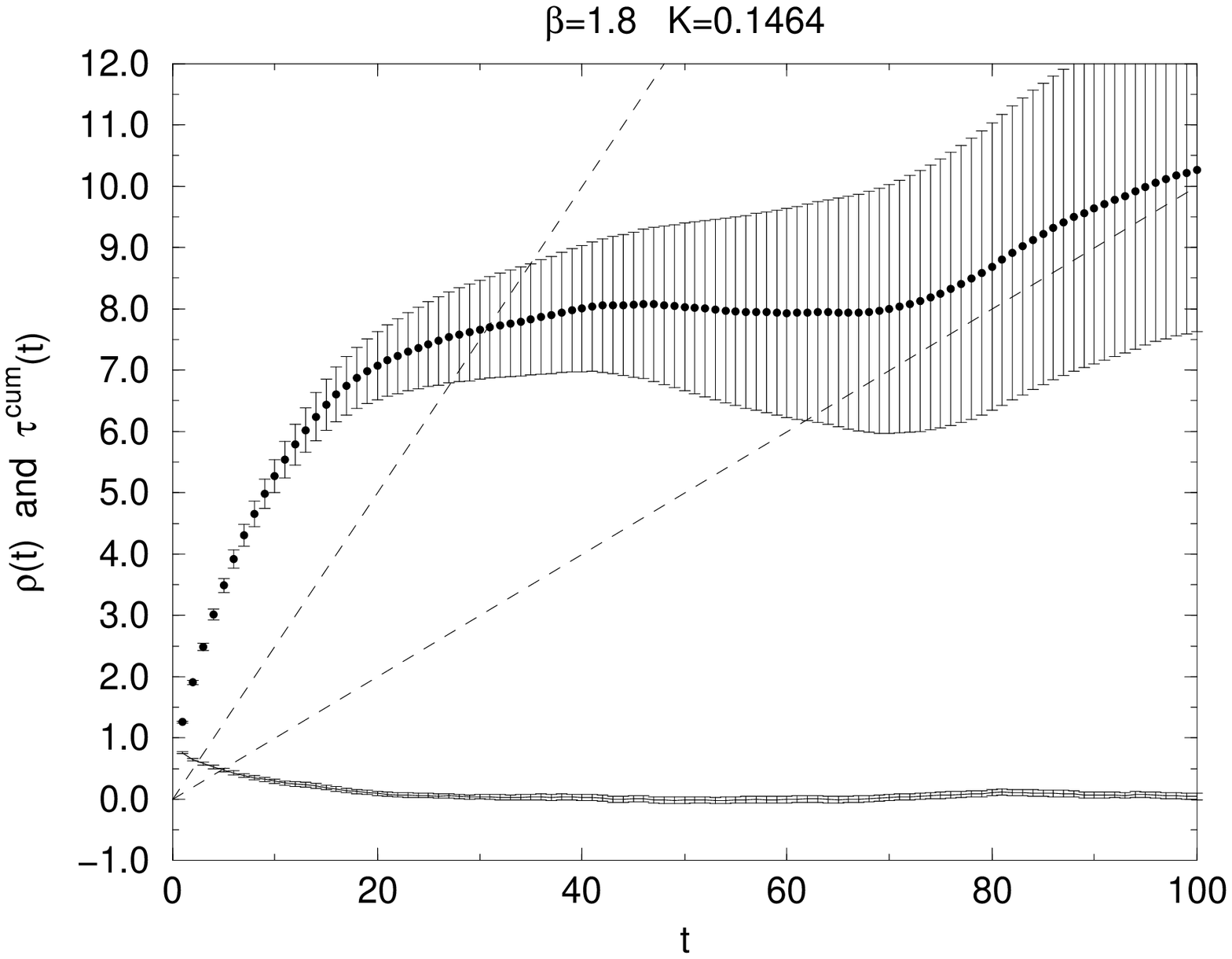}
\hspace{-8mm} \epsfxsize=8cm \epsfbox{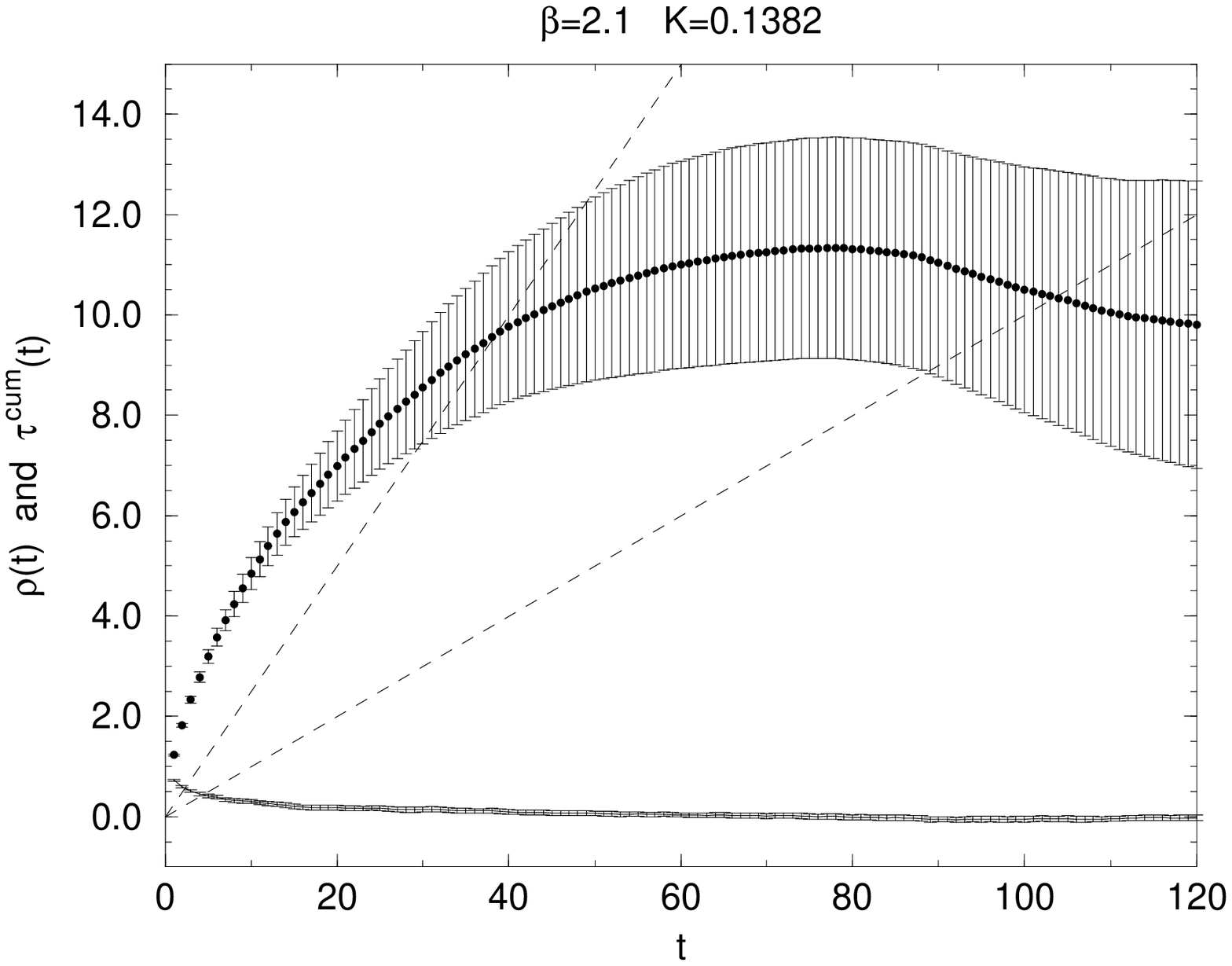}
}
\caption{Two examples of autocorrelation function (lower symbols) and
cumulative autocorrelation time (upper symbols) for $N_{\rm inv}$. Errors
are determined with the jack-knife method. Also plotted are two lines
$y(t)=t/4$ and $y(t)=t/10$ within which a plateau of $\tau^{\rm cum}(t)$ can
be observed.}
\label{fig:autocor}
\end{figure*}

Two examples for autocorrelation function and cumulative autocorrelation
time are shown in Fig.~\ref{fig:autocor}. The cumulative autocorrelation
time shows a plateau around the expected region from which we estimate the
integrated autocorrelation times $\tau_f^{\rm int}$ given in
Table~\ref{tab:autocor}.

Values of $\tau_f^{\rm int}$ are generally below 10 HMC trajectories for the
runs at $\beta \leq 2.10$. These numbers are significantly lower than
initial estimates for the HMC algorithm~\cite{LosAlamosFull} and also lower
than estimates from recent simulations with the Wilson or clover fermion
action~\cite{UKQCD-IMPfull,SESAM-autocor}. A possible reason might be
coarser lattice spacings of our simulations compared to the studies
mentioned above. It has also been noticed in Ref.~\cite{UKQCD-IMPfull}
that autocorrelations appear to be weaker on larger lattices. Our lattice
sizes in physical units are considerably larger than the ones in
Refs.~\cite{UKQCD-IMPfull,SESAM-autocor}.

Another point of interest is the size of increase of the autocorrelation
time with decreasing sea quark mass. For the gauge action and for $N_{\rm
inv}$ the autocorrelation time grows by about a factor of two in the range
of simulated sea quark masses, whereas for the effective pion mass the
situation is less clear. These observations are roughly consistent with the
findings in Refs.~\cite{UKQCD-IMPfull,SESAM-autocor}.

\begin{figure*}[p]
\centerline{
              \epsfxsize=7cm \epsfbox{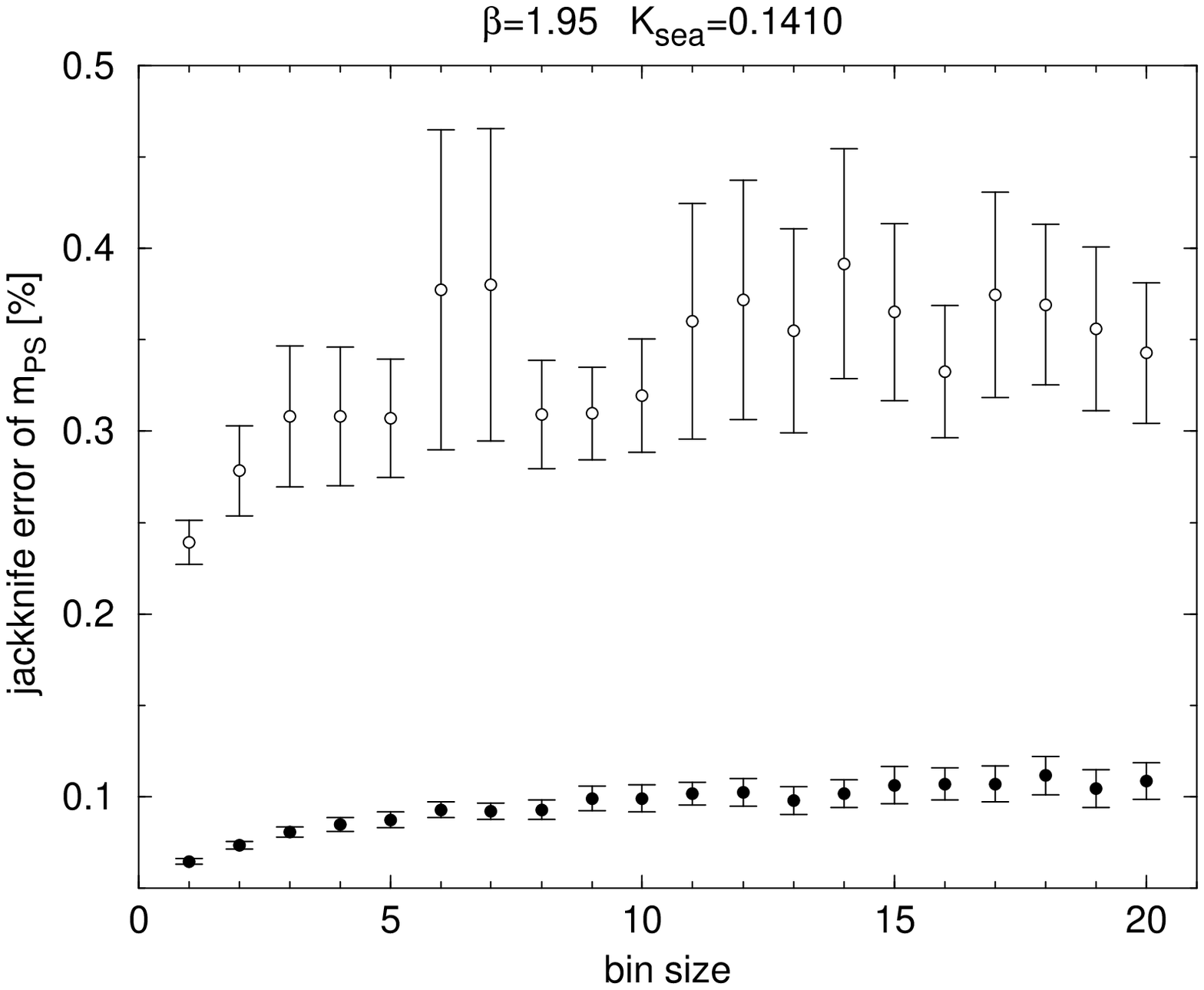}
\hspace{5mm}  \epsfxsize=7cm \epsfbox{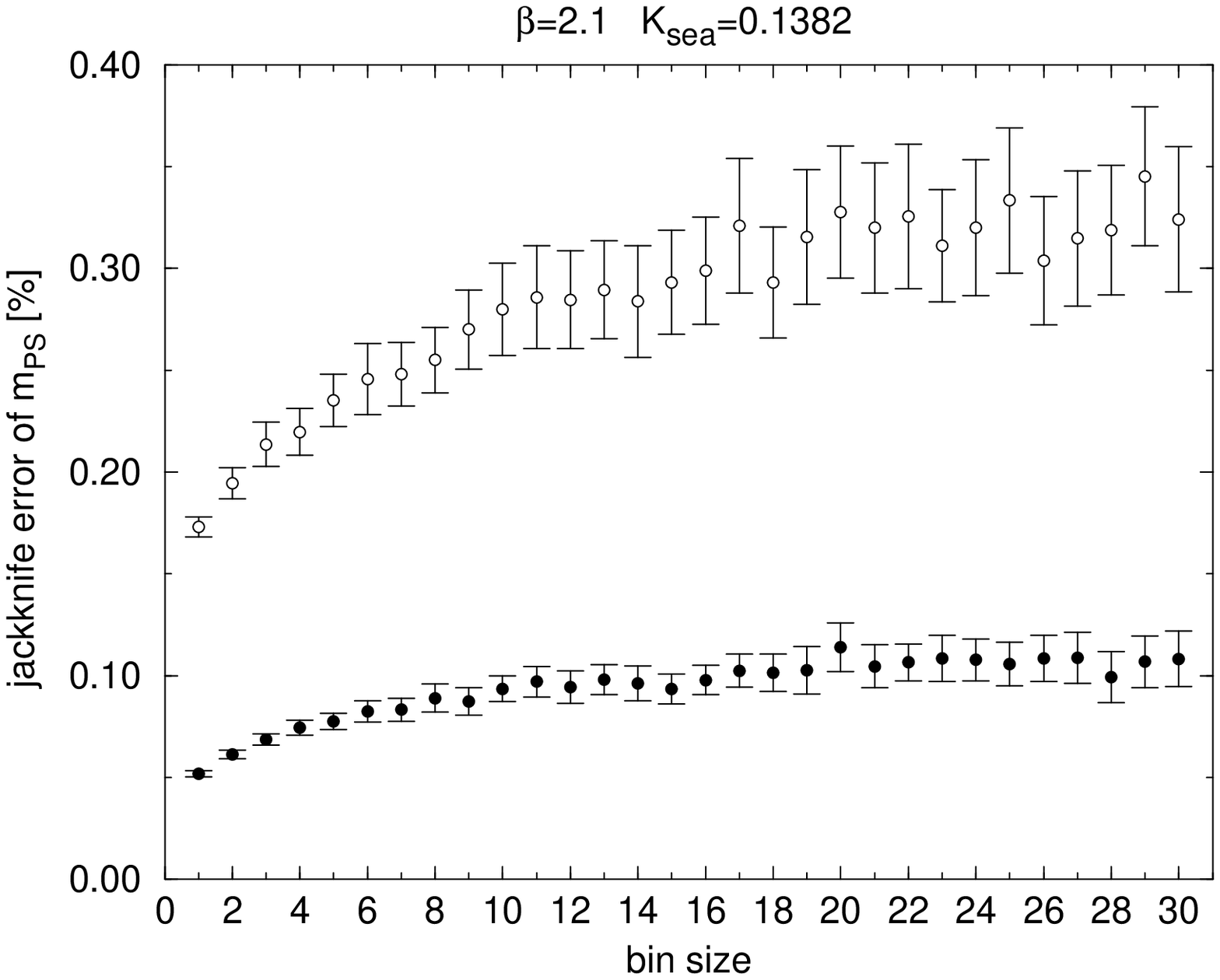}
}
\caption{Relative errors of the pseudoscalar meson mass as a function of
the bin size. Two examples are shown, each at the lightest sea quark mass
of $\mps/\mv \approx 0.6$. Data at the heaviest valence quark mass are
represented by filled symbols and the ones from the lightest valence quark
mass with open symbols.}
\label{fig:bin}
\end{figure*}

A practical way to take into account autocorrelations in error analyses is
to use the binning method. In Fig.~\ref{fig:bin} we show the increase of
the relative error of the pion mass as a function of the bin size. The
plotted error bars are determined by a jack-knife on jack-knife method. For
this plot we have used uncorrelated fits to the pion propagator, since for
larger bin sizes the number of configurations would not be large enough to
reliably determine the covariance matrix for correlated fits. We observe
that the error rises to a plateau which is about a factor $\sqrt{2\tau^{\rm
int}_f}$ larger than the naive error obtained with a unit bin size. From
these and similar figures at other simulation parameters we find that a bin
size of 10 configurations, equivalent to 50 HMC trajectories, covers all
the autocorrelations we have examined while leaving sufficient number of
bins to allow correlated fits. We therefore employ this bin size in all
error analyses.


\section{Chiral Extrapolations}
\label{sec:chiral}

The calculation of the physical hadron spectrum requires an extrapolation
from simulated quark masses to the physical point. In order to make these
extrapolations we have to fit hadron masses to a functional form chosen to
express their chiral behavior. Hadron masses are functions of $\kappa_{\rm
sea}$ and $\kappa_{\rm val}^{(i)}$, where $i=1,2,\ldots$ labels valence
quarks. We take this into account by performing combined fits to all
measured masses of a given channel.

The hopping parameter is not the only choice for the basic variable in
these fits. Pseudoscalar meson masses can be employed as well for vector
mesons and baryons. This has the advantage that only measured hadron masses
are involved, and we employ this way of parametrizing vector meson and
baryon masses. Pseudoscalar meson masses themselves, however, have to be
expressed in terms of quark masses in order to fix the physical point in
terms of quark masses.

\subsection{Pseudoscalar mesons}

Let us recall that the definition of quark mass suggested by a Ward 
identity for vector currents (VWI) has the form 
\begin{equation}
m^{\rm VWI} = \frac{1}{2}
   \left ( \frac{1}{\kappa} - \frac{1}{\kappa_c} \right ),
\end{equation}
where $\kappa_c$ is the critical hopping parameter at which pseudoscalar
meson mass vanishes. For a combined fit of pseudoscalar meson masses in
terms of this ``VWI'' quark mass, we define sea and valence quark masses
through
\begin{eqnarray}
m^{\rm VWI}_{\rm sea}    &=& \frac{1}{2} 
  \left (\frac{1}{\kappa_{\rm sea}}-\frac{1}{\kappa_c} \right ), \\
m^{\rm VWI}_{{\rm val}(i)} &=& \frac{1}{2} 
  \left (\frac{1}{\kappa_{\rm val}^{(i)}}-\frac{1}{\kappa_c} \right ),
\label{eq:VWIvalDef}
\end{eqnarray}
where $\kappa_{\rm val}^{(i)}$ denote for $i=1,2$ the hopping parameters of
the valence quark and antiquark which make the meson. In the leading order
of chiral perturbation theory the masses squared of pseudoscalar mesons are
linear functions of the average quark mass. We therefore define an average
valence quark mass through
\begin{equation}
m^{\rm VWI}_{\rm val}
= \frac{1}{2} 
  \left ( m^{\rm VWI}_{{\rm val}(1)}+m^{\rm VWI}_{{\rm val}(2)} \right )
= \frac{1}{2} 
  \left (\frac{1}{\kappa_{\rm val}}-\frac{1}{\kappa_c} \right ),
\qquad \frac{1}{\kappa_{\rm val}} = \frac{1}{2}
	\left(\frac{1}{\kappa_{\rm val}^{(1)}} + \frac{1}{\kappa_{\rm
val}^{(2)}}\right). 
\label{eq:VWIvalAv}
\end{equation}

\begin{figure*}[p]
\centerline{
              \epsfxsize=7cm \epsfbox{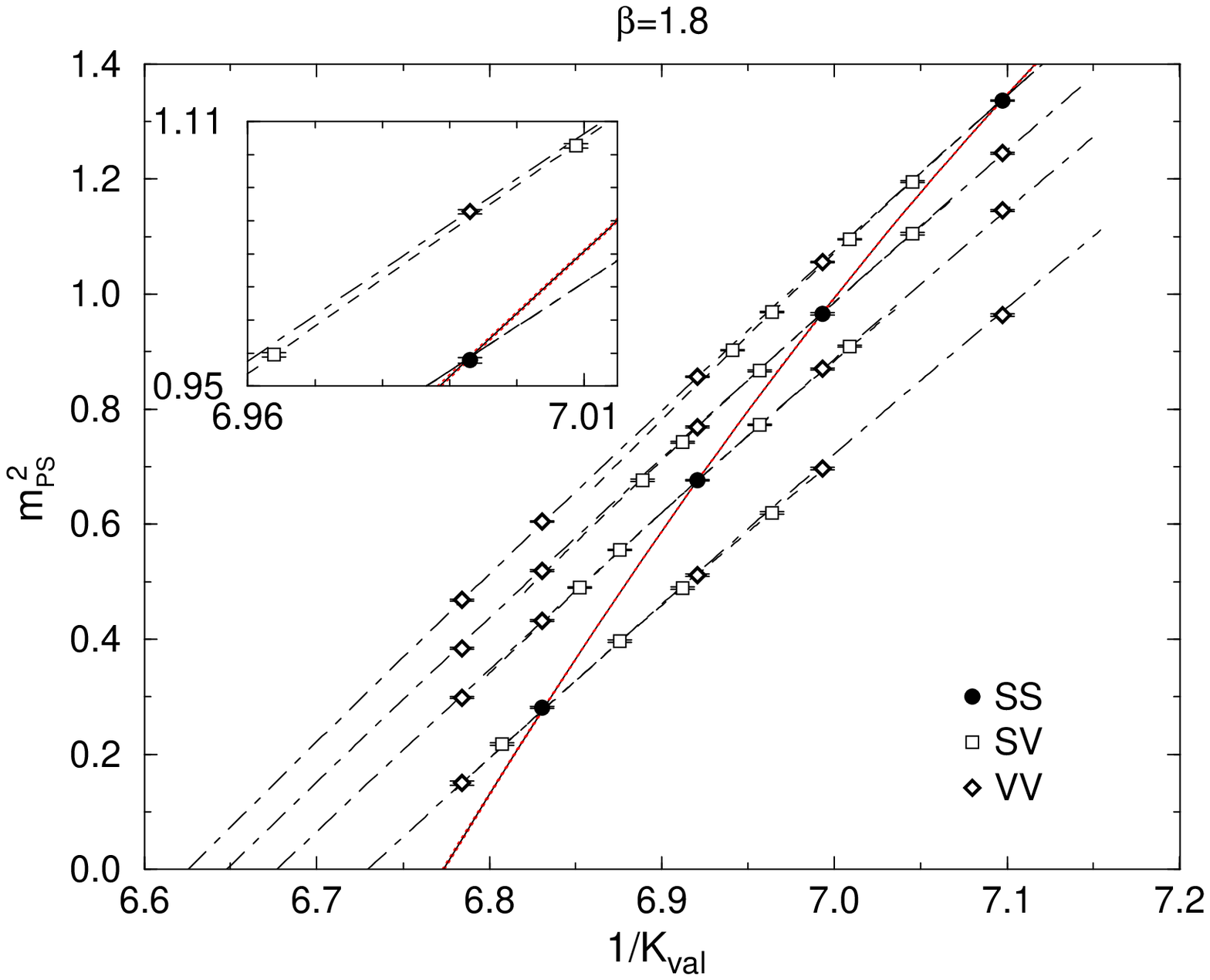}
\hspace{5mm}  \epsfxsize=7cm \epsfbox{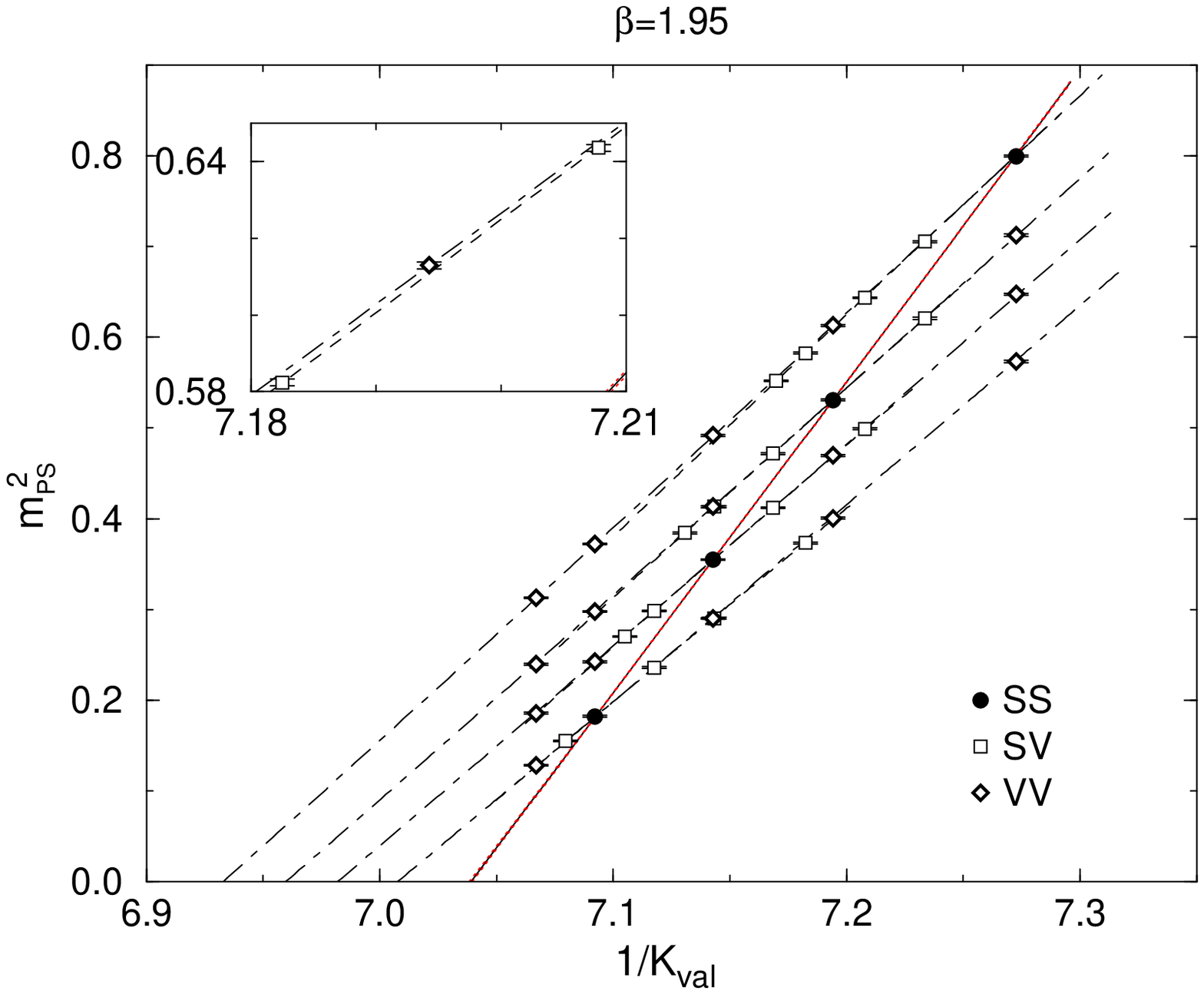}
}
\vspace{2mm}
\centerline{
              \epsfxsize=7cm \epsfbox{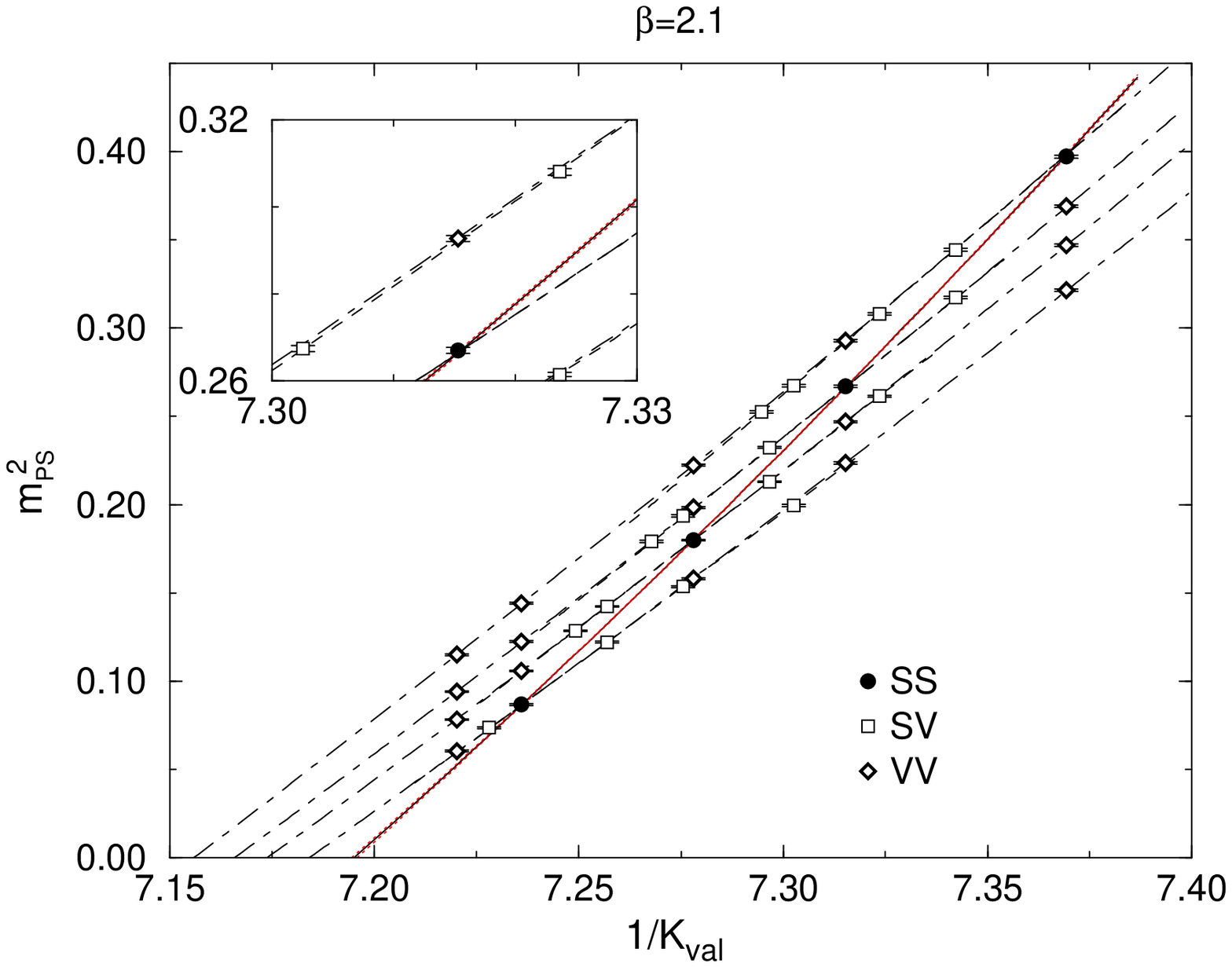}
\hspace{5mm}  \epsfxsize=7cm \epsfbox{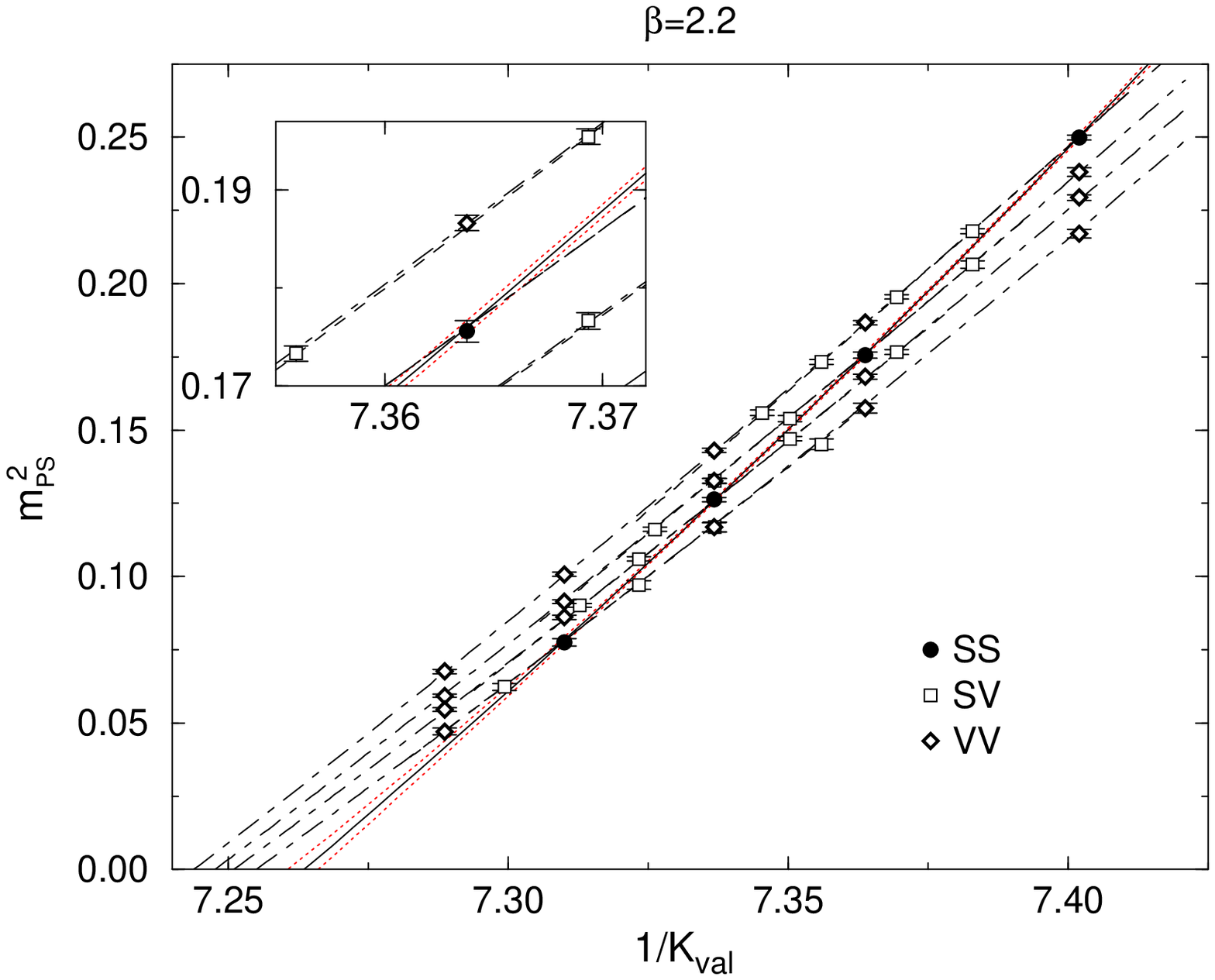}
}
\caption{Chiral extrapolations of pseudoscalar meson masses. S and V are
for valence quarks equal or different to the sea quark. Lines are from
combined quadratic fits with Eq.~(\ref{eq:ps-fit-full-VWI}).} 
\label{fig:chiralPS}
\end{figure*}

Figure~\ref{fig:chiralPS} shows pseudoscalar meson masses as functions of
$1/\kappa_{\rm val}$. We observe that partially quenched data ({\it i.e.,}
VV and SV) lie along clearly distinct lines when the hopping parameter of
sea quark $\kappa_{\rm sea}$ is varied. Each of the partially quenched
data are close to linear, but their slope shows a variation with
$\kappa_{\rm sea}$. As illustrated in the inlaid figures we also see that
the VV and SV masses lie along slightly different lines, which means that
masses depend on the individual valence quark masses $m_{{\rm val}(i)}^{\rm
VWI}$ in addition to their average.

These features of pseudoscalar meson mass data lead us to adopt a fit
ansatz which consists of general linear and quadratic terms in the valence
quark mass and in the sea quark mass given by,
\begin{eqnarray}
\mps^2 \left 
      (\kappa_{\rm sea};\kappa_{\rm val}^{(1)},\kappa_{\rm val}^{(2)}
             \right )  
&=&  b_s m^{\rm VWI}_{\rm sea} + b_v m^{\rm VWI}_{\rm val}
   + c_s \left (m^{\rm VWI}_{\rm sea} \right )^2 
   + c_v \left (m^{\rm VWI}_{\rm val} \right )^2 \nonumber \\
& & +\ c_{sv} m^{\rm VWI}_{\rm sea} m^{\rm VWI}_{\rm val}
   + c_{vv} m^{\rm VWI}_{{\rm val}(1)} m^{\rm VWI}_{{\rm val}(2)}.   
\label{eq:ps-fit-full-VWI}
\end{eqnarray}
Figure~\ref{fig:chiralPS} shows the fit with solid lines for the SS channel
and with dashed (SV) or dot-dashed (VV) lines for partially quenched
data. The lines follow the data well. We employ uncorrelated fits for
chiral extrapolations even though data with common $\kappa_{\rm sea}$ are
expected to be correlated. Obtained values of $\chi^2/N_{\rm DF}$ can
therefore only be considered as rough guidelines to judge the quality of
fits. Except for $\beta=1.8$ where $\chi^2/N_{\rm DF}=4$, we obtain values
which are smaller than 1. Fit parameters $\kappa_c$, $b$'s for linear terms
and $c$'s for quadratic terms and $\chi^2/N_{\rm DF}$ are given in
Table~\ref{tab:ps-fit-full}.

A different definition of quark mass suggested by a Ward identity for
axial vector currents is given by Eq.~(\ref{eq:awi-mass}). Since this is a
measured quantity derived from meson propagators it depends on three
hopping parameters, $\kappa_{\rm val}^{(i)} (i=1,2)$ of the valence quark
and antiquark, and $\kappa_{\rm sea}$ of the sea quark. We define
\begin{eqnarray}
m^{\rm AWI}_{{\rm val}(i)} &=& 
   m^{\rm AWI} \left 
    (\kappa_{\rm sea};\kappa_{\rm val}^{(i)},\kappa_{\rm val}^{(i)}
   \right ), \\
m^{\rm AWI}_{\rm val}    &=& 
   \frac{1}{2} \left ( m^{\rm AWI}_{{\rm val}(1)} 
                     + m^{\rm AWI}_{{\rm val}(2)}  
   \right ), 
\label{eq:AWIvalDef} \\
m^{\rm AWI}_{\rm sea}    &=&  
   m^{\rm AWI} \left 
   (\kappa_{\rm sea};\kappa_{\rm sea},\kappa_{\rm sea} \right ). 
\end{eqnarray}
Pseudoscalar meson masses are expressed in terms of these quantities
with the quadratic ansatz,
\begin{equation}
\mps^2 \left 
  (\kappa_{\rm sea};\kappa_{\rm val}^{(1)},\kappa_{\rm val}^{(2)} \right )
 = b'_v m^{\rm AWI}_{\rm val} +  c'_v \left (m^{\rm AWI}_{\rm val} \right )^2
 + c'_{sv} m^{\rm AWI}_{\rm sea} m^{\rm AWI}_{\rm val}.
\label{eq:ps-fit-full-AWI}
\end{equation}
In contrast to Eq.~(\ref{eq:ps-fit-full-VWI}) monomial terms in the sea
quark mass are absent since pseudoscalar masses vanish in the chiral limit
$m^{\rm AWI}_{\rm val} = 0$ for {\it each} value of the sea quark
mass. Data of different degeneracies lie on common lines and therefore we
have dropped the term with individual $m^{\rm AWI}_{{\rm val}(i)}$. Fit
parameters and $\chi^2/N_{\rm DF}$ are given in
Table~\ref{tab:ps-fit-full}.

\subsection{Vector mesons}
\label{sec:fit-vec}

\begin{figure*}[p]
\centerline{
              \epsfxsize=7cm \epsfbox{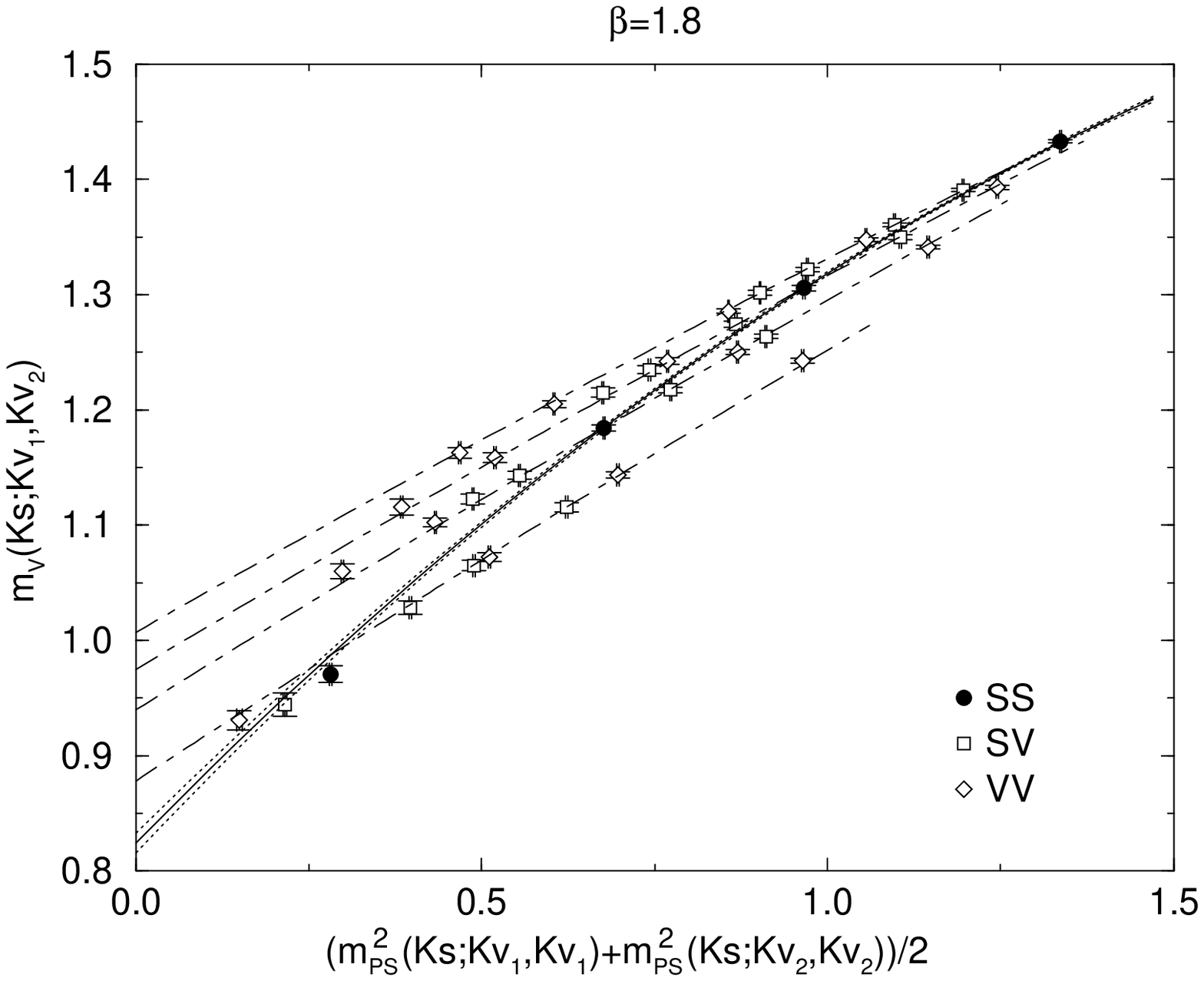}
\hspace{5mm}  \epsfxsize=7cm \epsfbox{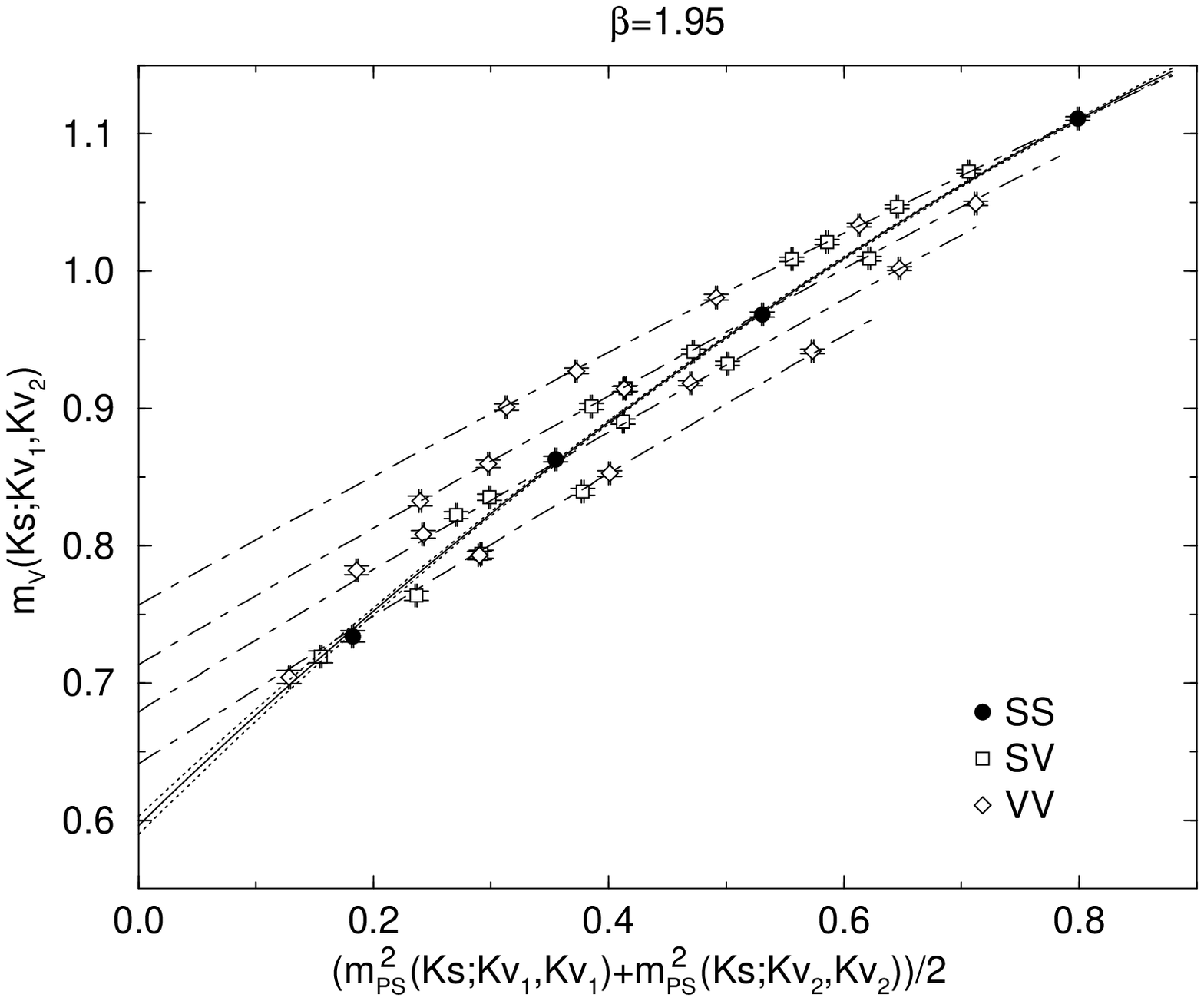}
}
\vspace{2mm}
\centerline{
              \epsfxsize=7cm \epsfbox{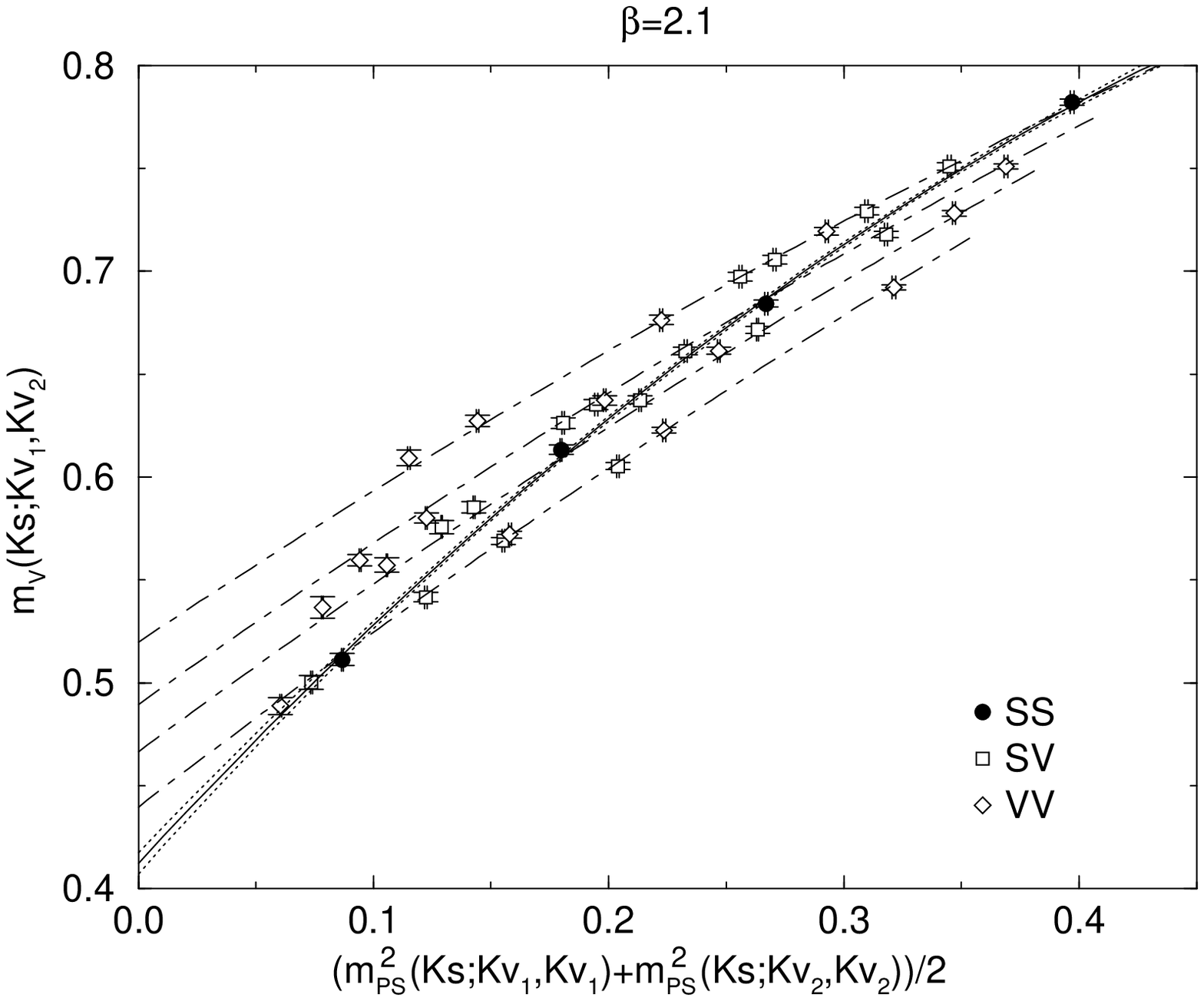}
\hspace{5mm}  \epsfxsize=7cm \epsfbox{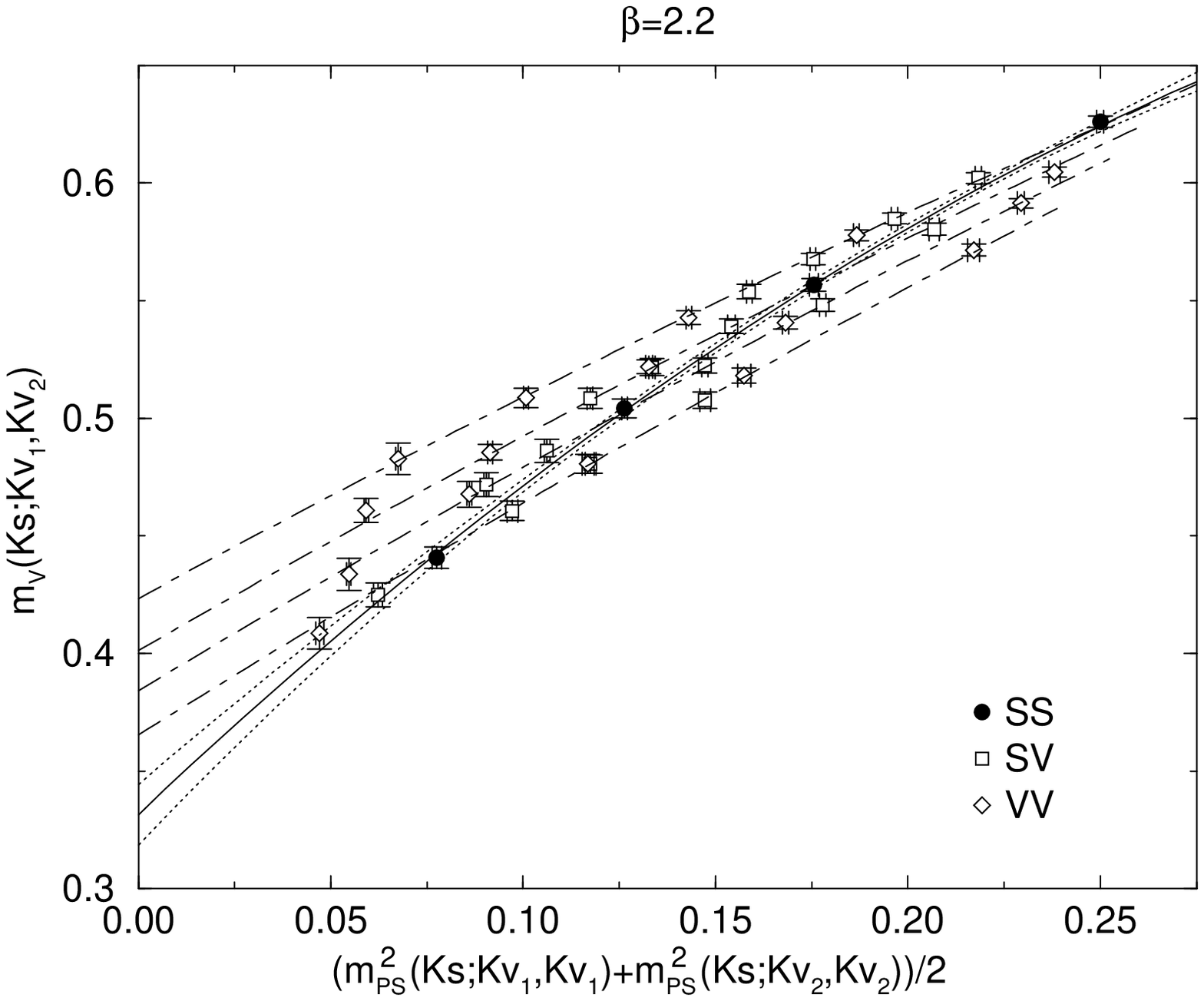}
}
\caption{Chiral extrapolations of vector meson masses. Lines are from fits with
Eq.~(\ref{eq:vec-fit-full}).}  
\label{fig:chiralVec}
\end{figure*}

Vector meson masses are fit in terms of measured pseudoscalar meson
masses. We define
\begin{eqnarray}
\mu_i     &=&  \mps^2 
   \left 
  (\kappa_{\rm sea};\kappa_{\rm val}^{(i)},\kappa_{\rm val}^{(i)} \right ), \\
\mu_{\rm val} &=& \frac{1}{2} \left ( \mu_1+\mu_2 \right ), 
\label{eq:MuValDef} \\
\mu_{\rm sea} &=& \mps^2
   \left ( \kappa_{\rm sea};\kappa_{\rm sea},\kappa_{\rm sea} \right ).
\end{eqnarray}
Vector meson masses as functions of $\mu_{\rm val}$ are shown in
Fig.~\ref{fig:chiralVec}. The general feature of data is similar to the one
for pseudoscalar mesons. We find, however, that the lines for VV and SV are
indistinguishable. Hence, vector meson masses do not require terms in
individual $\mu_i$'s. We therefore take a quadratic function in $\mu_{\rm
sea}$ and $\mu_{\rm val}$ of the form
\begin{equation}
\mv \left 
  (\kappa_{\rm sea};\kappa_{\rm val}^{(1)},\kappa_{\rm val}^{(2)} \right )
    =  A^V + B^V_s \mu_{\rm sea} + B^V_v \mu_{\rm val} 
       +  C^V_s \mu_{\rm sea}^2 + C^V_v \mu_{\rm val}^2 
       +  C^V_{sv} \mu_{\rm sea} \mu_{\rm val}.   
\label{eq:vec-fit-full}
\end{equation}
Fit lines describe data well as shown in Fig.~\ref{fig:chiralVec}, 
and $\chi^2/N_{\rm DF}$ is at most 1.4. 
Fit parameters and $\chi^2/N_{\rm DF}$ are given in 
Table~\ref{tab:vec-fit-full}.

Chiral perturbation theory predicts~\cite{VectorChPT} that the first
correction to the linear term in $\mu$ has a non-analytic $3/2$ power of
$\mu$. In order to examine if data show evidence for such a dependence, we
attempt a fit of the form
\begin{equation}
\mv \left 
  (\kappa_{\rm sea};\kappa_{\rm val}^{(1)},\kappa_{\rm val}^{(2)} \right ) 
    =  A^V + B^V_s \mu_{\rm sea} + B^V_v \mu_{\rm val} 
       +  D^V_s \mu_{\rm sea}^{3/2} + D^V_v \mu_{\rm val}^{3/2} 
       +  D^V_{sv} \mu_{\rm sea} \mu_{\rm val}^{1/2}.   
\label{eq:vec-fit-full-15}
\end{equation}
The cross-term of the form $\mu_{\rm sea}\mu_{\rm val}^{1/2}$ gives rise to
a term proportional to $\mu_{\rm val}^{1/2}$ for the partially quenched
case where $\mu_{\rm sea}$ is kept constant. This is similar to quenched
QCD. Terms proportional to $\mu_{\rm sea}^{1/2}$ are expected to be
absent~\cite{maarten}.

\begin{figure*}[p]
\centerline{
              \epsfxsize=7cm \epsfbox{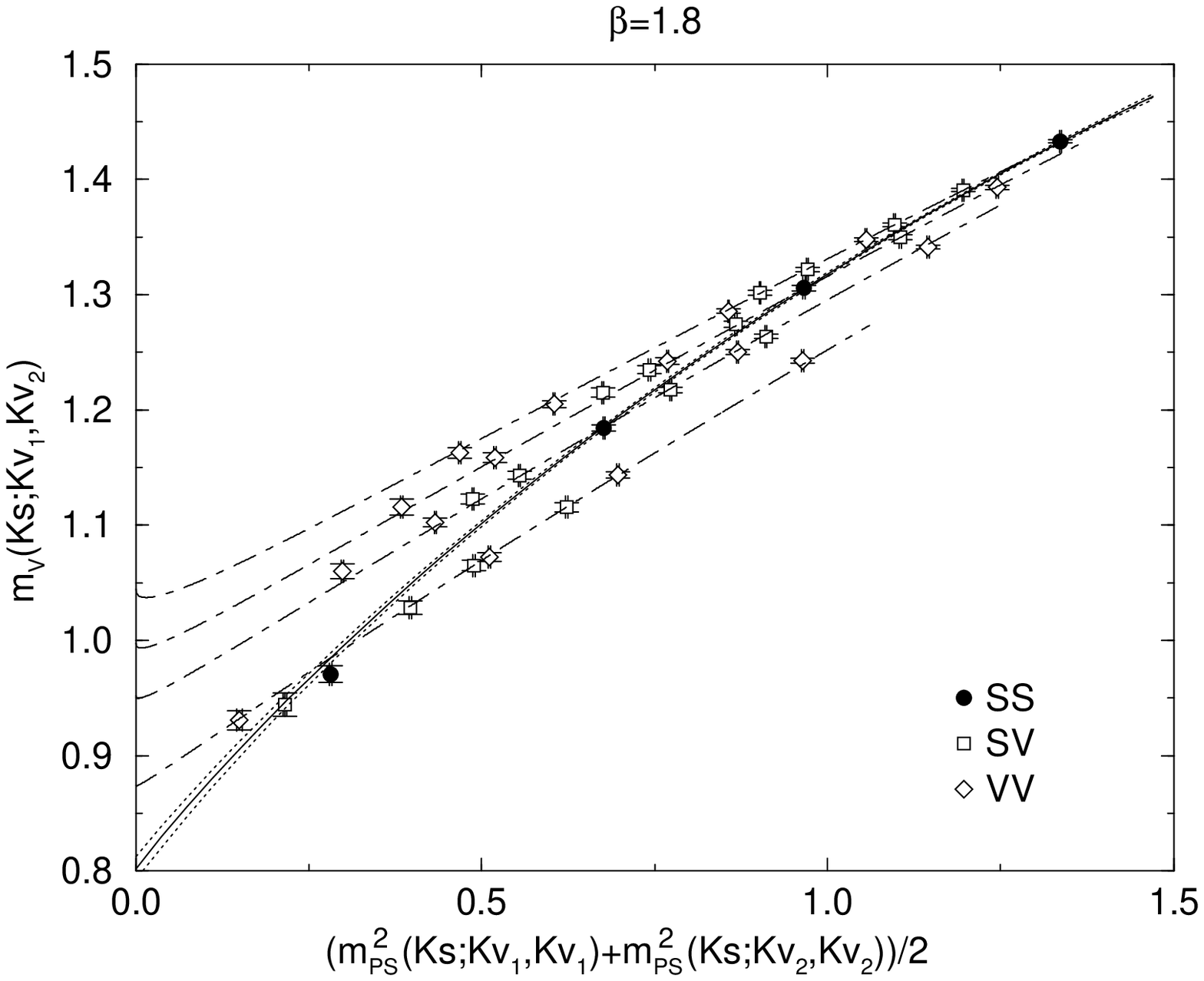}
\hspace{5mm}  \epsfxsize=7cm \epsfbox{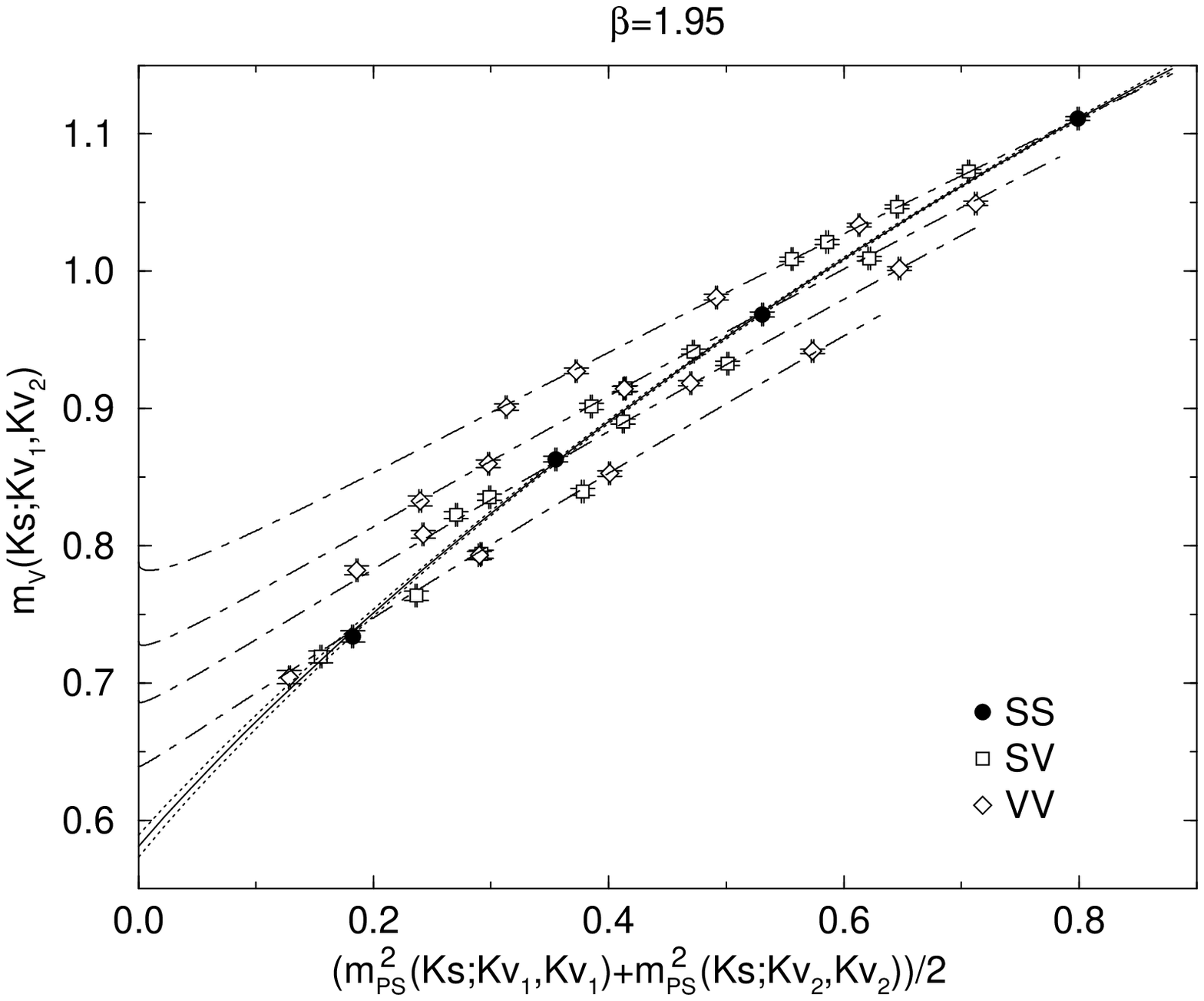}
}
\vspace{2mm}
\centerline{
              \epsfxsize=7cm \epsfbox{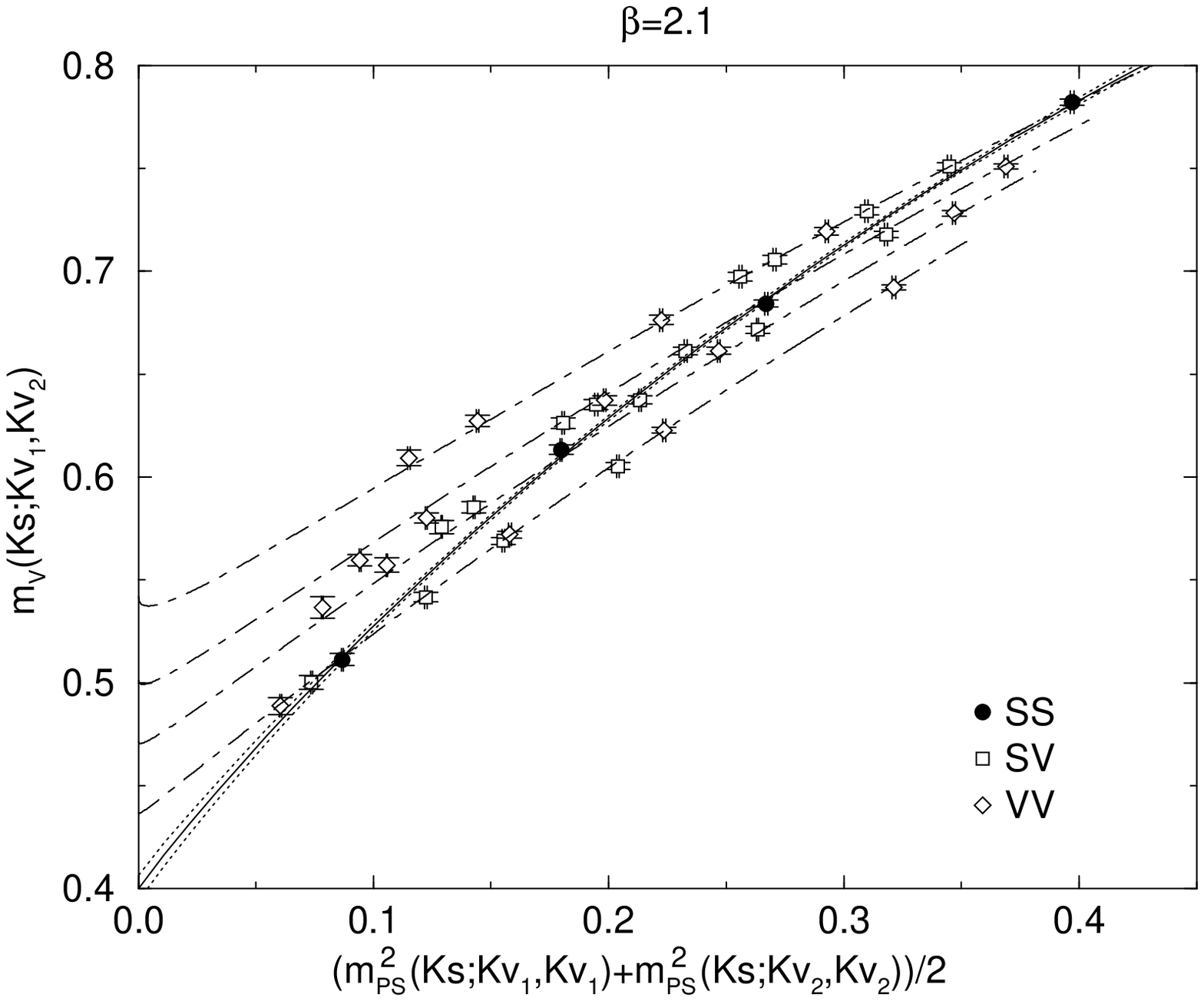}
\hspace{5mm}  \epsfxsize=7cm \epsfbox{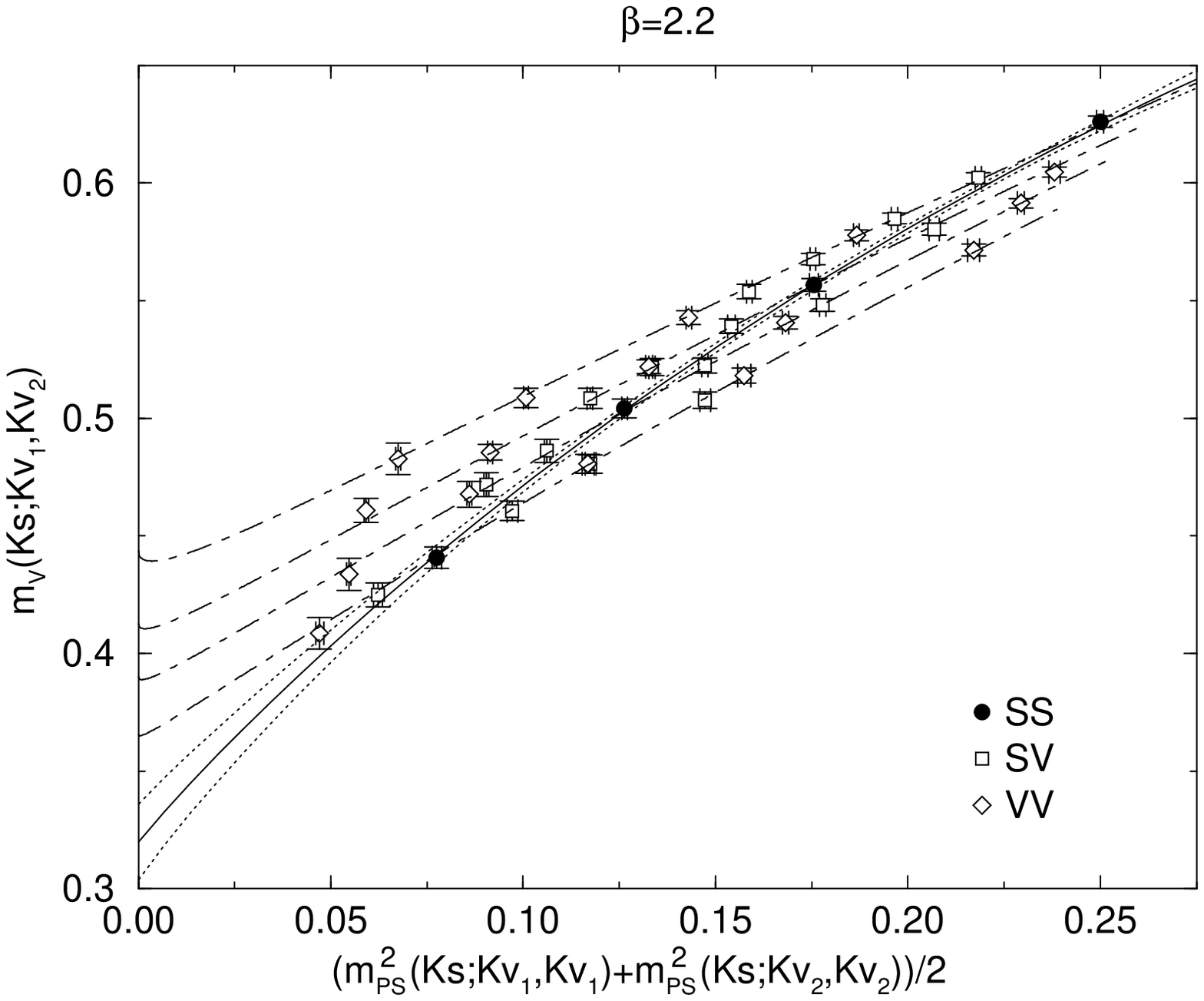}
}
\caption{Chiral extrapolations of vector meson masses. Mass data
are the same as in Fig.~\ref{fig:chiralVec} but fit lines are from
the alternative fits with
Eq.~(\ref{eq:vec-fit-full-15}).}
\label{fig:chiralVec15}
\end{figure*}

In Fig.~\ref{fig:chiralVec15} we show lines for this alternative fit
together with measured data. Due to the presence of the $\mu_{\rm
val}^{1/2}$ term, fit lines show a small increase close to the chiral limit
of valence quark when the difference between sea and valence quark is
large. This is similar to the behavior observed for quenched QCD in
Ref.~\cite{quench.CPPACS}. The amount of increase becomes smaller when sea
and valence quarks have values closer to each other, and vanishes for full
QCD.

Fit parameters and $\chi^2/N_{\rm DF}$ are given in
Table~\ref{tab:vec-fit-full}. $\chi^2/N_{\rm DF}$ is slightly smaller for
the fit with Eq.~(\ref{eq:vec-fit-full-15}) than the one with
Eq.~(\ref{eq:vec-fit-full}) but the difference between the two is not
significant. We can therefore not answer the question whether a fit with
power $3/2$ or 2 is preferred. We employ Eq.~(\ref{eq:vec-fit-full}) for
main results and use Eq.~(\ref{eq:vec-fit-full-15}) to estimate systematic
differences arising from the choice of chiral fit form.

\subsection{Baryons}

Baryons are made from three valence quarks and hence their masses are
expressed in terms of the three $\mu_i$'s and $\mu_{\rm sea}$. In the
measurements described in Sec.~\ref{sec:params}, however, at least two
valence quarks are degenerate. We use $\mu_2$ to stand for the pair of
degenerate valence quarks and $\mu_1$ for the third valence quark.

\begin{figure*}[p]
\centerline{
              \epsfxsize=7cm \epsfbox{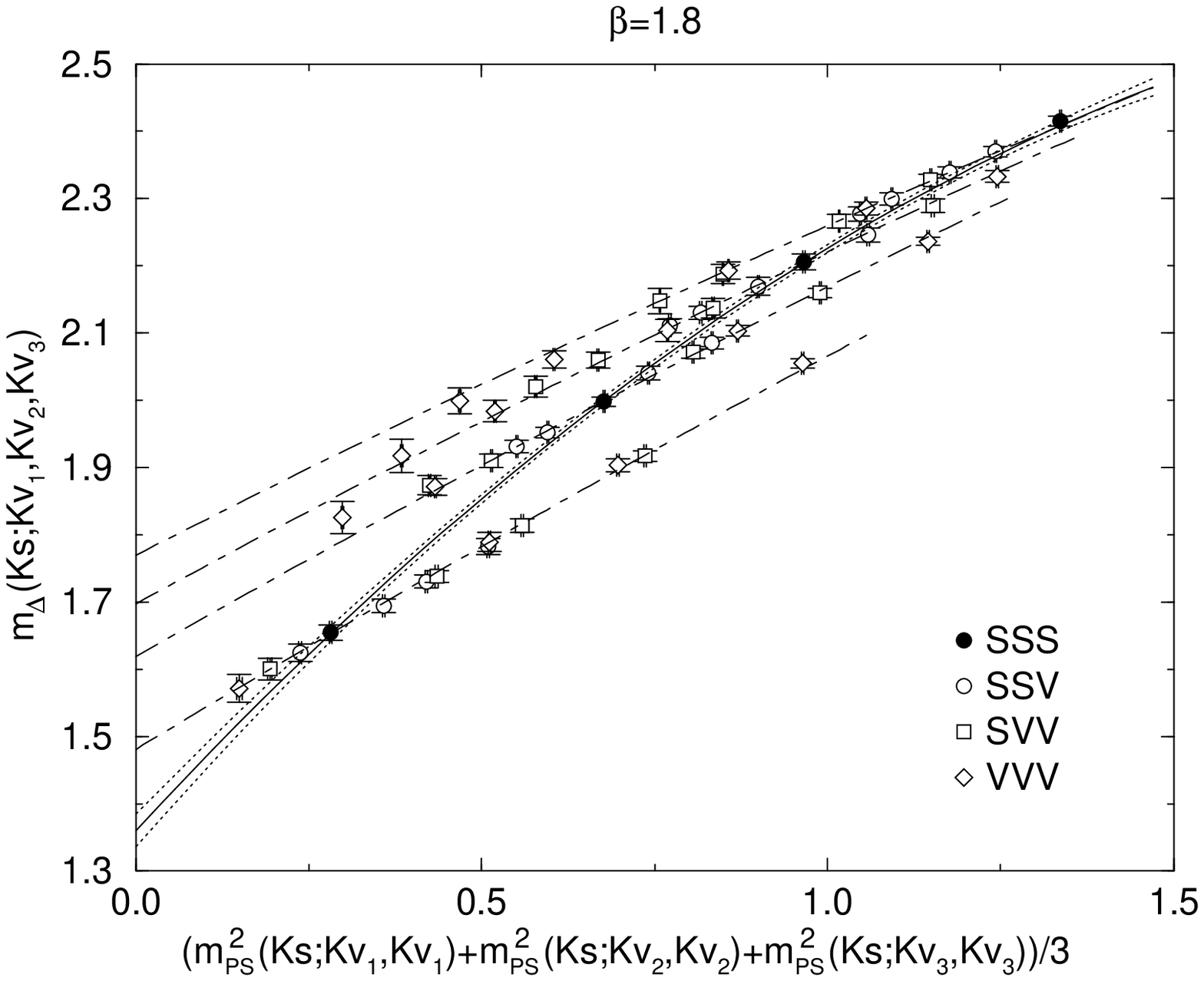}
\hspace{5mm}  \epsfxsize=7cm \epsfbox{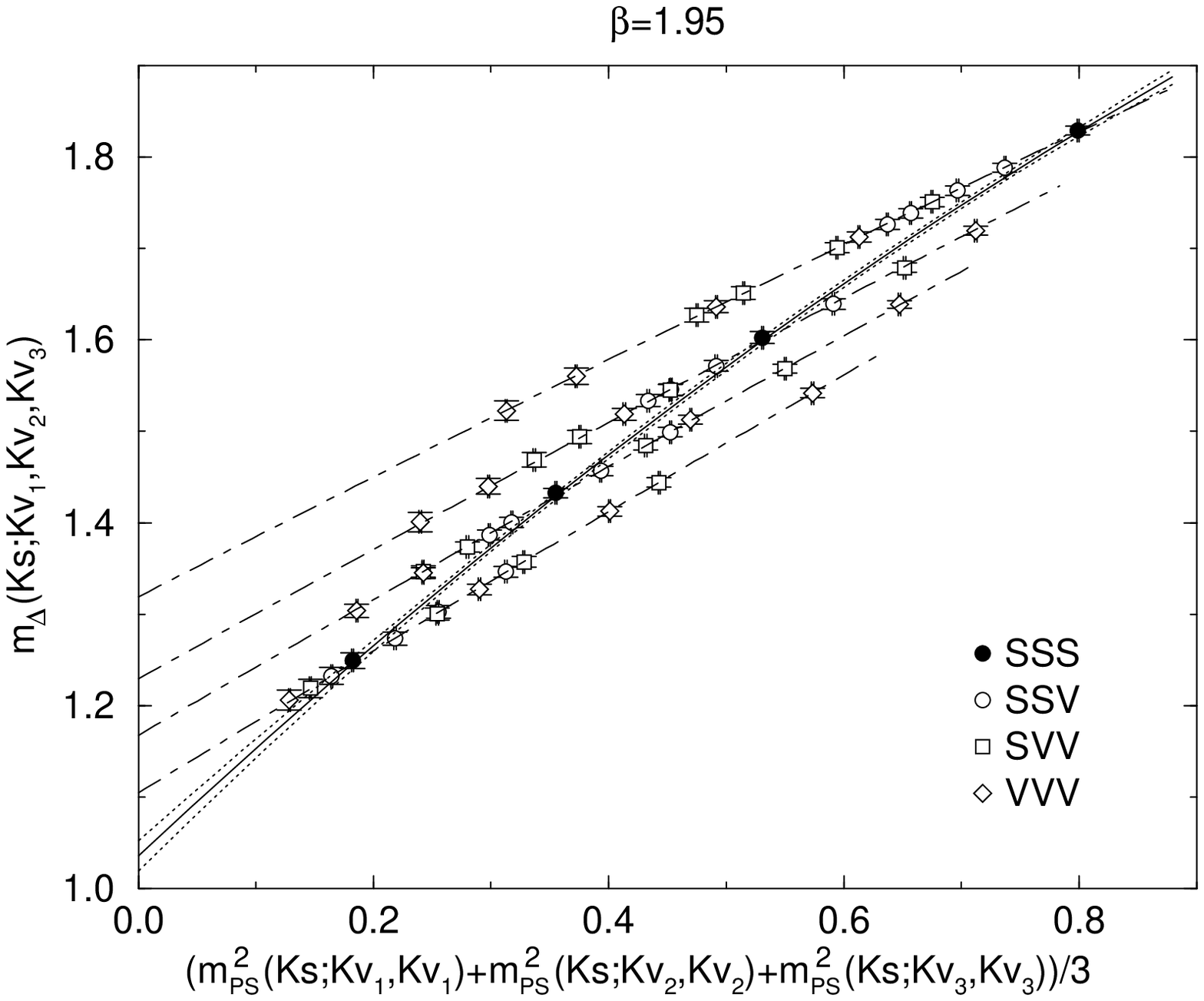}
}
\vspace{2mm}
\centerline{
              \epsfxsize=7cm \epsfbox{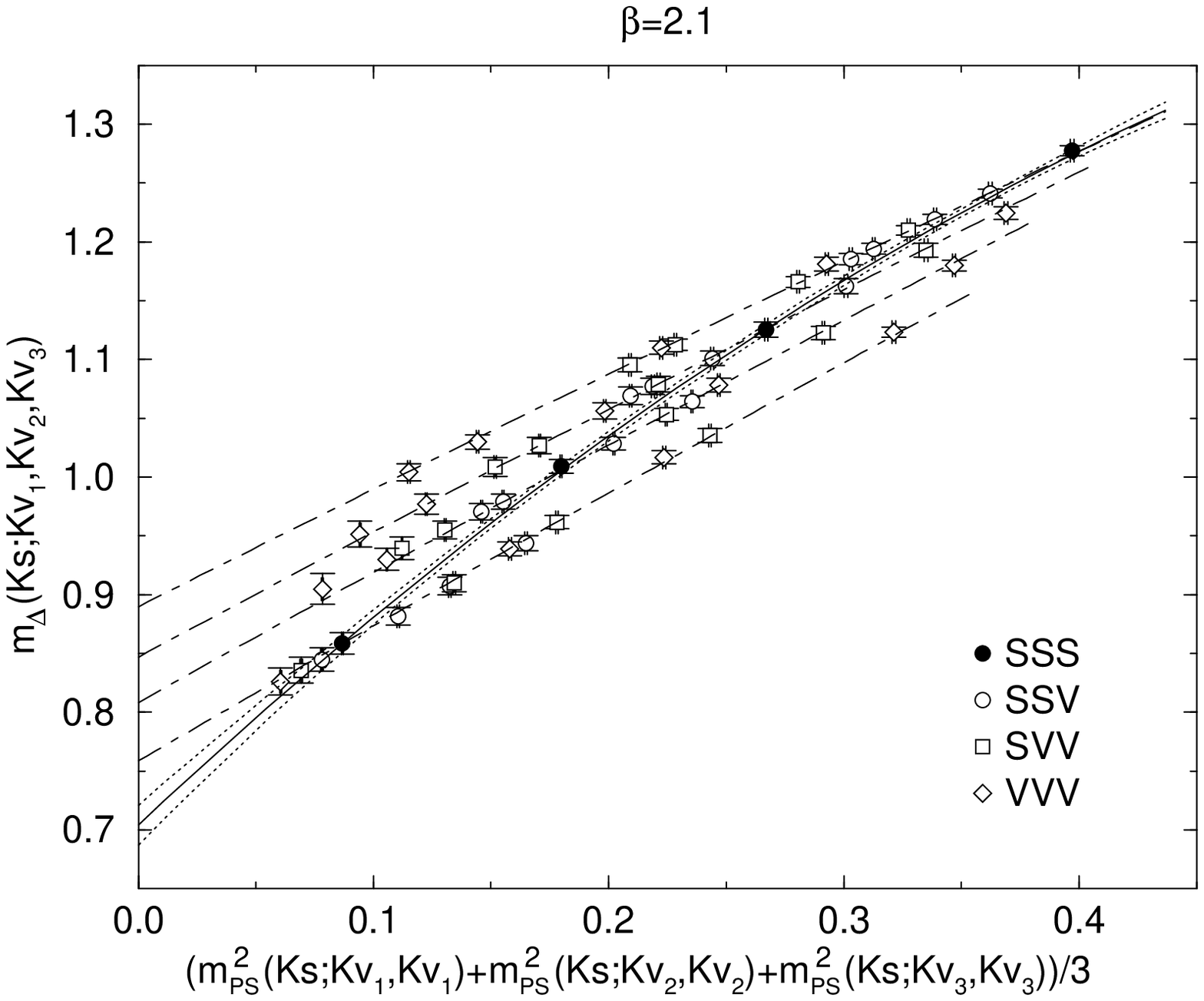}
\hspace{5mm}  \epsfxsize=7cm \epsfbox{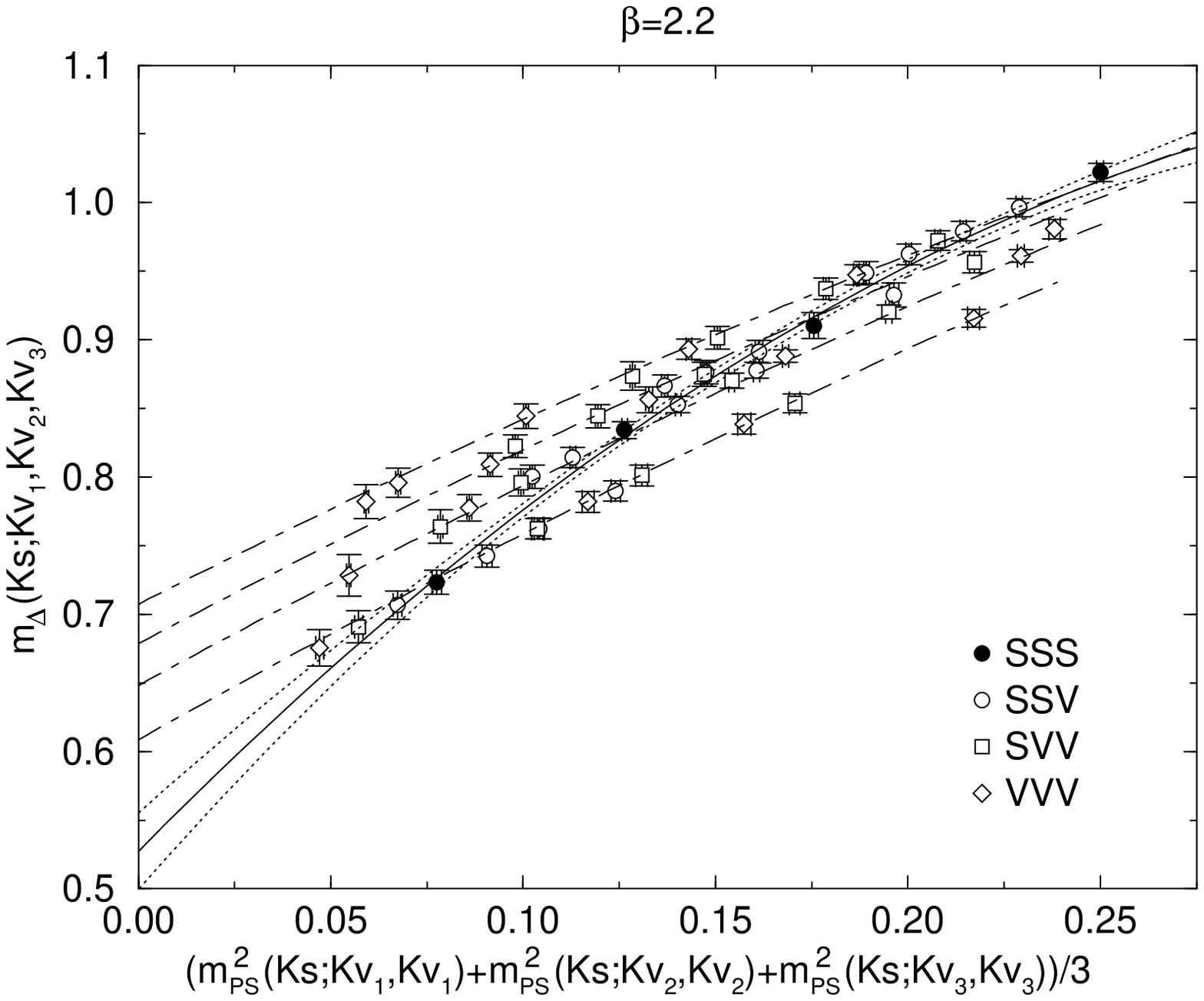}
}
\caption{Chiral extrapolations of decuplet baryon masses. Lines are from
fits with Eq.~(\ref{eq:dec-fit-full}).} 
\label{fig:chiralDec}
\end{figure*}

For the decuplet baryons masses can be expressed as a function of the
average valence quark mass. Hence we define
\begin{equation}
\mu_{\rm val}= \frac{1}{3} \left ( \mu_1+\mu_2+\mu_2 \right ),
\label{eq:MuValOct}
\end{equation}
and plot decuplet baryon masses as a function of $\mu_{\rm val}$ in
Fig.~\ref{fig:chiralDec}. The behavior of mass data is very similar to the
one observed for vector meson masses with clearly distinguishable lines of
variable slope for partially quenched data and stronger curvature for full
QCD data. We therefore employ an ansatz of the same structure as for vector
mesons which takes the form
\begin{equation}
m_{\rm D} \left
     (\kappa_{\rm sea};
      \kappa_{\rm val}^{(1)},\kappa_{\rm val}^{(2)},\kappa_{\rm val}^{(2)}
    \right )
     =  A^D + B^D_s \mu_{\rm sea} + B^D_v \mu_{\rm val}
       +  C^D_s \mu_{\rm sea}^2 + C^D_v \mu_{\rm val}^2
       +  C^D_{sv} \mu_{\rm sea} \mu_{\rm val}.   
\label{eq:dec-fit-full}
\end{equation}
As shown in Fig.~\ref{fig:chiralDec} data are fitted well with
$\chi^2/N_{\rm DF}$ of at most 0.35. Fit parameters and $\chi^2/N_{\rm DF}$
are given in Table~\ref{tab:dec-fit-full}.

\begin{figure*}[p]
\centerline{
              \epsfxsize=7cm \epsfbox{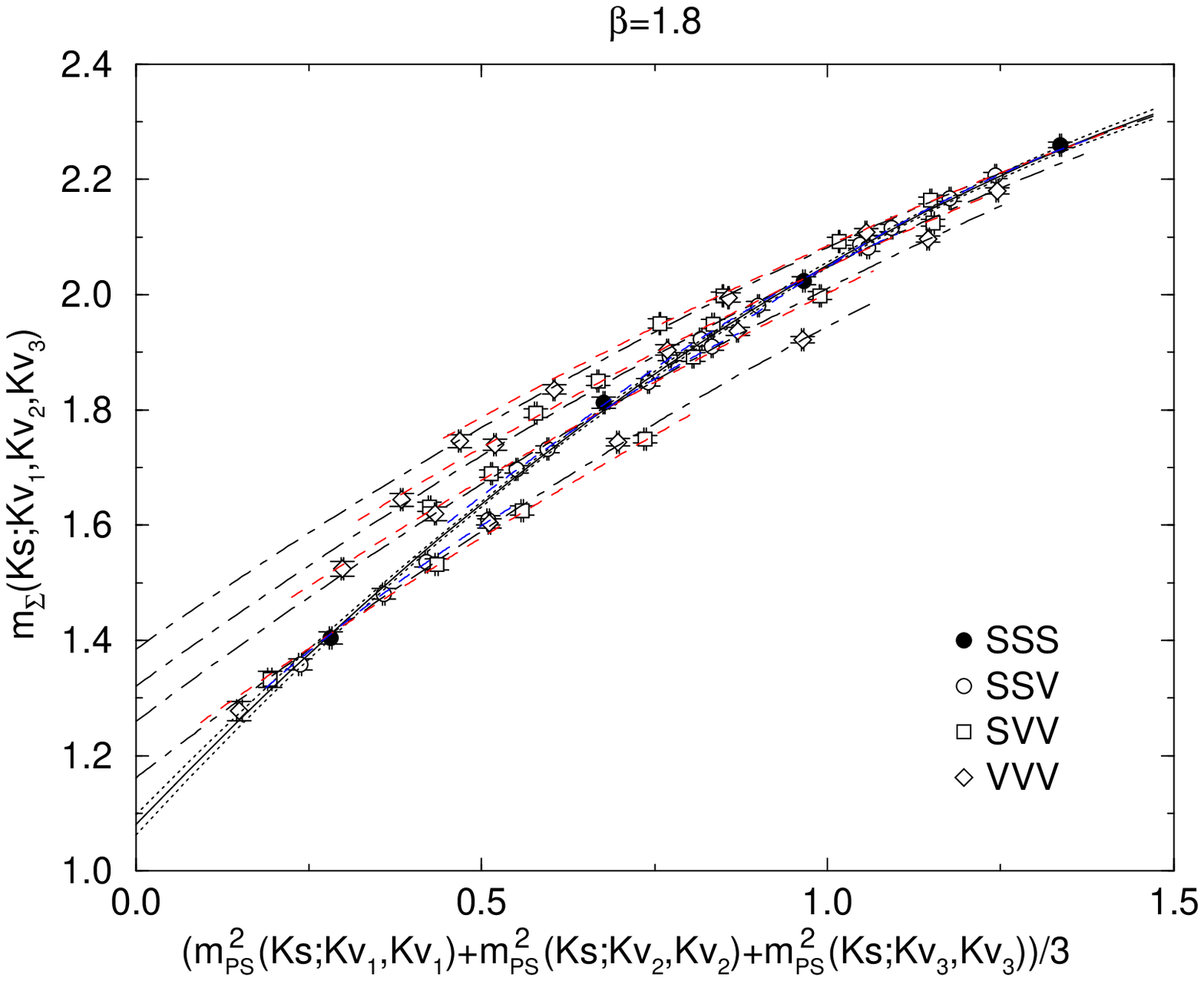}
\hspace{5mm}  \epsfxsize=7cm \epsfbox{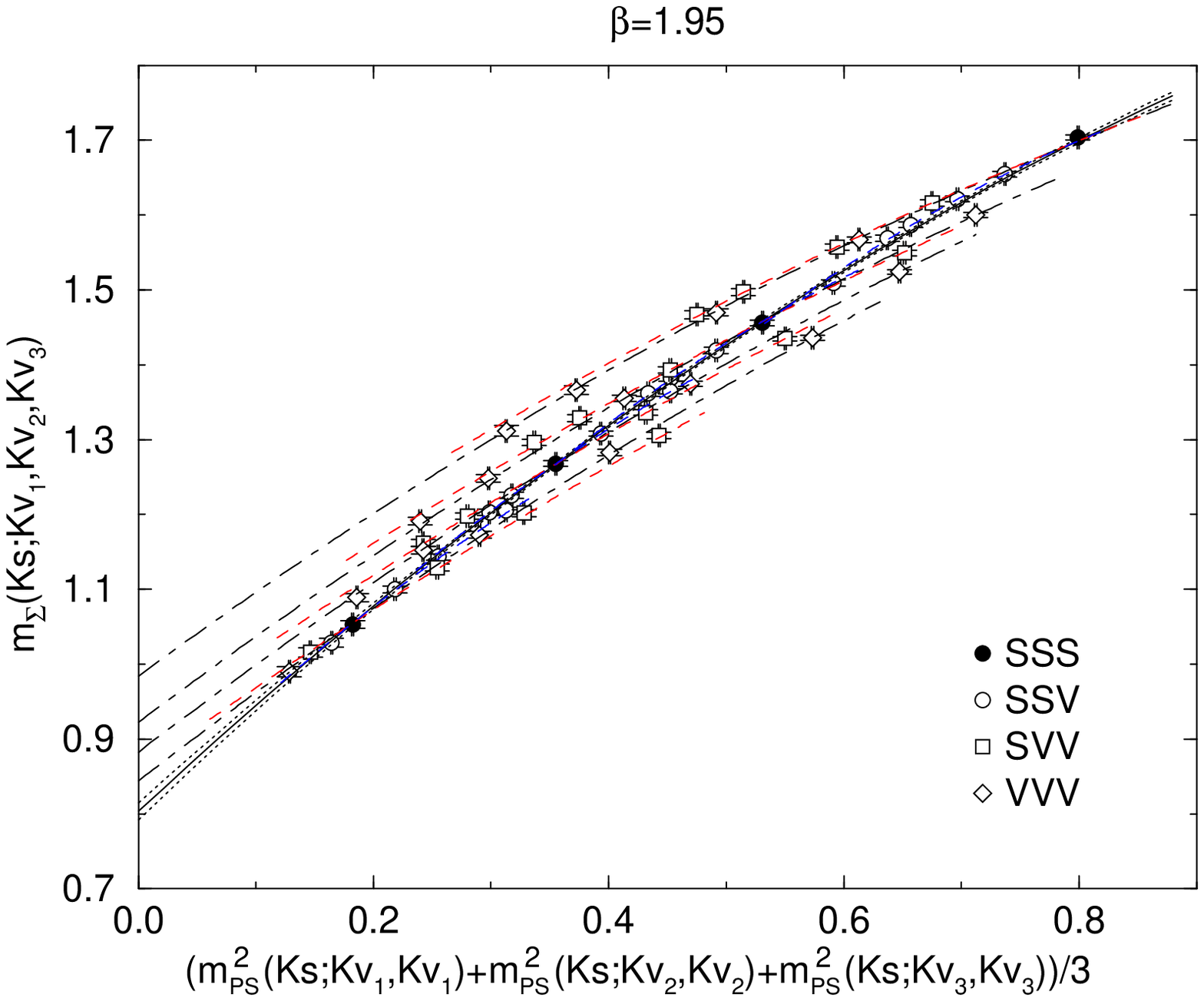}
}
\vspace{2mm}
\centerline{
              \epsfxsize=7cm \epsfbox{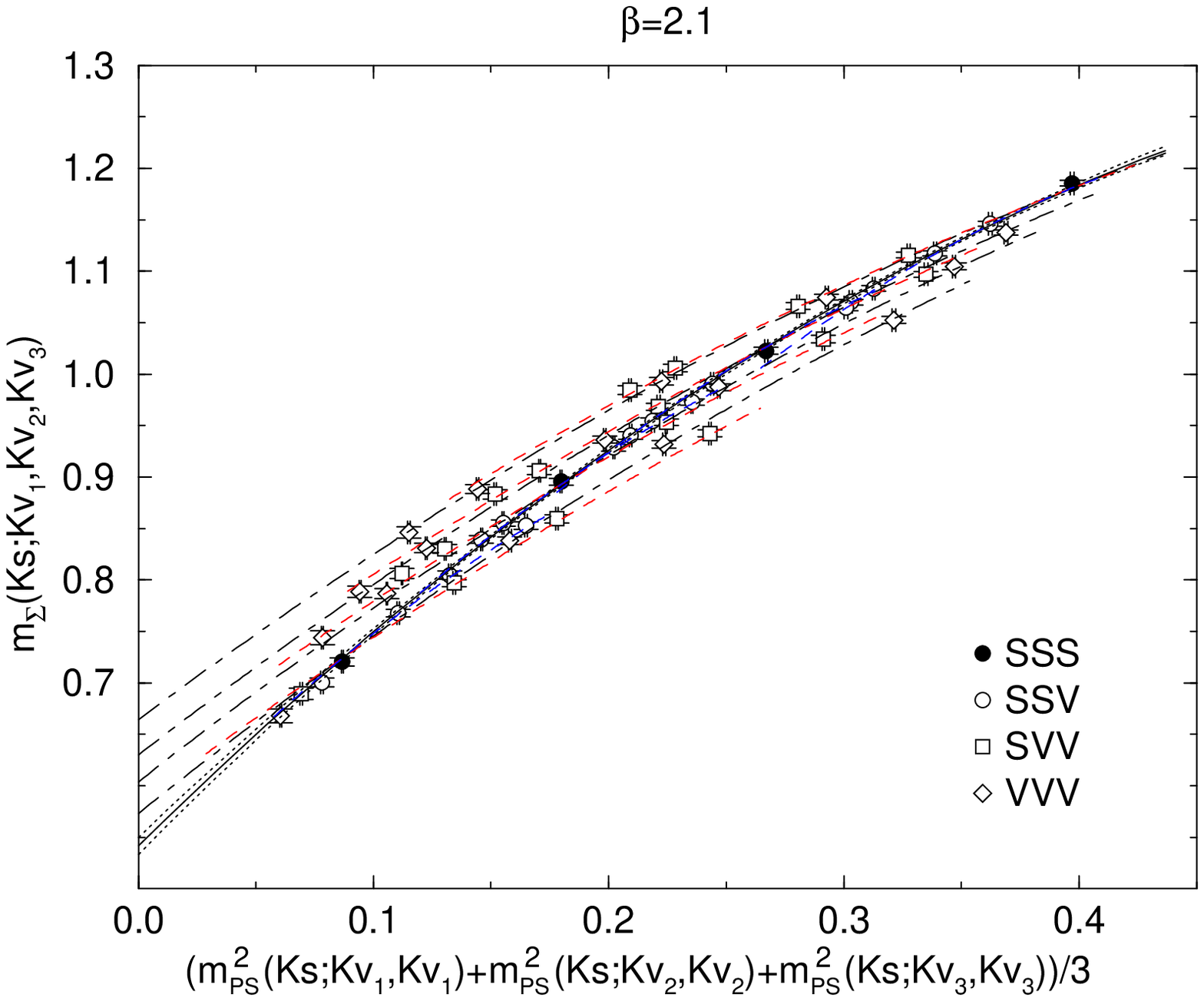}
\hspace{5mm}  \epsfxsize=7cm \epsfbox{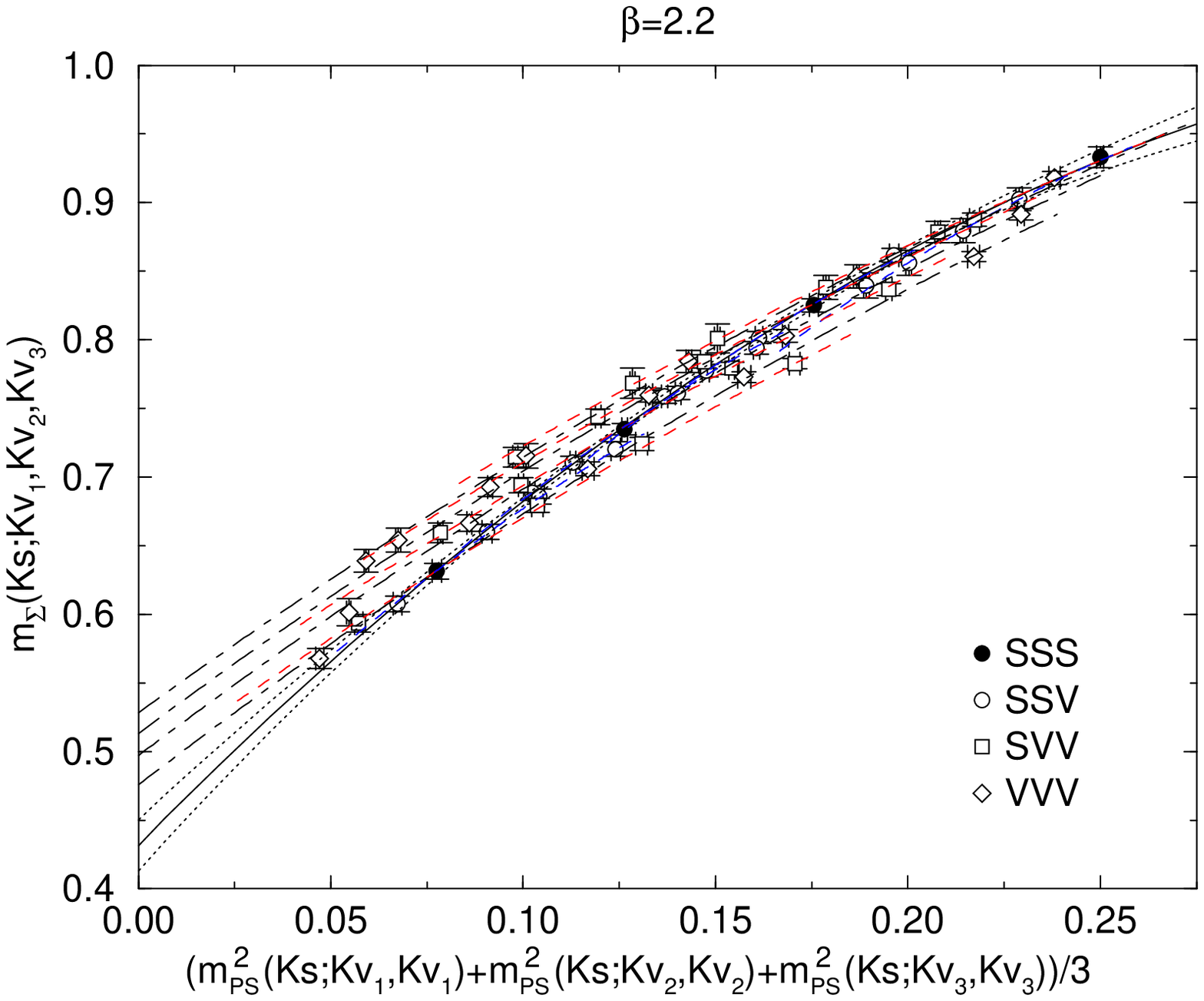}
}
\caption{Chiral extrapolations of octet baryon masses. Plots only show
$\Sigma$-like octet baryons. Lines are from fits with
Eq.~(\ref{eq:ol-fit-full}).}  
\label{fig:chiralOct}
\end{figure*}

Octet baryon masses are not simple functions of the average valence quark
mass. This can be seen in Fig.~\ref{fig:chiralOct} where we plot masses of
$\Sigma$-like octet baryons as a function of $\mu_{\rm val}$ defined in
Eq.~(\ref{eq:MuValOct}). The three sets of partially quenched data VVV, SVV
and SSV lie along different lines. We also see a clear distinction between
results for different sea quark masses.

We analyze octet baryon masses by using a formula inspired by chiral
perturbation theory~\cite{BaryonChPT}. In the leading order $\Sigma$-like
and $\Lambda$-like octet baryon masses are parametrized as a function of
quark masses with two constants $b_D$ and $b_0$. We use these expressions
for terms linear in the valence quark mass. For convenience we use a
slightly different notation; the parameters $F_v^O$ and $D_v^O$ are related
to those of Ref.~\cite{BaryonChPT} through $F_v^O = -2(b_D+b_0)$ and
$D_v^O=-2b_0$. In order to describe the dependence on the sea quark mass we
add linear terms in the sea quark mass, and terms quadratic in the sea and
valence quark mass to incorporate curvature seen in mass data. The number
of additional terms introduced by this procedure is limited by the
requirement that $m_\Lambda = m_\Sigma$ when $\mu_1 = \mu_2$. This leads
to expressions for $\Sigma$-like and $\Lambda$-like baryons of the form
\begin{eqnarray}
m_\Sigma \left
  (\kappa_{\rm sea};
   \kappa_{\rm val}^{(1)},\kappa_{\rm val}^{(2)},\kappa_{\rm val}^{(2)}
  \right )
  & = & A^O + B_s^O \mu_{\rm sea} + \left (F_v^O-D_v^O \right ) \mu_1 
        + 2F_v^O \mu_2  \nonumber \\  
  &   & +\ C_s^O \mu_{\rm sea}^2 + C_{vv}^O \mu_1 \mu_2
        + \left (C_v^O+C_v^\Sigma \right ) \mu_1^2 
        + \left (C_v^O-C_v^\Sigma \right ) \mu_2^2 
        \nonumber \\
  &   & +\ \left (C_{sv}^O+C_{sv}^\Sigma \right ) \mu_{\rm sea} \mu_1 
        +  \left (C_{sv}^O-C_{sv}^\Sigma \right ) \mu_{\rm sea} \mu_2, 
\label{eq:ol-fit-full} \\
m_\Lambda \left
  (\kappa_{\rm sea};
   \kappa_{\rm val}^{(1)},\kappa_{\rm val}^{(2)},\kappa_{\rm val}^{(2)}
  \right ) 
  & = & A^O + B_s^O \mu_{\rm sea} 
        + \left (F_v^O+\frac{D_v^O}{3} \right ) \mu_1 
        + 2 \left (F_v^O-\frac{2D_v^O}{3} \right ) \mu_2 \nonumber \\
  &   & +\ C_s^O \mu_{\rm sea}^2 + C_{vv}^O \mu_1 \mu_2
        + \left (C_v^O+C_v^\Lambda \right ) \mu_1^2 
        + \left (C_v^O-C_v^\Lambda \right ) \mu_2^2 
        \nonumber \\
  &   & +\ \left (C_{sv}^O+C_{sv}^\Lambda \right ) \mu_{\rm sea} \mu_1 
        +  \left (C_{sv}^O-C_{sv}^\Lambda \right ) \mu_{\rm sea} \mu_2. 
        \nonumber
\end{eqnarray}
Figure~\ref{fig:chiralOct} shows masses and fit for $\Sigma$-like octet
baryons. Different line styles are used for the three types of partially
quenched data, VVV, SVV and SSV. They do not fall onto each other because
of the presence of monomial terms in $\mu_i$ in Eq.~(\ref{eq:ol-fit-full}).
Fit parameters and $\chi^2/N_{\rm DF}$ are given in
Table~\ref{tab:ol-fit-full}.

\subsection{String tension and Sommer scale}

In full QCD, gluonic quantities are still subject to chiral extrapolations
through their indirect dependence on sea quark masses. We therefore perform
such extrapolations on the parameters describing the static quark
potential.

\begin{figure*}[p]
\centerline{
              \epsfxsize=7cm \epsfbox{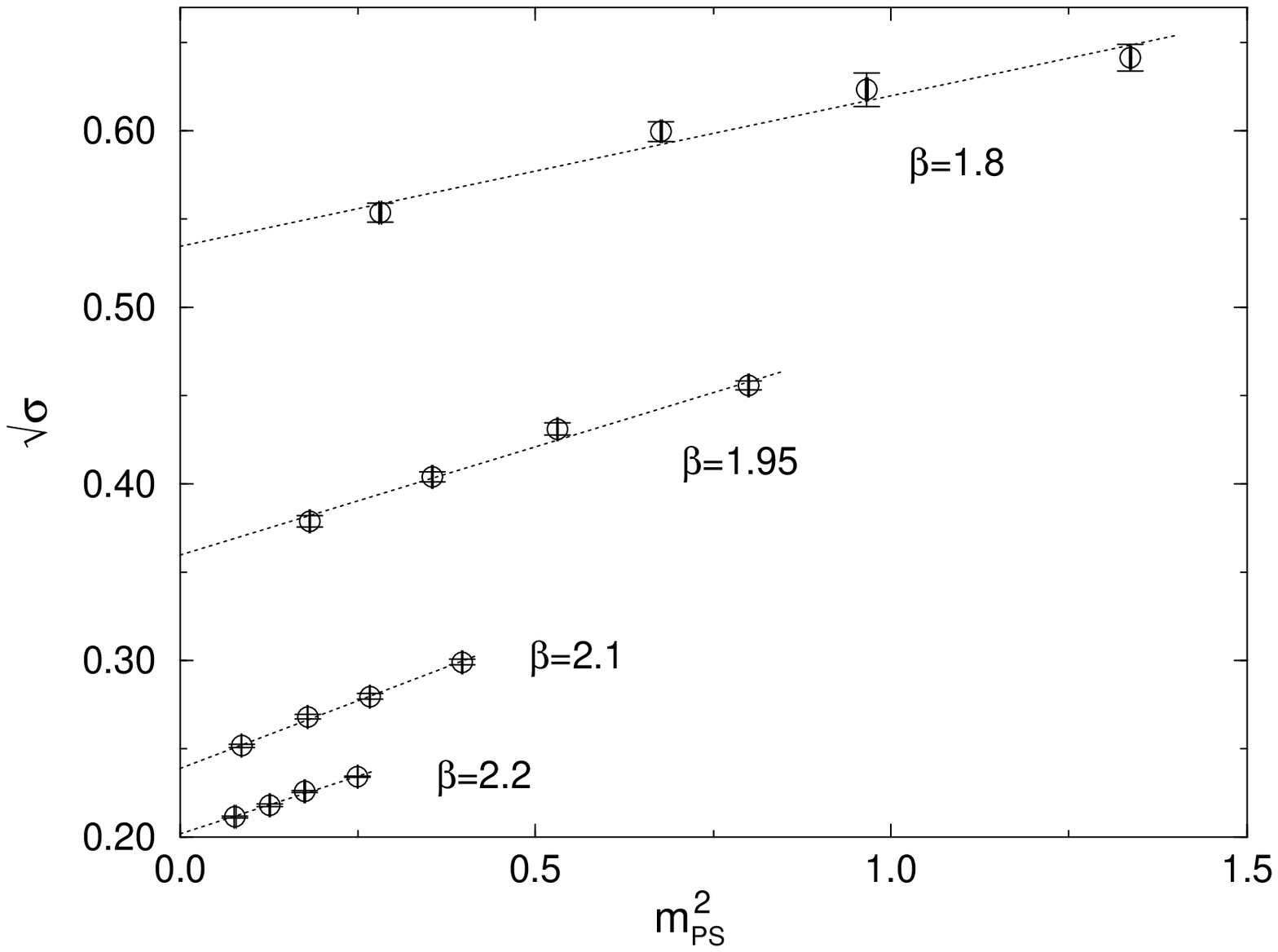}
\hspace{5mm}  \epsfxsize=7cm \epsfbox{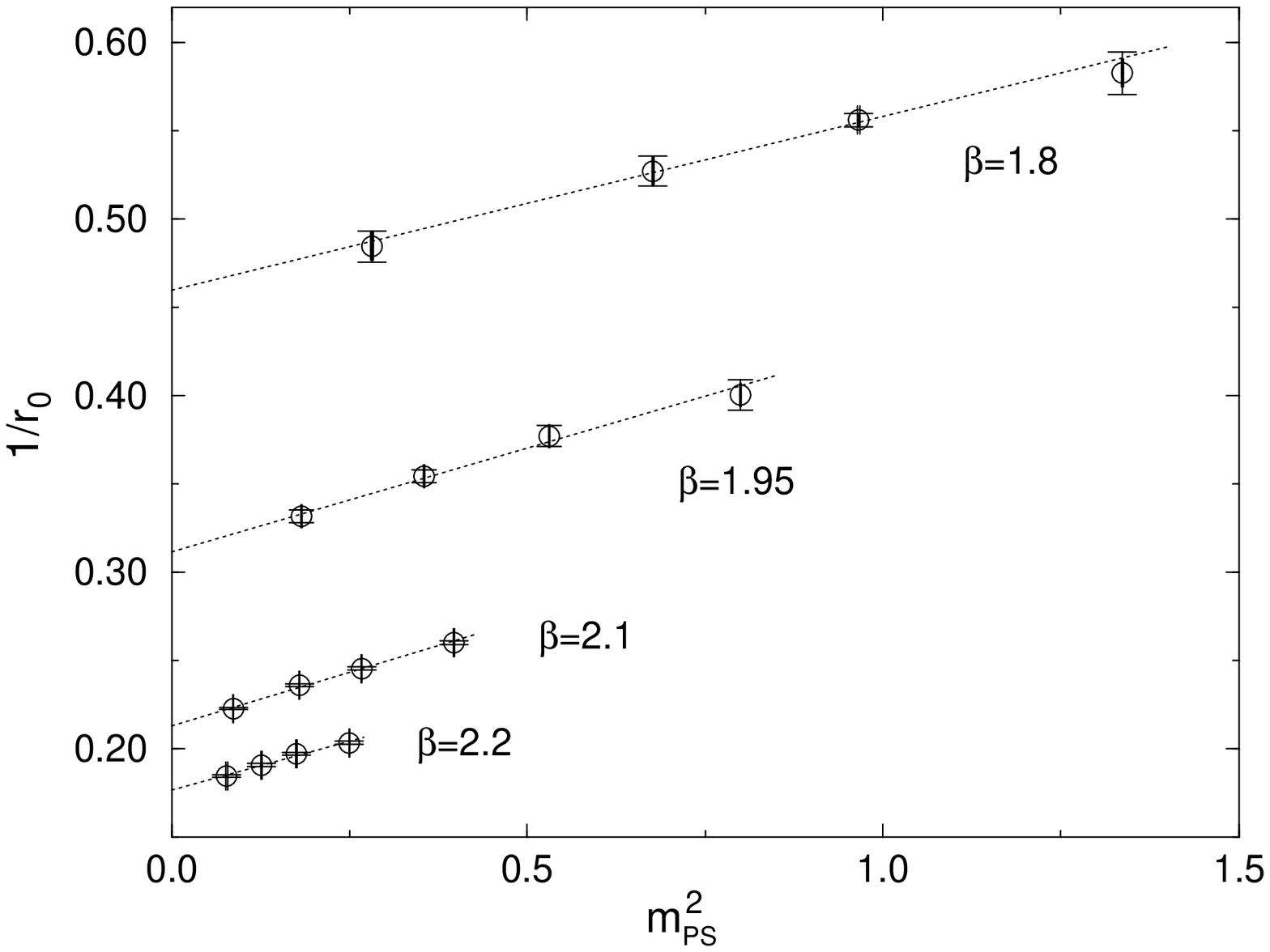}
}
\caption{Chiral extrapolations of string tension and Sommer scale.}
\label{fig:string}
\end{figure*}

In Fig.~\ref{fig:string} we show $\sqrt{\sigma}$ and $1/r_0$ obtained from
the analysis described in Sec.~\ref{sec:potmeas}, as a function of the
squared pseudoscalar meson mass with valence quarks equal to the sea quark.
The sea quark mass dependence of both quantities is approximately linear.
Therefore we apply fits of the form
\begin{equation}
\sqrt{\sigma} \left (\kappa_{\rm sea} \right ) 
   = \sqrt{\sigma^{\chi}} 
   + B_{\sigma} \mps^2 \left ( 
     \kappa_{\rm sea};\kappa_{\rm sea},\kappa_{\rm sea} \right ).
\label{eq:stringfit}
\end{equation}
and
\begin{equation}
\frac{1}{r_0} \left (\kappa_{\rm sea} \right ) = \frac{1}{r_0^{\chi}} 
  + B_{r_0} \mps^2\left ( 
    \kappa_{\rm sea};\kappa_{\rm sea},\kappa_{\rm sea} \right ).
\label{eq:R0fit}
\end{equation}
for extrapolations to the chiral limit. $\sigma^{\chi}$ and
$1/r_0^{\chi}$ in the chiral limit are given in Table~\ref{tab:sig}.


\section{Full QCD light hadron spectrum}
\label{sec:spectrum}

\subsection{Determination of the physical points}
\label{sec:fix}

Using the chiral fits of Sect.~\ref{sec:chiral} we determine the physical
point of quark masses and the lattice spacing for each $\beta$. As
experimental input we use $M_\pi=0.1350$~GeV and $M_\rho=0.7684$~GeV for
the up-down quark sector. For the strange quark sector, we compare the two
experimental inputs $M_K=0.4977$~GeV and $M_\phi=1.0194$~GeV.

The two flavors of dynamical quarks in our simulation represent 
up and down quarks which are taken as degenerate. Hence we set 
$\mu_{\rm val} = \mu_{\rm sea}$ in Eq.~(\ref{eq:vec-fit-full}) and determine 
the pion mass $m_{\pi}$ in lattice units by solving the equation 
\begin{equation}
\frac{m_{\pi}}
{A^V + (B^V_s + B^V_v)m_{\pi}^2 +  (C^V_s + C^V_v +  C^V_{sv})m_{\pi}^4}    
= \frac{M_\pi}{M_\rho}
\label{eq:fixmpi}
\end{equation}
for $m_{\pi}$. The rho meson mass in lattice units $m_{\rho}$ is then found
by inserting $m_{\pi}$ into Eq.~(\ref{eq:vec-fit-full}). The error is
determined with the jack-knife procedure described in
Appendix~\ref{app:jack}. The result of $m_{\rho}$ is used to set the
lattice spacing $a$ by identification with the physical value
$M_{\rho}$. Lattice spacings obtained in this way are given in
Table~\ref{tab:parameters}. Inserting $m_{\pi}$ obtained just above into
Eqs.~(\ref{eq:dec-fit-full}) and (\ref{eq:ol-fit-full}) with $\mu_i =
\mu_{\rm sea} = m_{\pi}^2$ the masses of non-strange baryons $N$ and
$\Delta$ are determined.

We calculate the strange spectrum in two ways, using either the mass of $K$
or $\phi$ meson as input. As a preparation, we determine the hopping
parameter of up and down quarks $\kappa_{ud}$ by solving the equation
$\mps^2(\kappa_{ud};\kappa_{ud},\kappa_{ud})=m_{\pi}^2$ applying the chiral
formula Eq.~(\ref{eq:ps-fit-full-VWI}) and substituting $m_{\pi}$ obtained
above. The hopping parameter corresponding to the strange point $\kappa_s$
is then fixed by the relation
$\mps^2(\kappa_{ud};\kappa_{ud},\kappa_s)/m_{\pi}^2 =
M_{K}^2/M_{\pi}^2$. In the next step, $\kappa_s$ is used to determine the
mass of the $\eta_{ss}$, a fictitious pseudoscalar meson consisting of two
strange quarks, through
$m_{\eta_{ss}}^2=\mps^2(\kappa_{ud};\kappa_s,\kappa_s)$. Finally, values of
$m_{\pi}^2$ and $m_{\eta_{ss}}^2$ are inserted into
Eqs.~(\ref{eq:vec-fit-full}), (\ref{eq:dec-fit-full}) and
(\ref{eq:ol-fit-full}) to obtain the rest of the spectrum.

In an alternative determination using the $\phi$ meson mass as input, we
first calculate the mass of the $\eta_{ss}$ meson by using
Eq.~(\ref{eq:vec-fit-full}) and solving the equation
\begin{equation}
\frac{A^V + B^V_s m_{\pi}^2 + B^V_v m_{\eta_{ss}}^2 
  +  C^V_s m_{\pi}^4 + C^V_v m_{\eta_{ss}}^4 
  +  C^V_{sv} m_{\pi}^2 m_{\eta_{ss}}^2}{m_{\rho}} 
= \frac{M_{\phi}}{M_{\rho}}
\end{equation}
for $m_{\eta_{ss}}$. Substituting $m_{\pi}^2$ and $m_{\eta_{ss}}^2$ the
spectrum can be calculated as above, except for the $K$ meson, for which
first $\kappa_s$ is determined from
$m_{\eta_{ss}}^2=\mps^2(\kappa_{ud};\kappa_s,\kappa_s)$ and then inserted
into $m_K^2=\mps^2(\kappa_{ud};\kappa_{ud},\kappa_s)$.

\begin{figure*}[p]
\centerline{
\epsfxsize=8cm \epsfbox{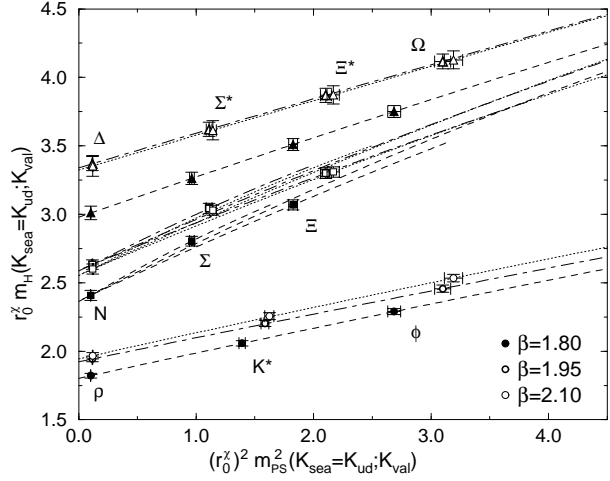}
}
\caption{Partially quenched spectrum at the physical sea quark mass. Lines
are obtained from 
Eqs.~(\ref{eq:vec-fit-full}), (\ref{eq:dec-fit-full}) and
(\ref{eq:ol-fit-full}) by fixing $\mu_{\rm sea} = m_{\pi}^2$. 
The strange spectrum, marked with
symbols on the lines, is obtained using $M_K$ as input.}
\label{fig:PQspectrum}
\end{figure*}

We list lattice spacings and the hopping parameters $\kappa_{ud}$ and
$\kappa_s$ in Table~\ref{tab:parameters}. Results for the hadron spectrum
are given in Table~\ref{tab:physical}. In Fig.~\ref{fig:PQspectrum} hadron
masses are plotted as a function of the valence quark mass $\mu_{\rm val}$. 
For this figure a normalization in terms of the Sommer scale $r_0^{\chi}$
is used to plot data at different lattice spacings together. Lines,
obtained from Eqs.~(\ref{eq:vec-fit-full}), (\ref{eq:dec-fit-full}) and
(\ref{eq:ol-fit-full}), correspond to a partially quenched world with sea
quarks equal to the physical up and down quarks.

\subsection{Continuum extrapolation}
\label{sec:cont-ex}

\begin{figure*}[p]
\centerline{
              \epsfxsize=7cm \epsfbox{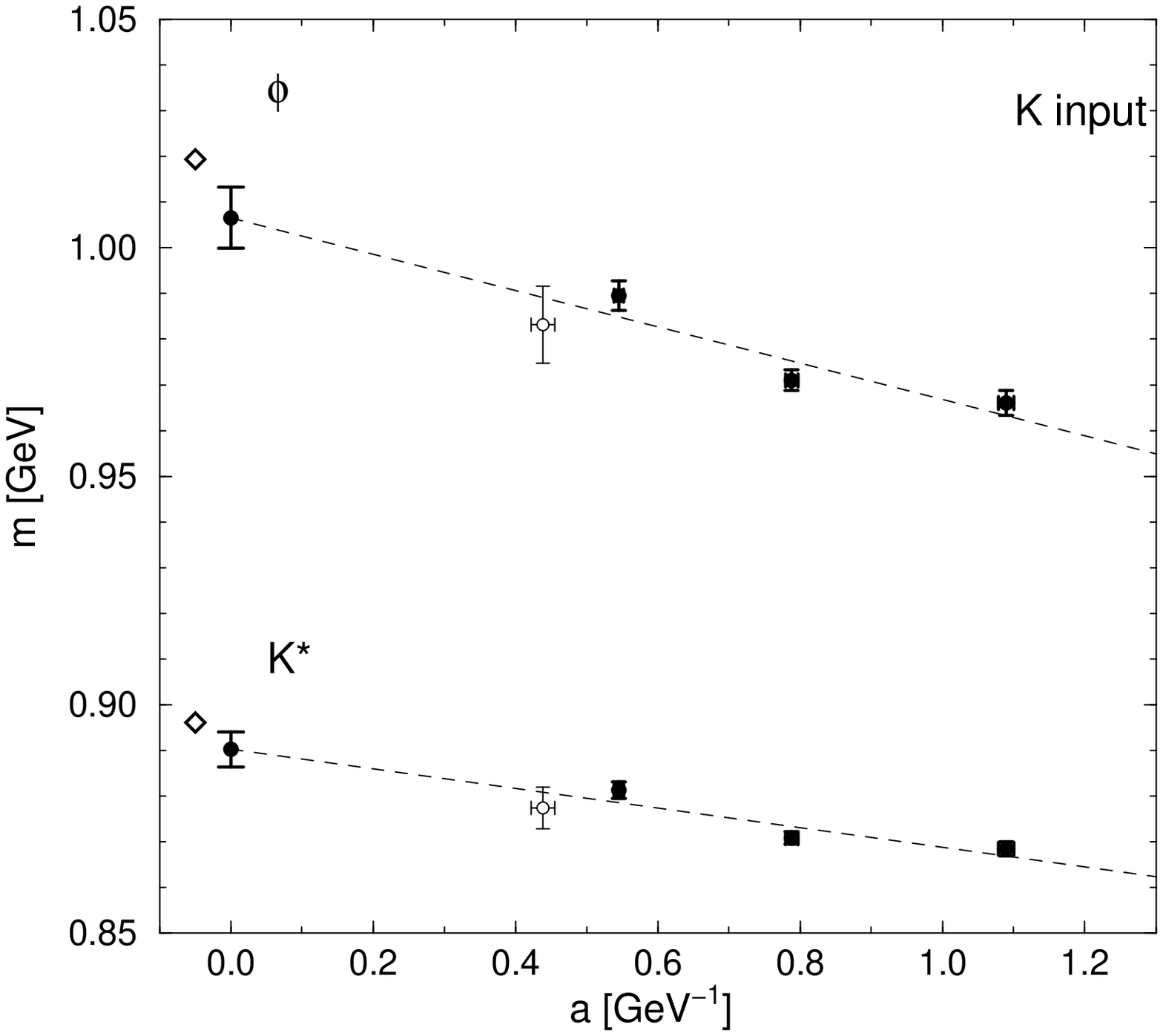}
\hspace{5mm}  \epsfxsize=7cm \epsfbox{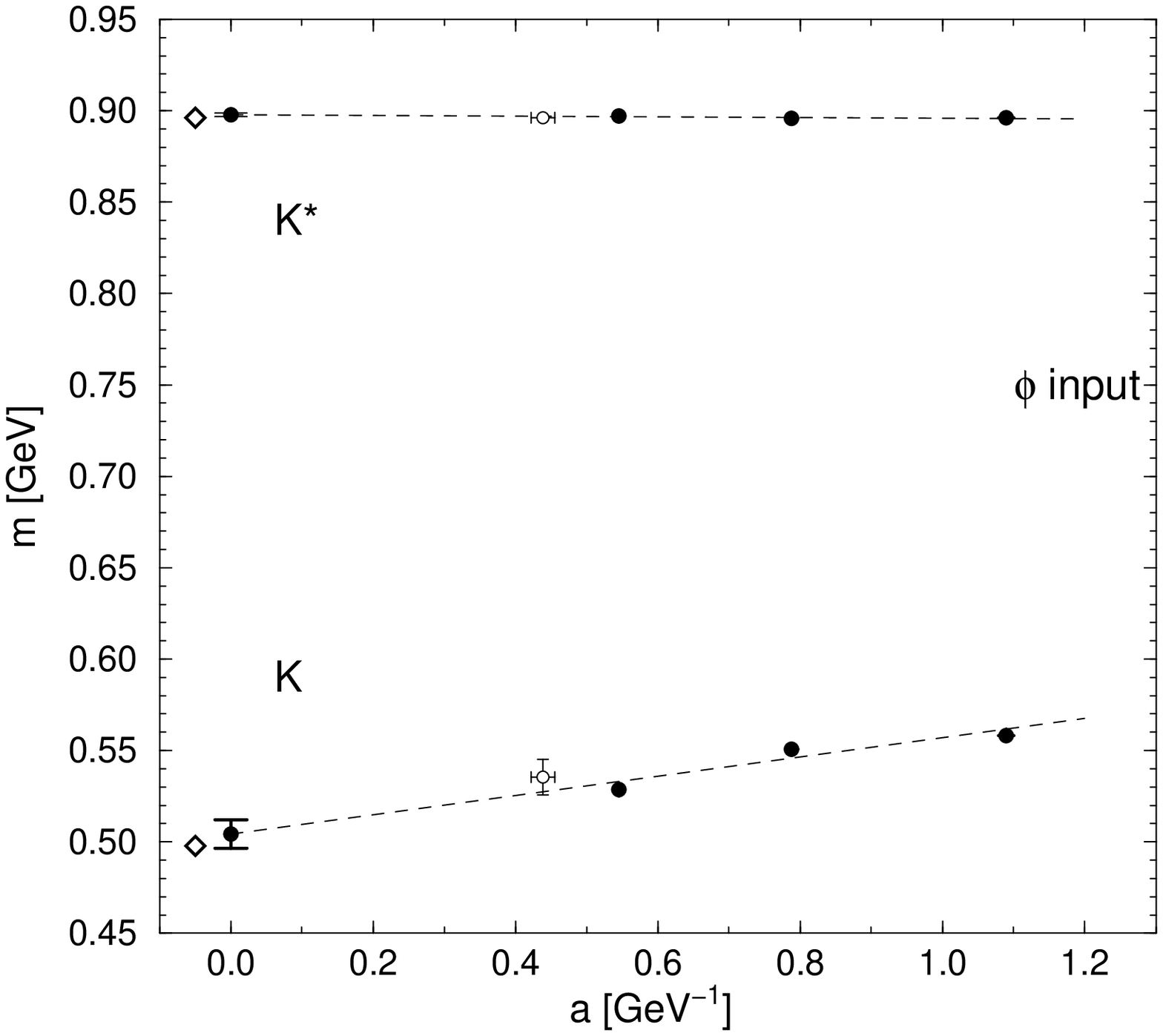}
}
\caption{Meson masses in full QCD as function of the lattice
spacing. Masses in the left hand figures have been obtained using the $K$
meson mass as input while the ones in the right hand figure have been
determined using the mass of the $\phi$ meson as input. Experimental values
are indicated with diamonds. Masses from the additional run at $\beta=2.2$
are shown with open symbols. Continuum values and extrapolation lines are
from a linear fit to the main data at three lattice spacings.}
\label{fig:mesonContExtNf2}
\end{figure*}

\begin{figure*}[p]
\centerline{
              \epsfxsize=7cm \epsfbox{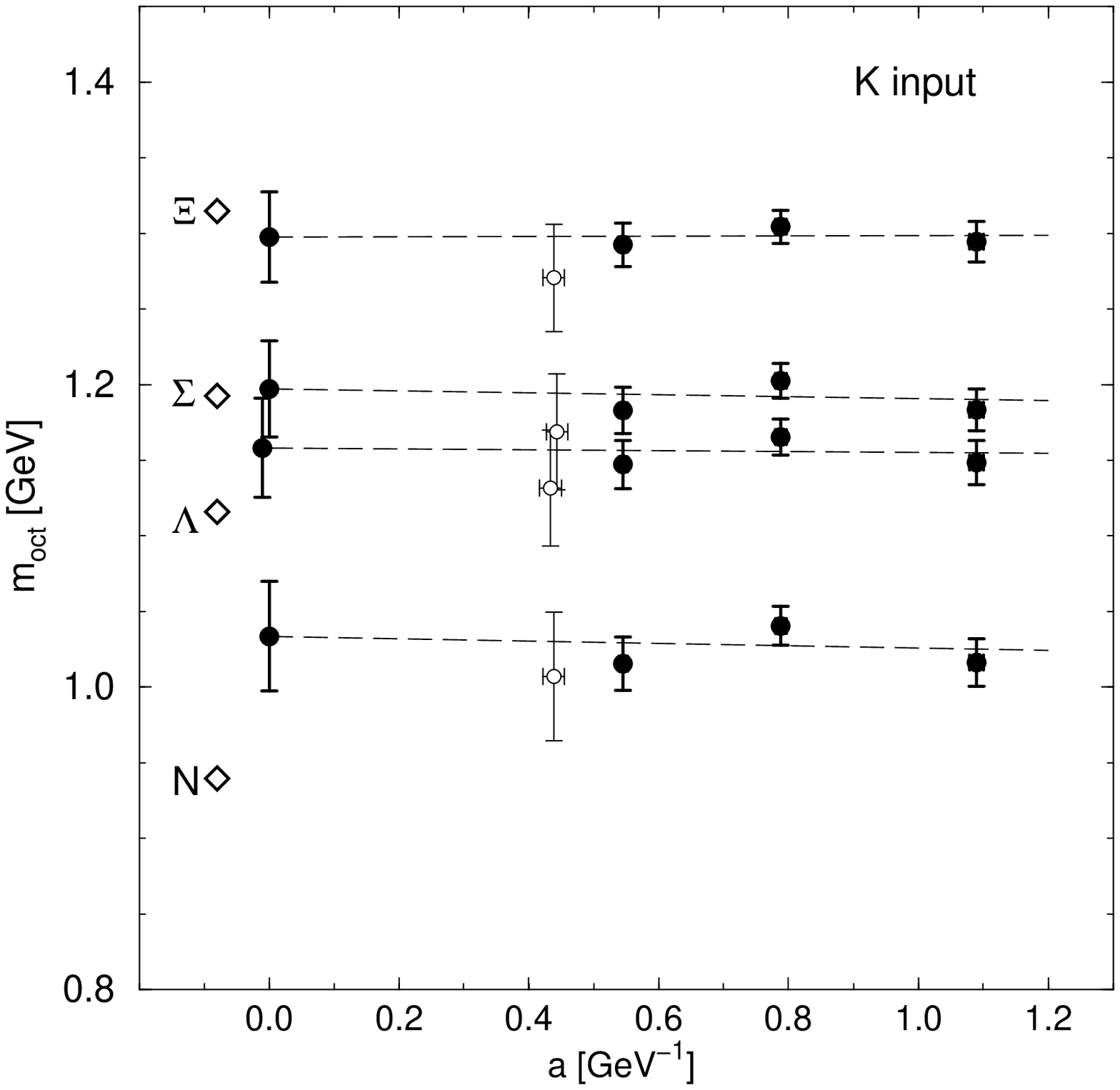}
\hspace{5mm}  \epsfxsize=7cm \epsfbox{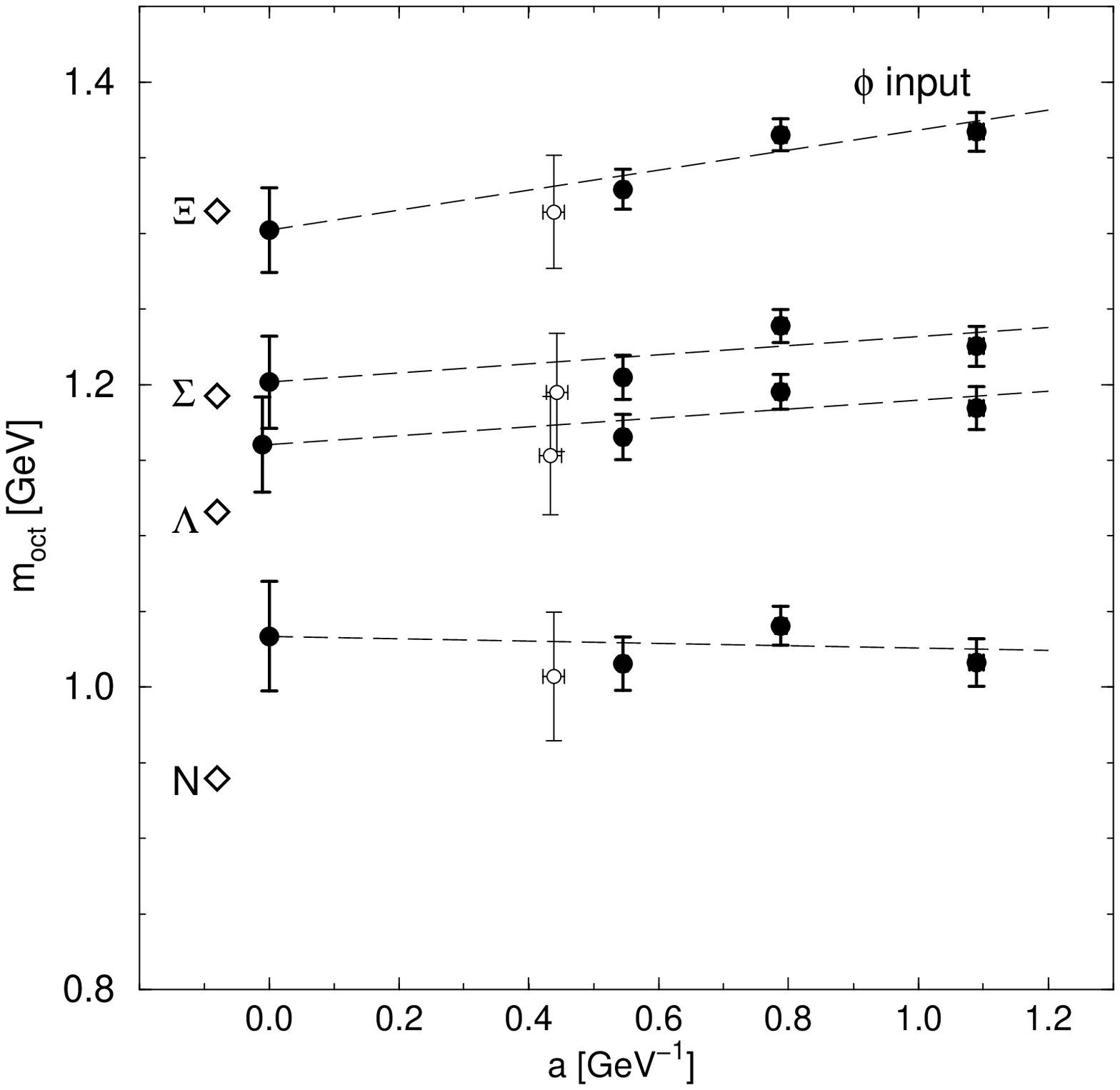}
}
\caption{Full QCD octet baryon masses as function of the lattice
spacing. The strange spectrum is determined with $K$ input (left figure) or 
$\phi$ input (right figure). Data represented with open symbols are from
the run at $\beta=2.2$.}
\label{fig:OctBarContExtNf2}
\end{figure*}

\begin{figure*}[p]
\centerline{
              \epsfxsize=7cm \epsfbox{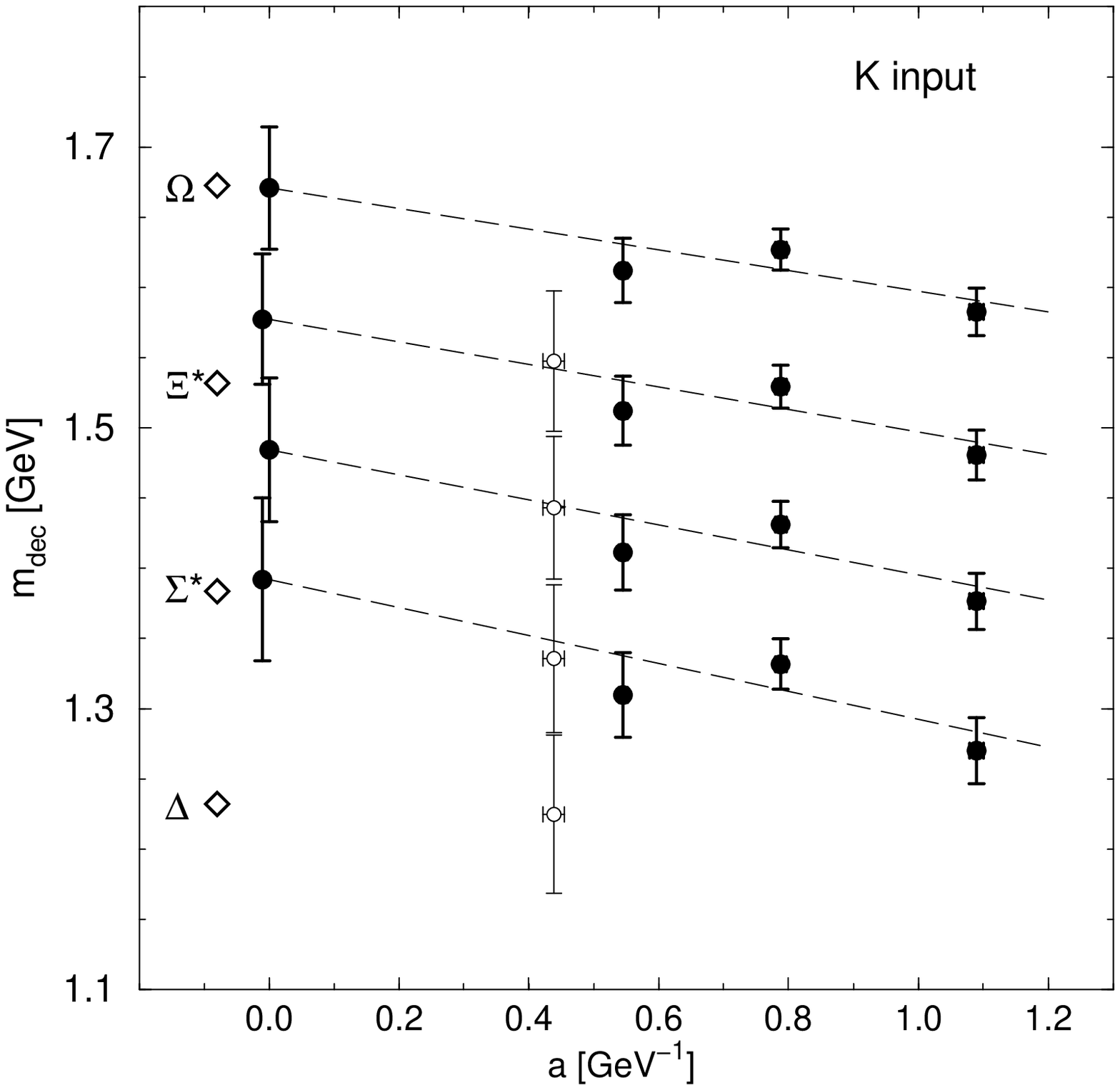}
\hspace{5mm}  \epsfxsize=7cm \epsfbox{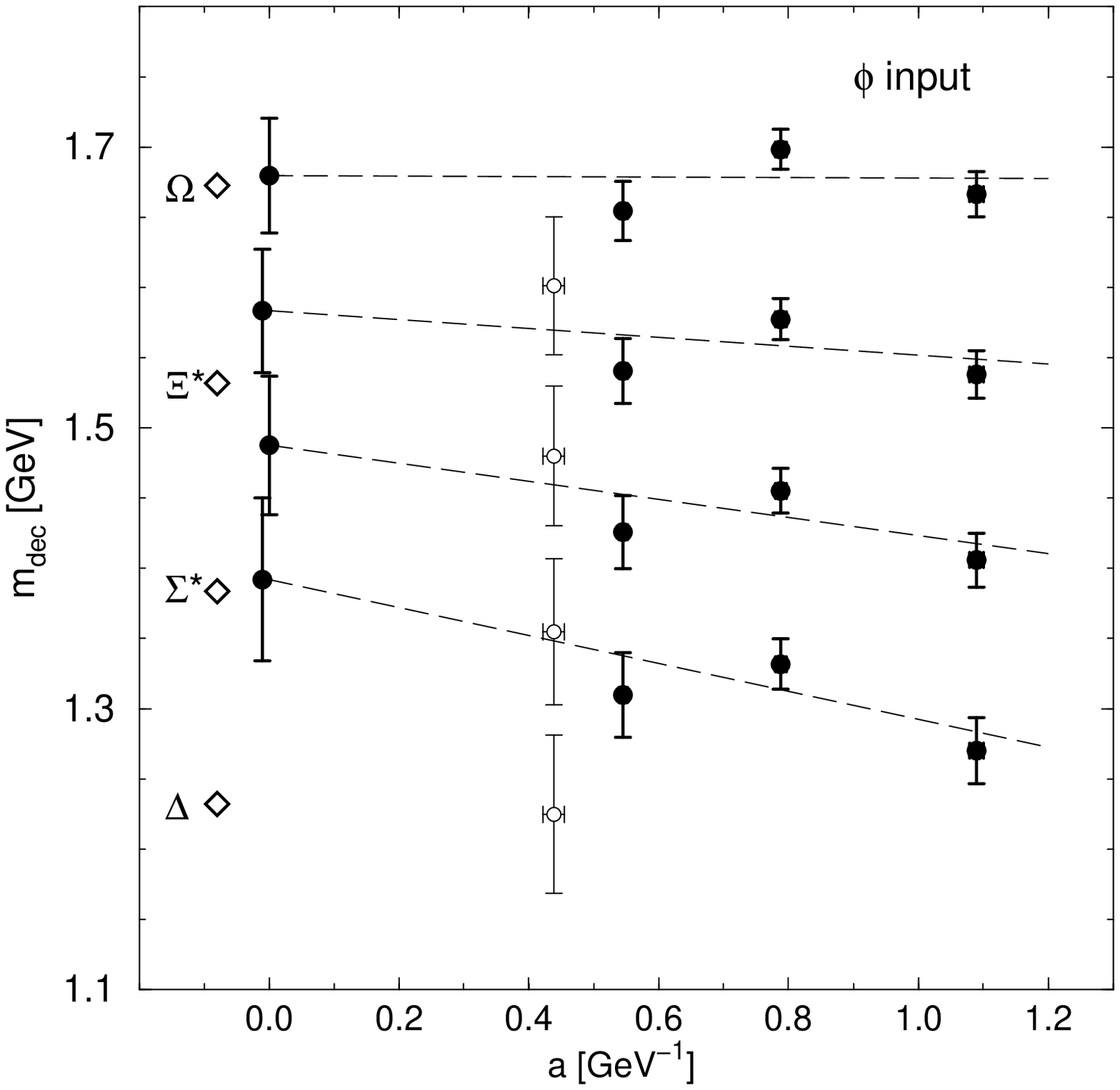}
}
\caption{Full QCD decuplet baryon masses as function of the lattice
spacing. The strange spectrum is determined with $K$ input (left figure) or 
$\phi$ input (right figure). Data represented with open symbols are from
the run at $\beta=2.2$.}
\label{fig:DecBarContExtNf2}
\end{figure*}

In Fig.~\ref{fig:mesonContExtNf2} we show meson masses as functions of the
lattice spacing. Baryon masses are plotted in
Figs.~\ref{fig:OctBarContExtNf2} and \ref{fig:DecBarContExtNf2}. Solid
symbols represent our main results at three lattice spacings with a
constant physical lattice size. Additional masses at $\beta=2.2$ with a
smaller lattice size are depicted with open symbols.

We find that scaling violations are contained within acceptable limits. The
largest scaling violation for mesons is observed in the $K$ meson mass
(using $\phi$ as input), which changes by 6\% between $a=0.22$~fm and
$a=0.11$~fm. The largest difference in baryon masses between these two
lattice spacings occurs with $\Delta$ for decuplet baryons and with $\Xi$
(with $K$ as input) for octet baryons, both amounting to 3\%.

The RG-improved gluon action leads to scaling violation which starts with
$O(a^2)$. With our quark action, since the clover coefficient $c_{\rm SW}$
is not tuned exactly at one-loop order, the leading scaling violation is
$O(g^2a)$. Here $g^2$ is the renormalized coupling constant
$g_{\overline{\rm MS}}^2(\mu)$~\cite{Aoki00} evaluated at a fixed scale
$\mu$, which is a constant. Higher order perturbative corrections of order
$g^4 a\log a$ can be neglected because in our short range of lattice
spacings $\log a$ is almost constant. Accordingly, we attempt continuum
extrapolations by applying linear fits to the main data at three lattice
spacings. We do not include results at $\beta=2.2$ because of its smaller
lattice size compared to the other runs. Lines from linear fits are plotted
in Figs.~\ref{fig:mesonContExtNf2}, \ref{fig:OctBarContExtNf2} and
\ref{fig:DecBarContExtNf2}. The slopes of the fits are small; parametrizing
the dependence on the lattice spacing as $m = m_{\rm cont}(1- \alpha a)$,
we find, using $m_K$ as input, typical values of $\alpha \approx
0.02$--0.04~GeV for mesons, $\alpha \approx -0.005$~GeV for octet baryons
and $\alpha \approx 0.04$--0.07~GeV for decuplet baryons.

The values of $\chi^2$ for these fits are $\chi^2/N_{\rm DF} \approx 5$--7
for mesons, resulting in a goodness of fit $Q\approx 1$--2\%. The quality
of fits is therefore marginal. Partly due to larger error bars, fits for
baryons are better with $\chi^2/N_{\rm DF} \approx 2$ corresponding to $Q
\approx 15\%$. Having only three data points does not allow us to explore
the magnitude of possible higher order terms of scaling violations. Hadron
masses extrapolated to the continuum limit with linear fits are listed in
Table~\ref{tab:physical}.

\subsection{Hadron spectrum in the continuum limit}

We observe that meson masses in the continuum limit are quite close to
experiment. When using $K$ as input, the differences for $K^*$ and $\phi$
are 0.7\% or 1.3\%, respectively, which amount to 1.6~$\sigma$ or
1.9~$\sigma$ in terms of the statistical error. When using $\phi$ as
input, the mass of the $K^*$ is within 0.2\% of experiment while the $K$
mass differs by 1.3\% which is still within the statistical error. As we
discuss in more detail in Sec.~\ref{sec:effects}, these results are
markedly improved from those of quenched QCD~\cite{quench.CPPACS} which
show deviation of about 10\% from experiment.

The situation is different for baryon masses. As is seen with $\Xi$ and
$\Sigma$ in the octet in Fig.~\ref{fig:OctBarContExtNf2} and with $\Omega$
in the decuplet in Fig.~\ref{fig:DecBarContExtNf2}, there is good agreement
with experimental masses when the strange quark content is high. The
difference from experiment increases as strange quarks are replaced with
up-down quarks, and the largest difference is observed for non-strange
baryons; the nucleon mass is larger than experiment by 10\% or
2.6~$\sigma$, and the difference for the $\Delta$ is 13\% or 2.8~$\sigma$.

This pattern of disagreement with experiment appears to be present already
at finite lattice spacings. Hence it is likely to be a systematic effect
rather than a statistical fluctuation. A possible reason behind this are
finite size effects arising from the lattice size of $La\approx 2.5$~fm.
We expect lighter baryons made of lighter quarks to be affected more from
these effects, which is consistent with the pattern we observe. A detailed
investigation is needed, however, since finite size effects in full QCD can
be quite complicated, arising from both sea and valence quarks wrapping
around the lattice in the spatial directions.

We add a remark for strange baryons. Masses obtained using either $K$ or
$\phi$ as input (left and right panels in Figs.~\ref{fig:OctBarContExtNf2}
and \ref{fig:DecBarContExtNf2}) differ at coarse lattice spacings. The
difference decreases with lattice spacing, however, and almost disappears
toward the continuum limit. This reassuring finding is connected with a
good agreement of the strange meson spectrum with experiment in the
continuum limit.


\section{Quenched QCD with improved actions}
\label{sec:quenched}

\subsection{Purpose}

Up to this stage we have discussed the two-flavor full QCD hadron spectrum.
In order to analyze how dynamical sea quarks manifest their presence in the
spectrum, we need to compare full QCD results with those of quenched QCD.

The quenched hadron spectrum has been examined in detail in
Ref.~\cite{quench.CPPACS}. Systematics of simulations in
Ref.~\cite{quench.CPPACS} differ, however, from those of two-flavor QCD in
the present work. The standard plaquette gluon action and the Wilson quark
action are used in Ref.~\cite{quench.CPPACS}, and the continuum
extrapolation is made from a finer range of lattice spacing $a\approx
0.1$--0.05~fm in~\cite{quench.CPPACS} as compared to $a\approx 0.2$--0.1~fm
in the present work. The lightest valence quark mass is pushed down to
$\mps/\mv \approx 0.4$ for quenched QCD while it only reaches $\mps/\mv
\approx 0.5$ in full QCD, and the physical lattice sizes are $La \approx
3$~fm for quenched QCD and $La \approx 2.5$~fm for full QCD.

We consider that a more direct comparison with a common choice of actions
over similar range of lattice parameters is desirable. Therefore we carry
out a new set of quenched simulations with the same set of improved actions
as employed for two-flavor full QCD.

\subsection{Matching quenched and full QCD simulations} 

We use the string tension to match the scale of quenched QCD with that of
full QCD, {\it i.e.,} for each value of $\beta$ and $\kappa_{\rm sea}$ at
which full QCD simulations are made, we make a corresponding quenched run
with $\beta$ chosen such that the string tension $\sigma$ in lattice units
takes the same value.

This is carried out at four values of $\kappa_{\rm sea}$ at $\beta=1.95$
and at $2.1$, and also at the chiral limit $\kappa_{\rm sea}=\kappa_c$ at
the two values of $\beta$ of full QCD. A summary of the 10 gauge couplings
used for quenched simulations is given in Table~\ref{tab:quench}. In the
same table we list measured string tensions, to be compared to the ones for
full QCD in Table~\ref{tab:sig}. We also quote lattice spacings obtained
using the rho meson mass as input.

Simulations are carried out using the same lattice size as the
corresponding full QCD runs, namely $16^3\times 32$ and $24^3 \times
48$. Physical lattice sizes vary therefore between $La \approx 2.6$~fm and
$La \approx 3.5$~fm.

\subsection{Simulation details}

Gauge configurations are generated with a combination of the 5-hit
pseudo-heat-bath algorithm with two SU(2) sub-matrices and the
over-relaxation algorithm. The two algorithms are mixed in the ratio of 1:4
and the combination is called an iteration. For vectorization and
parallelization of the simulation code, a 16-color algorithm is developed
for the RG-improved gauge action.

We skip 100 iterations between two configurations for hadron propagator
measurements. We check that this number of iterations is sufficient to
regard the configurations to be independent. We calculate hadron
propagators over 200 configurations per gauge coupling. These statistics
are comparable to the number of independent configurations in the full QCD
runs.

The measurement procedure parallels the one for full QCD. Hopping
parameters are chosen so that ratios $\mps/m_{\rm V}$ for degenerate mesons
match the ones of the corresponding full QCD run. For the quark matrix
inversion we use the same set of stopping conditions and smearing
parameters as the ones for corresponding full QCD runs. Masses are
extracted from hadron propagators with smeared sources using correlated
fits and fit ranges similar to those used for full QCD.

For chiral extrapolations we follow the strategy of fitting vector and
baryon masses as a function of measured pseudoscalar masses, and these in
turn as a function of valence quark masses. To be specific, we fit
pseudoscalar meson masses to the formula
\begin{equation}
\mps^2 \left (\kappa_{\rm val}^{(1)},\kappa_{\rm val}^{(2)} \right ) 
=  b_v m^{\rm VWI}_{\rm val}
   + c_v \left (m^{\rm VWI}_{\rm val} \right )^2
   + c_{vv} m^{\rm VWI}_{{\rm val}(1)} m^{\rm VWI}_{{\rm val}(2)},    
\label{eq:ps-fit-quench-VWI}
\end{equation}
where variables are defined as in Eqs.~(\ref{eq:VWIvalDef}) and
(\ref{eq:VWIvalAv}). This is the quenched analogy of
Eq.~(\ref{eq:ps-fit-full-VWI}) with terms containing $m^{\rm VWI}_{\rm
sea}$ dropped. Similarly, when making fits as a function of AWI quark
masses we employ the formula
\begin{equation}
\mps^2 \left (\kappa_{\rm val}^{(1)},\kappa_{\rm val}^{(2)} \right )
 = b'_v m^{\rm AWI}_{\rm val} +  c'_v \left (m^{\rm AWI}_{\rm val} \right )^2,
\label{eq:ps-fit-quench-AWI}
\end{equation}
which corresponds to Eq.~(\ref{eq:ps-fit-full-AWI}) for full QCD.

\begin{figure*}[p]
\centerline{
\epsfxsize=8cm \epsfbox{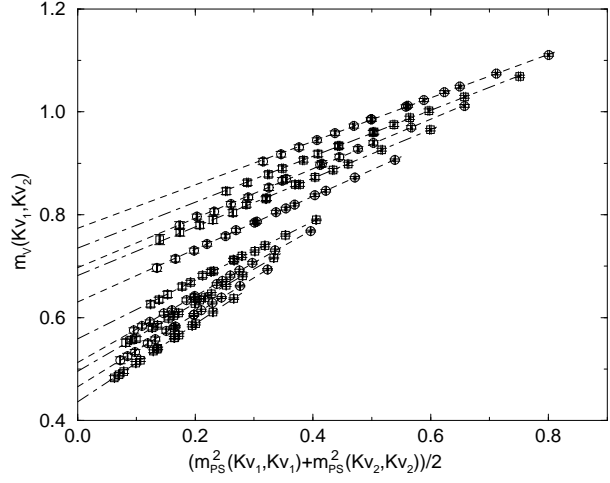}
}
\caption{Chiral extrapolations of vector meson masses in quenched
QCD. Lines are from linear fits with Eq.~(\ref{eq:vec-fit-quench}).}
\label{fig:chiralVecQ}
\end{figure*}

For vector mesons an inspection of mass data, plotted in
Fig.~\ref{fig:chiralVecQ}, shows that they are well described by a linear
function. If we nevertheless perform a quadratic fit the coefficient of the
quadratic term is ill defined with large error bars. We therefore employ
fits with
\begin{equation}
\mv \left (\kappa_{\rm val}^{(1)},\kappa_{\rm val}^{(2)} \right )
    =  A^V + B^V_v \mu_{\rm val}, 
\label{eq:vec-fit-quench}
\end{equation}
as shown in Fig.~\ref{fig:chiralVecQ}.
Parameters of chiral fits to mesons with 
Eqs.~(\ref{eq:ps-fit-quench-VWI}), (\ref{eq:ps-fit-quench-AWI}) 
and (\ref{eq:vec-fit-quench}) are given in Table~\ref{tab:meson-fit-quench}.

\begin{figure*}[p]
\centerline{
\epsfxsize=8cm \epsfbox{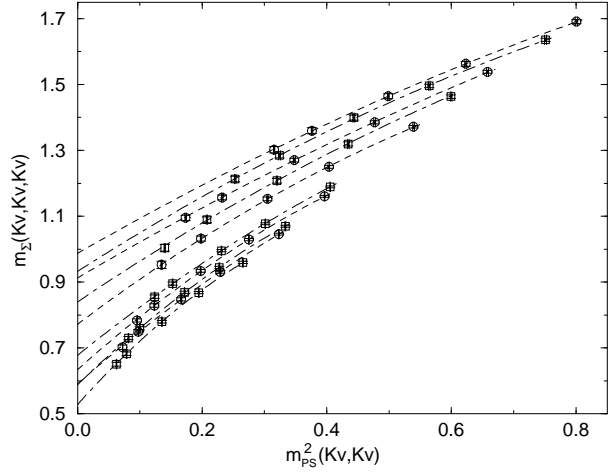}
}
\caption{Chiral extrapolations of octet baryon masses in quenched
QCD. While fits have been made to $\Sigma$ and $\Lambda$ type baryons of
all degeneracies together, only data and lines for degenerate masses are
plotted for the sake of clarity.}
\label{fig:chiralOctQ}
\end{figure*}

\begin{figure*}[p]
\centerline{
\epsfxsize=8cm \epsfbox{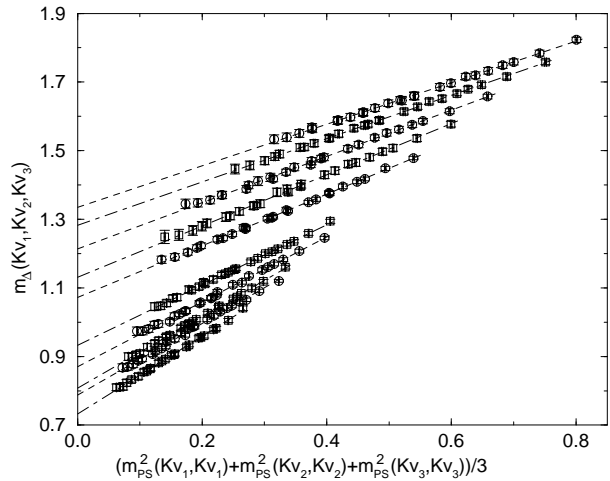}
}
\caption{Chiral extrapolations of decuplet baryon masses in quenched
QCD. Lines are from linear fits as described in the text.}
\label{fig:chiralDecQ}
\end{figure*}

Analysis of baryon masses proceeds in a similar way. For decuplet baryons
we again find quadratic terms in quark masses to be unnecessary. Data for
baryon masses together with fits are plotted in Figs.~\ref{fig:chiralOctQ}
and \ref{fig:chiralDecQ}.

\subsection{Results}

\begin{figure*}[p]
\centerline{
              \epsfxsize=7cm \epsfbox{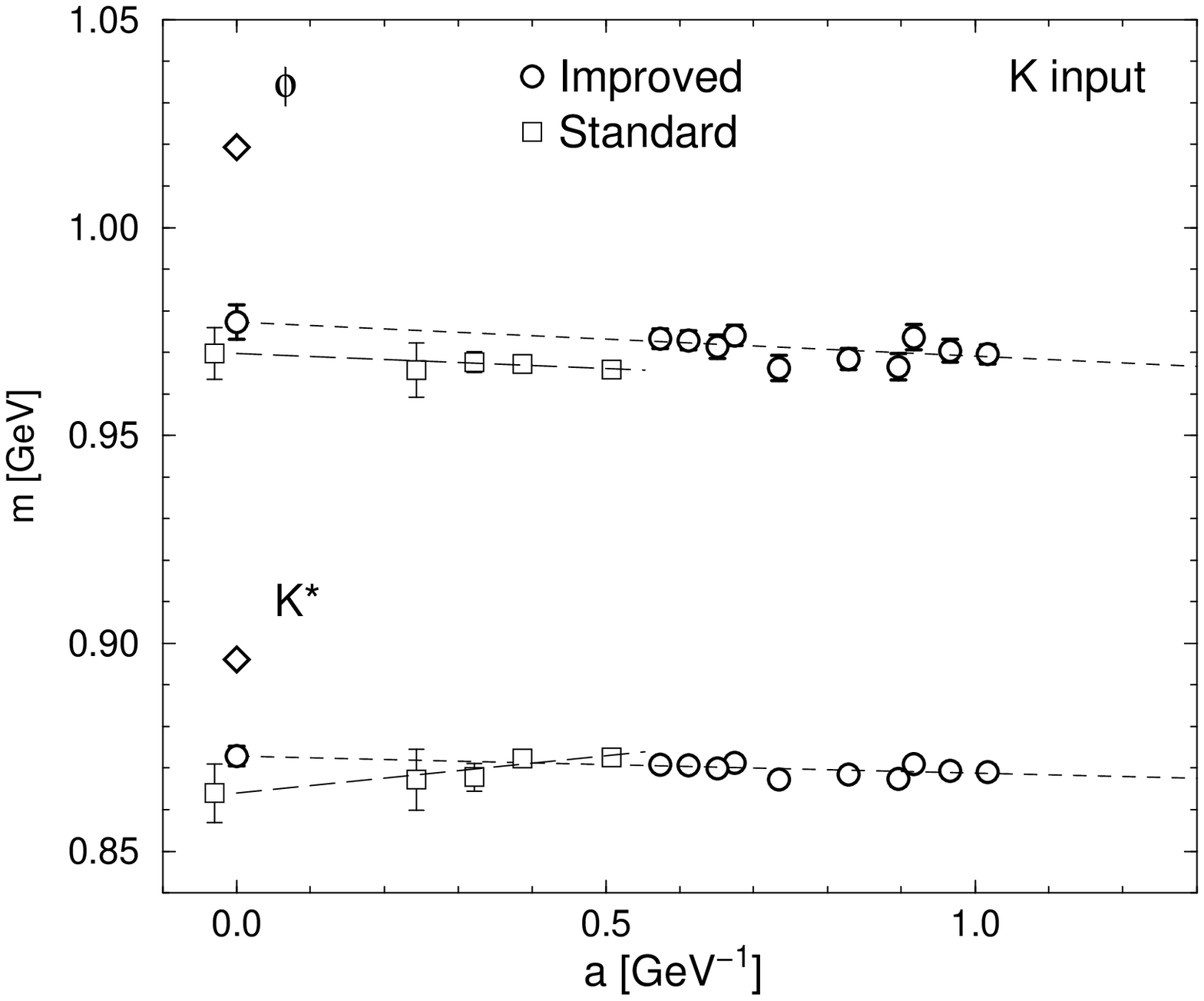}
\hspace{5mm}  \epsfxsize=7cm \epsfbox{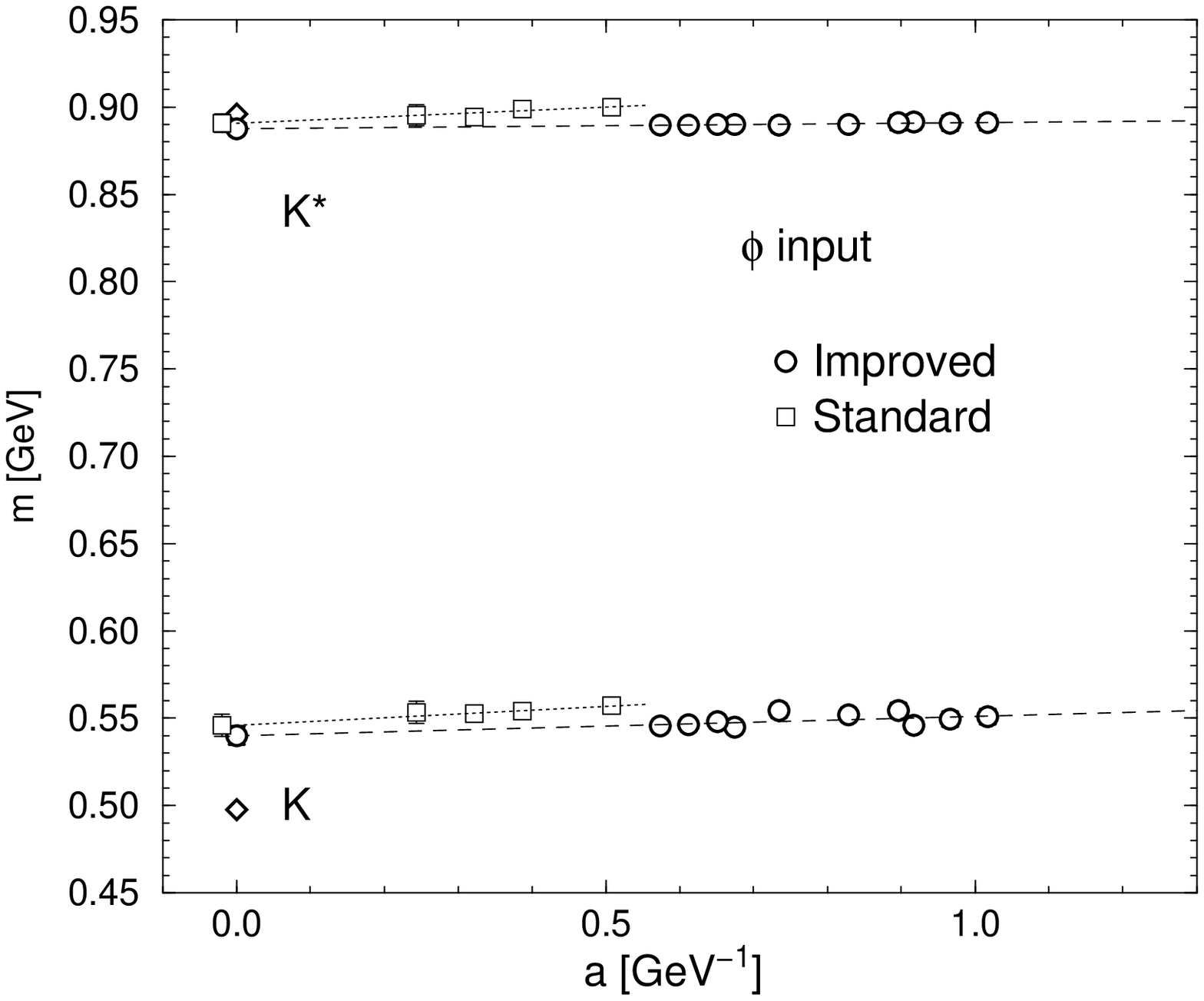}
}
\caption{Meson masses in quenched QCD with improved and standard actions.}
\label{fig:mesonContExtQuench}
\end{figure*}

\begin{figure*}[p]
\centerline{
              \epsfxsize=7cm \epsfbox{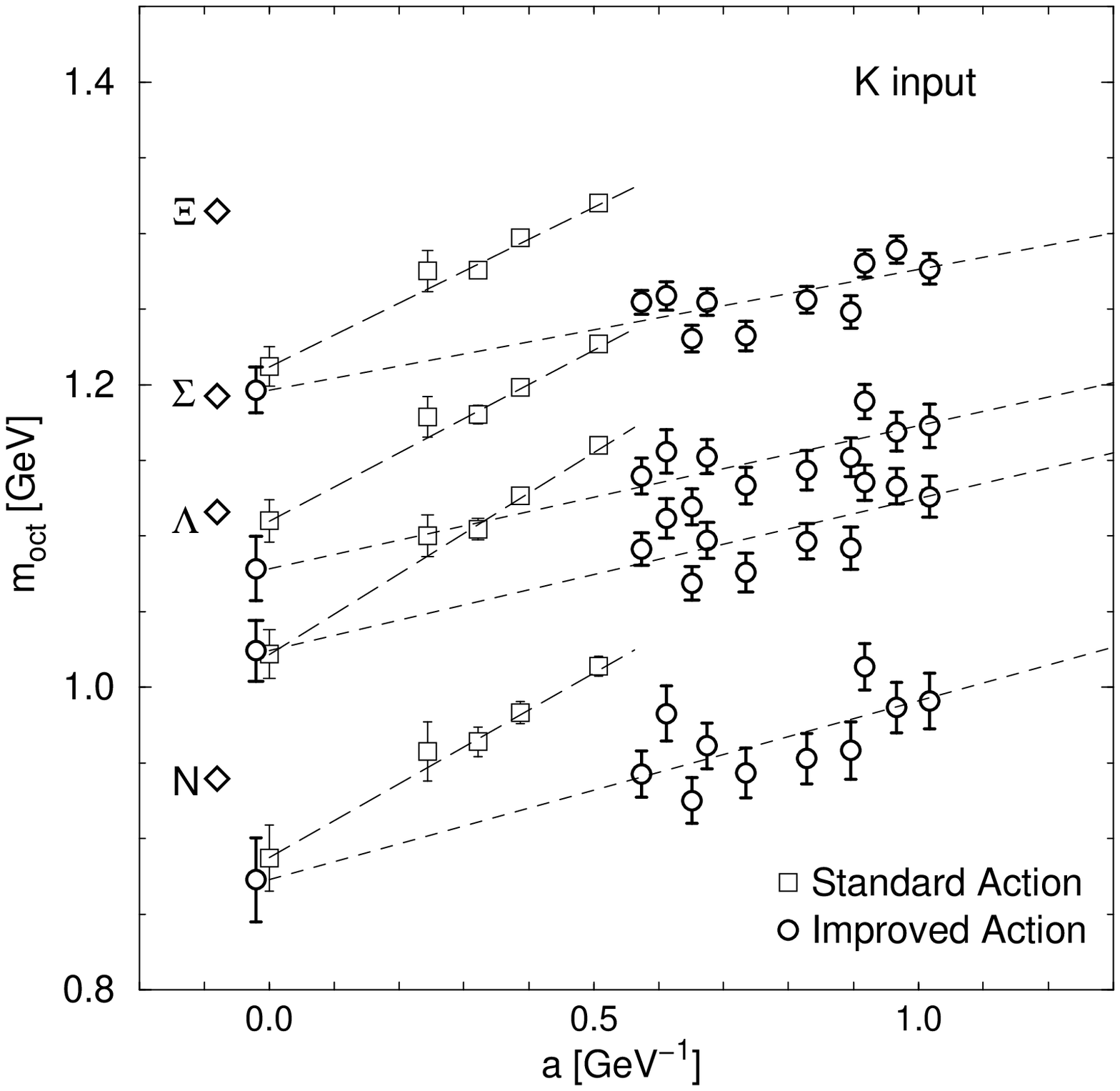}
\hspace{5mm}  \epsfxsize=7cm \epsfbox{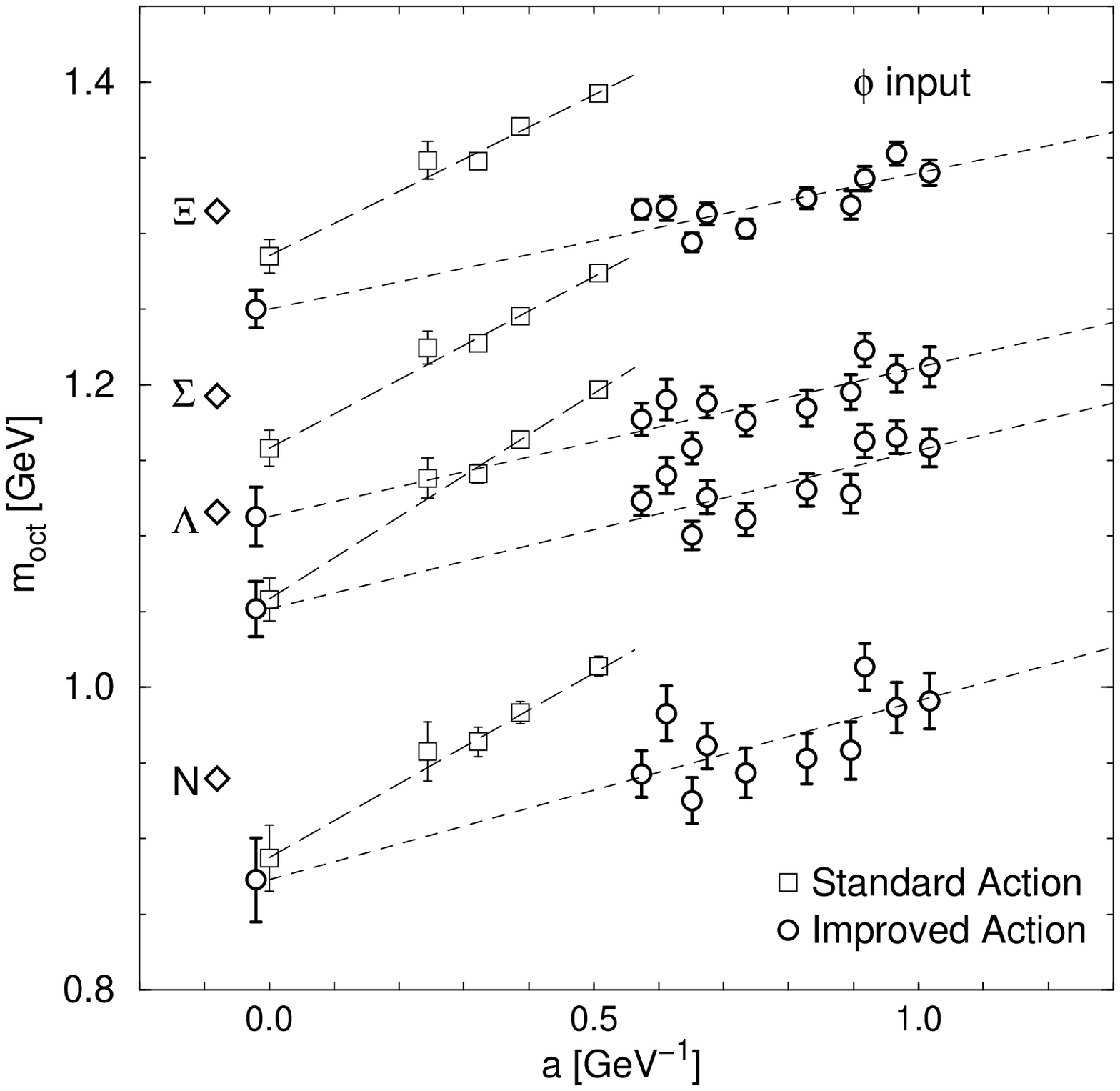}
}
\caption{Octet baryon masses in quenched QCD with improved and standard
actions. The strange spectrum is determined with $K$ input (left figure) or 
$\phi$ input (right figure).} 
\label{fig:OctBarContExtQuench}
\end{figure*}

\begin{figure*}[p]
\centerline{
              \epsfxsize=7cm \epsfbox{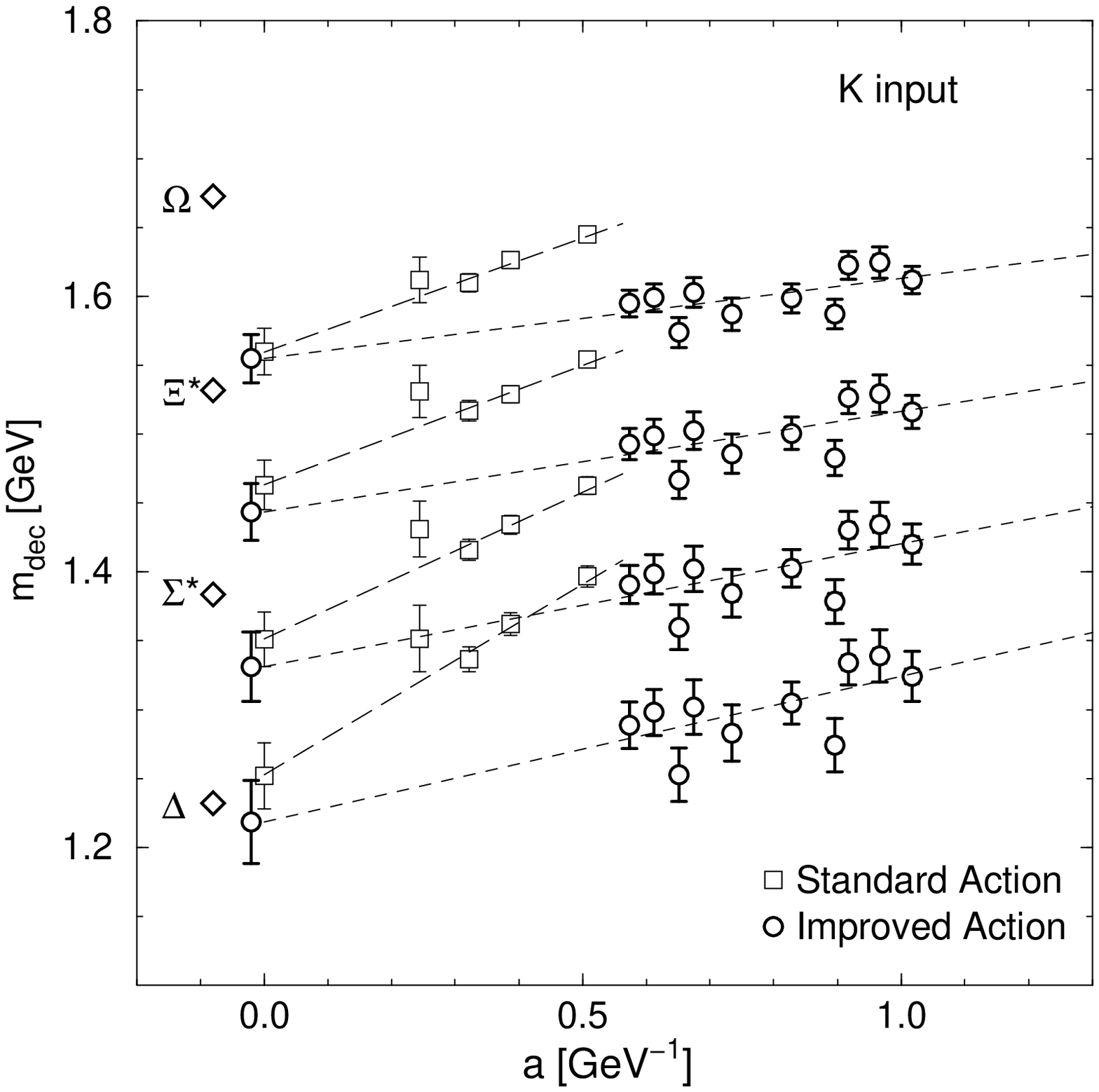}
\hspace{5mm}  \epsfxsize=7cm \epsfbox{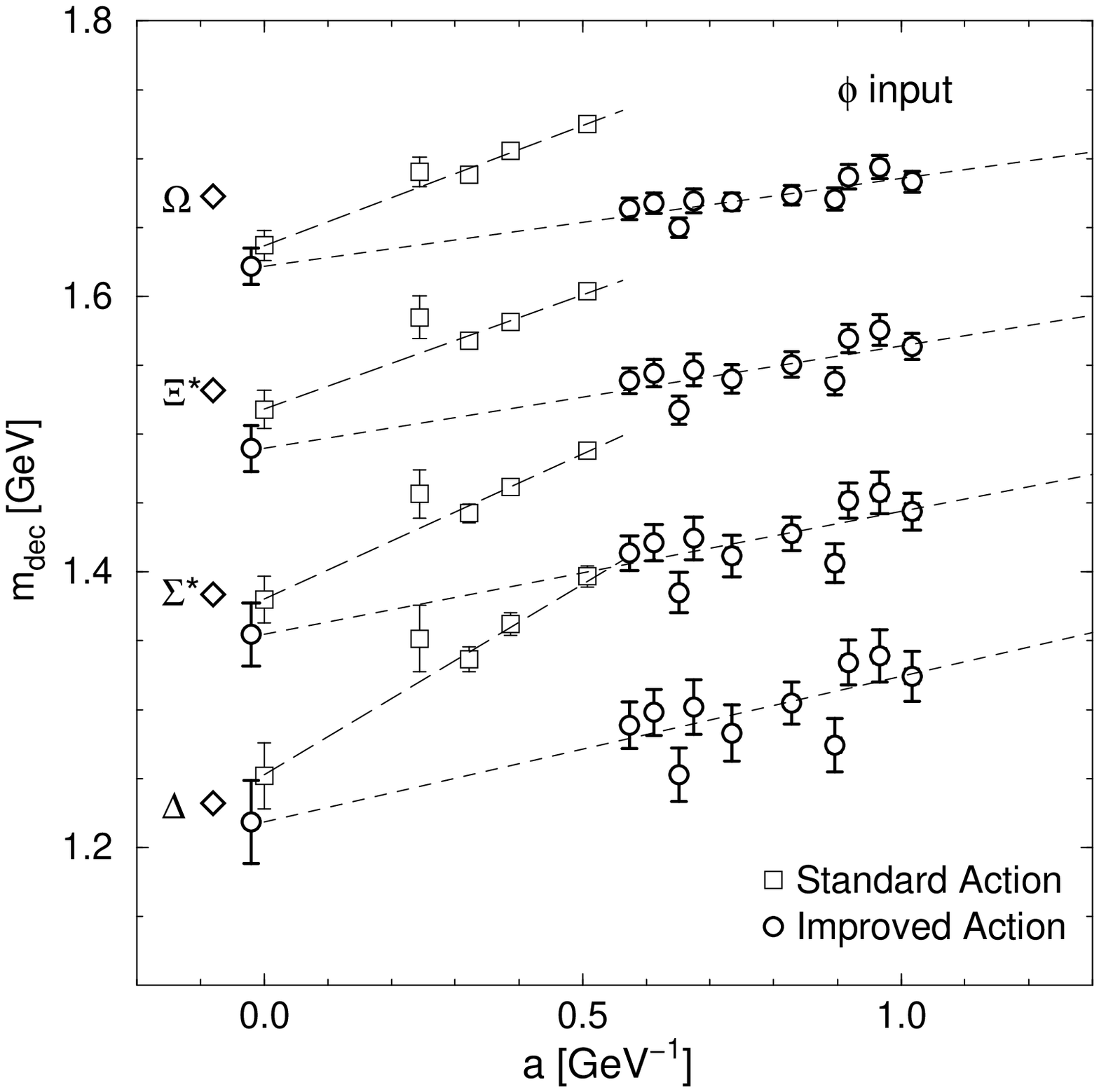}
}
\caption{Decuplet baryon masses in quenched QCD with improved and standard
actions. The strange spectrum is determined with $K$ input (left figure) or
$\phi$ input (right figure).}
\label{fig:DecBarContExtQuench}
\end{figure*}

Physical hadron masses are summarized in Table~\ref{tab:physicalNf0}. They
are plotted as a function of the lattice spacing in
Fig.~\ref{fig:mesonContExtQuench} for mesons and in
Figs.~\ref{fig:OctBarContExtQuench} and \ref{fig:DecBarContExtQuench} for
baryons.

In the same figures we also plot hadron masses obtained with the standard
action in Ref.~\cite{quench.CPPACS}. In this work, the analysis was made
with two sets of functions for chiral extrapolation. The main method used
functional forms predicted from quenched chiral perturbation theory. As an
alternative method polynomial fits were also employed. It was found that
results from the two methods are consistent with each other within errors
after the continuum extrapolation. In particular, conclusions on the
deviation of the quenched spectrum were not altered by two different
methods. Since in this work we use polynomial fits for the analysis, we
take hadron masses from polynomial fits in Ref.~\cite{quench.CPPACS} for a
comparison within quenched QCD.

We perform continuum extrapolations of hadron masses for the improved
action linearly in the lattice spacing in accordance with the leading
scaling violation discussed in Sec.~\ref{sec:cont-ex}. Good $\chi^2/N_{\rm
DF} \approx 1$ are obtained for meson masses. Baryon mass data exhibit some
scatter and as a result larger $\chi^2/N_{\rm DF}$ are observed. The
largest value, reached for the $\Xi$ baryon, is $\chi^2/N_{\rm DF} = 2.8$;
hence we consider the scatter to be still within the limits of statistical
fluctuations.

Comparing masses in the continuum limit a good agreement is found between
calculations with the standard and improved actions. All results are
consistent within the statistical accuracy. This is a confirmation that the
quenched light hadron spectrum deviates from
experiment~\cite{quench.CPPACS}.

Meson masses from the two choices of actions both show very good scaling,
and they are already in agreement even at finite lattice spacings. For
baryons scaling behavior is improved for the improved action. This is in
accordance with our initial study of action improvement~\cite{comparative},
notwithstanding that this study was carried out for full QCD. The largest
scaling violation in improved baryon masses is observed for the nucleon
with a difference of 14\% between $a^{-1}\approx 1$~GeV and the continuum
limit.


\section{Sea quark effects in the light hadron spectrum}
\label{sec:effects}

\subsection{Light meson spectrum}
\label{sec:effects-meson}

\begin{figure*}[p]
\centerline{
              \epsfxsize=7cm \epsfbox{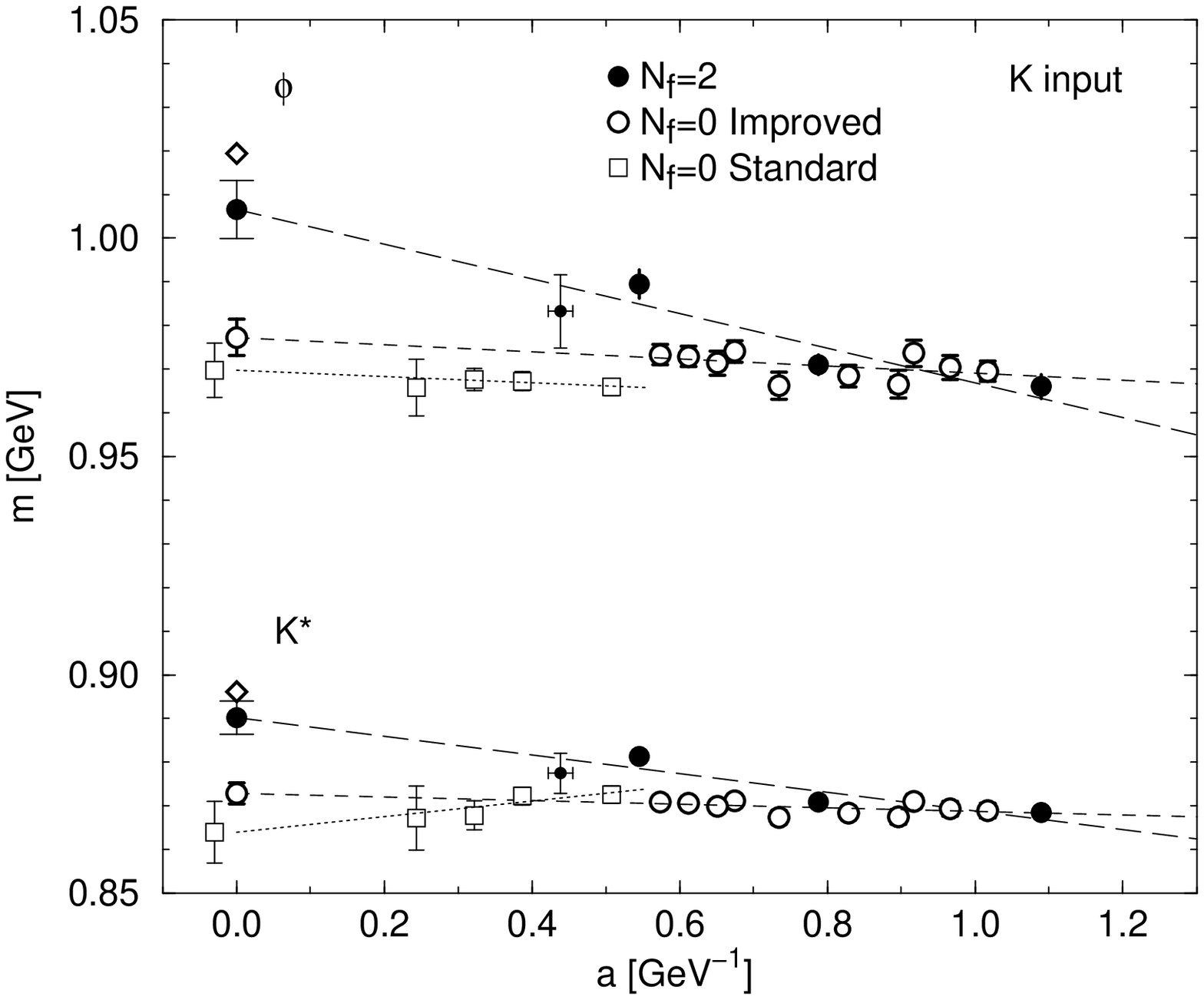}
\hspace{5mm}  \epsfxsize=7cm \epsfbox{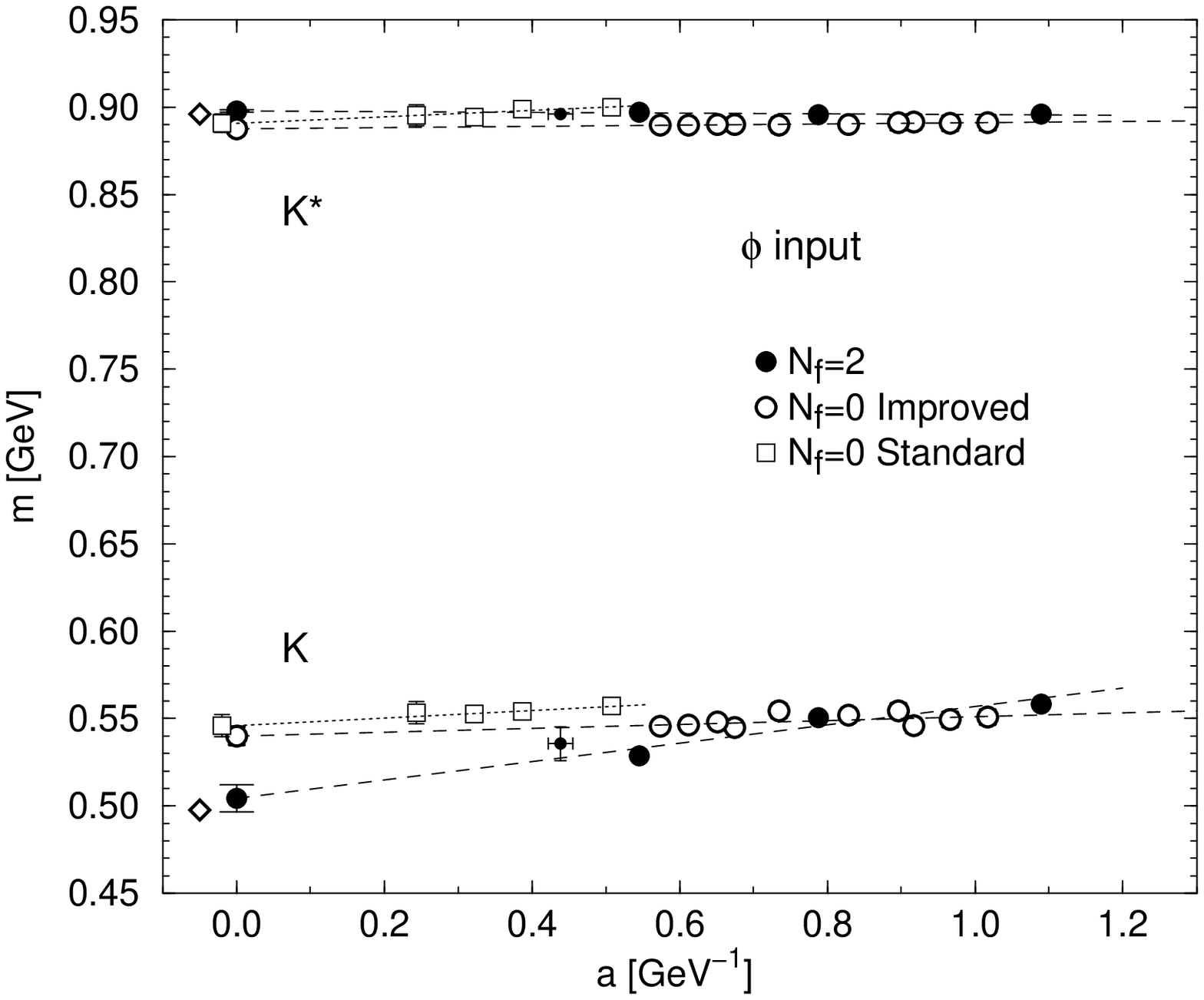}
}
\caption{Comparison of meson masses in full and quenched QCD. Data from
the additional full QCD run at $\beta=2.2$ are shown with small filled
circles.}
\label{fig:mesonFullQuench}
\end{figure*}

In Fig.~\ref{fig:mesonFullQuench} we compare the continuum extrapolation of
vector meson masses using the $K$ or $\phi$ meson mass as input for full
QCD and for the two quenched calculations. The deviation of the quenched
spectrum from experiment is considerably reduced in full QCD. For the $K^*$
meson the deviation is reduced from 2.6\% (3.1\% with the standard action)
to 0.7\%, and for the $\phi$ meson from 4.1\% (4.9\%) to 1.3\%, if the $K$
meson mass is used as input. Using the $\phi$ meson mass as input, the
difference in the $K^*$ meson is less than 1\% for both quenched and full
QCD, while the deviation for the $K$ meson is reduced from 8.5\% (9.7\%) in
quenched QCD to 1.3\% in full QCD. We consider this improvement in the
meson spectrum to be a manifestation of sea quark effects.

An important factor in reaching this conclusion is the continuum
extrapolation. At finite lattice spacings the difference between full and
quenched QCD is not obvious. At two coarse lattice spacings in particular,
the two sets of data are roughly consistent. However, the trend towards the
continuum limit is different. Full QCD leads to an increase for the $K^*$
and $\phi$ meson mass (decreasing for the $K$ meson mass) in contrast to a
flatter behavior in the quenched masses. A support that these trends are
not just fluctuations is provided by the additional calculation at
$\beta=2.2$, showing higher (lower) lying values, as can be seen from small
filled circles in Fig.~\ref{fig:mesonFullQuench}.

\begin{figure*}[p]
\centerline{
\epsfxsize=7cm \epsfbox{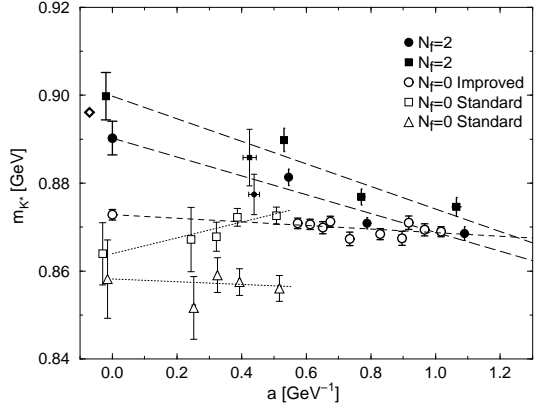}
}
\caption{Influence of choice of functional form for chiral extrapolation on
the $K^*$ mass. Filled symbols are for full QCD where for chiral
extrapolations Eq.(\ref{eq:vec-fit-full}) (circles) or
Eq.(\ref{eq:vec-fit-full-15}) (squares) are used. Data 
at $\beta=2.2$ are shown with small filled
symbols. Masses in quenched QCD with
the standard action are shown with open squares for polynomial chiral fits
or with open triangles for fits based on quenched chiral perturbation theory.}
\label{fig:KstSyst}
\end{figure*}

Let us discuss systematic errors which are relevant for this conclusion. In
Fig.~\ref{fig:KstSyst} we show how the $K^*$ meson mass changes when
different functional forms are used for chiral extrapolation. Filled
squares represent masses obtained using the fit with
Eq.~(\ref{eq:vec-fit-full-15}) instead of our standard analysis plotted
with filled circles. There is a noticeable effect on the $K^*$ mass, which
increases by 1\% in the continuum limit. A similar effect is seen for the
quenched data where we show results of Ref.~\cite{quench.CPPACS} for two
ways of chiral extrapolation. The trend remains, however, that the
continuum value for full QCD lies much closer to experiment than in
quenched QCD.

Another source of systematic errors is the continuum extrapolation. Within
the small number of data points available for full QCD, we may estimate the
upper error by making an extrapolation from the two points at $\beta=1.95$
and 2.1, and the lower error by taking the value at $\beta=2.1$. For the
$K^*$ meson mass this yields $m_{K^*}=0.890(4)^{+15}_{-9}$~GeV where the
second error represents the systematic error estimated in this way. For a
complementary analysis in the quenched simulation with the improved action,
we make a linear fit to the five points with fine lattice spacings
corresponding to the full QCD point at $\beta=2.1$ for the upper error, and
take the left-most point with the finest lattice spacing for the lower
error. We then obtain $m_{K^*}=0.873(2)^{+8}_{-2}$~GeV.

Similar analyses lead to $m_{\phi}=1.007(7)^{+25}_{-17}$~GeV and
$m_{K}=0.504(8)^{+25}_{-25}$~GeV for full QCD compared to
$m_{\phi}=0.977(4)^{+16}_{-4}$~GeV and $m_{K}=0.540(5)^{+6}_{-18}$~GeV for
quenched QCD with improved actions. Hence systematics of the continuum
extrapolation are unlikely to annul a closer agreement of full QCD masses
with experiment compared to quenched QCD.

In summary we find that effects of dynamical sea quarks are present beyond
the systematic as well as statistical uncertainties in strange meson
masses.

\subsection{$J$ parameter}

A useful quantity to quantify sea quark effects in the meson sector is the
$J$ parameter~\cite{UKQCDJ} defined by
\begin{equation}
J = \left . \mv \frac{d\,\mv}{d\,\mps^2} \right |_{\mv/\mps=M_{K^*}/M_K=1.8},
\label{eq:J}
\end{equation}
where only valence quark masses are to be varied in the differentiation. In
the real world this corresponds to a comparison between strange and
non-strange mesons. The derivative in Eq.~(\ref{eq:J}) can be replaced by a
finite difference and an ``experimental'' value for $J$ is then obtained as
\begin{equation}
J^{\rm exp} = M_{K^*} \frac{M_{K^*} - M_\rho}{M^2_{K} - M^2_\pi} = 0.48.
\label{eq:Jexp}
\end{equation}

\begin{figure*}[p]
\centerline{
\epsfxsize=10cm \epsfbox{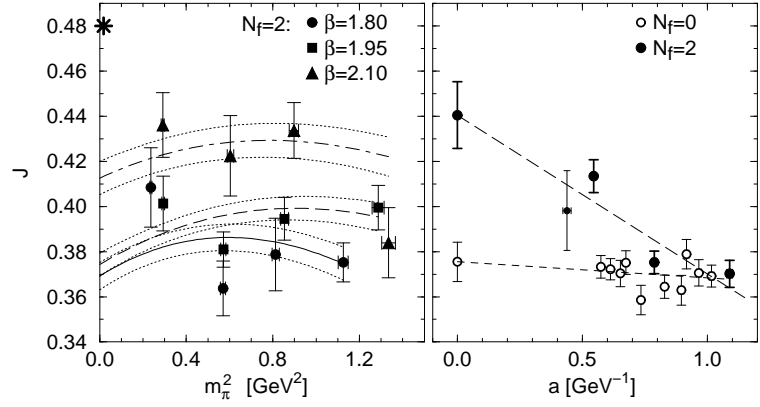}
}
\caption{The parameter $J$ in full QCD (left figure) as a
function of the sea pion mass and as a function of the lattice spacing in
quenched and full QCD (right figure). Individual points in the left figure
are from separate partially quenched fits while lines are from
combined fits. The star denotes the experimental value. Points for full QCD 
in the right figure are at the physical pion mass.}
\label{fig:J}
\end{figure*}

We calculate $J$ from fits to vector mesons as functions of pseudoscalar
mesons in two different ways. In the first one we use combined fits with
Eq.~(\ref{eq:vec-fit-full}), keep $\mu_{\rm sea}$ fixed and calculate
derivatives with respect to $\mu_{\rm val}$. This leads to the curves
shown on the left side of Fig.~\ref{fig:J}. For the second method we employ
separate partially quenched fits for each simulated sea quark. We use
quadratic fit functions obtained from dropping all terms containing
$\mu_{\rm sea}$ in Eq.~(\ref{eq:vec-fit-full}). Results are plotted with
filled symbols in Fig.~\ref{fig:J}. They tend to scatter more since, in
contrast to combined fits, no smoothness in the sea quark mass is imposed
for separate fits. The two methods yield consistent results within at most
two standard deviations, showing a trend of increase as the lattice spacing
is reduced. At fixed lattice spacing, on the other hand, we do not see a
clear dependence as function of the sea quark mass.

On the right hand side of Fig.~\ref{fig:J} we plot $J$ at the physical
point for quenched and two-flavor full QCD as a function of lattice
spacing. For quenched QCD, the values do not show much variation, and a
linear extrapolation to the continuum limit gives $J=0.375(9)^{+38}_{-2}$
where the second error represents the systematic error estimated in the
same way as in Sec.~\ref{sec:effects-meson}. This is consistent with
earlier observations of a too small value of $J$ in quenched QCD.

Full QCD data at $\beta=1.8$ and 1.95 do not differ much from this value.
It is intriguing, however, that at $\beta=2.1$ (and also $\beta=2.2$) $J$
is sizably larger. Consequently the continuum value of
$J=0.440(15)^{+59}_{-27}$, estimated by a linear extrapolation, lies much
closer to experiment.

\subsection{Sea quark mass dependence}

An interesting question with dynamical sea quark effects is how their
magnitude depends on sea quark mass. We examine this point by calculating
the mass ratio $m_{K^*}/m_\rho$ for fixed valence quark masses as a
function of sea quark mass.

The analysis proceeds in the following steps. We leave the sea quark mass
parametrized by $\mu_{\rm sea}$ as a free parameter, and first determine
the valence pion mass ``$m_\pi$'' and the rho meson mass ``$m_\rho$''
corresponding to a given ratio $\mps / \mv =$``$m_{\pi}$''$/$``$m_{\rho}$''
which may be different from the physical one, {\it e.g.,} $\mps / \mv =
0.5$ in an example shown below. In the next step the strange pseudoscalar
meson mass ``$m_{\eta_{ss}}$'' is fixed by a phenomenological ratio
``$m_{\eta_{ss}}$''$/$``$m_{\phi}$''$=\sqrt{2M_K^2-M_\pi^2}/M_{\phi}=0.674$. To
be specific, for full QCD an interpolation to this ratio consists of
solving the equation
\begin{equation}
\frac{\mbox{``}m_{\eta_{ss}}\mbox{''}}{A^V + B^V_s \mu_{\rm sea} 
  + B^V_v \mbox{``}m_{\eta_{ss}}^2\mbox{''} 
  + C^V_s \mu_{\rm sea}^2
  + C^V_v \mbox{``}m_{\eta_{ss}}^4\mbox{''} +
  C^V_{sv} \mu_{\rm sea} \mbox{``}m_{\eta_{ss}}^2\mbox{''}} 
  = 0.674
\end{equation}
for ``$m_{\eta_{ss}}$''. Finally using ``$m_\pi$'' and ``$m_{\eta_{ss}}$''
determined above, and setting $\mu_{\rm val}= (\mbox{``}m_\pi^2\mbox{''} +
\mbox{``}m_{\eta_{ss}}^2\mbox{''})/2$ in Eqs.~(\ref{eq:vec-fit-full}) or
(\ref{eq:vec-fit-quench}) we obtain the mass ``$m_{K^*}$'' of a fictitious
$K^*$ meson.  In this setup ``$m_{\rho}$'' is again used to set the scale
by calculating the mass ratio ``$m_{K^*}$''$/$``$m_{\rho}$''. As a measure
for the lattice spacing ``$m_{\rho}$'' in lattice units is used for
continuum extrapolation.

\begin{figure*}[p]
\centerline{
\epsfxsize=7cm \epsfbox{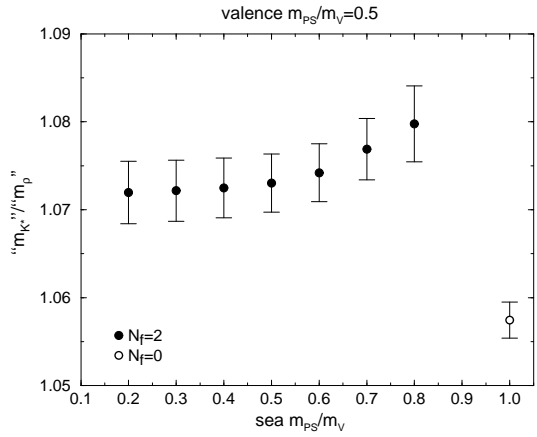}
}
\caption{Sea quark mass dependence of fictitious mass ratio
``$m_{K^*}$''$/$``$m_{\rho}$'' in the continuum limit.} 
\label{fig:KstFictContPQ}
\end{figure*}

In Fig.~\ref{fig:KstFictContPQ} we illustrate the ratio
``$m_{K^*}$''$/$``$m_{\rho}$'' as a function of $\mps / \mv$ of sea quarks
when $\mps / \mv$ of the valence quarks is fixed to 0.5. Naively we would
expect the points to be a smoothly decreasing function of $\mps / \mv$,
reaching the quenched value at $\mps / \mv =1$ corresponding to infinitely
heavy sea quark. In contrast to this expectation, but consistent with the
findings for the $J$ parameter, sea quark effects are almost constant up to
$\mps / \mv\approx 0.7$--0.8, which roughly corresponds to the strange
quark. This may be an indication that sea quark effects turn on rather
rapidly when sea quark mass decreases below a typical QCD scale of a few
hundred MeV.


\section{Light quark masses}
\label{sec:quarkmass}

Hadron mass calculations in lattice QCD provide us with a unique and
model-independent way to obtain quark masses. The main findings of our
light quark mass calculation have been presented in
Ref.~\cite{QuarkLetter}. We give here a more detailed account of the
analysis and results.

\subsection{Extraction of quark masses}

Quark masses can be calculated by inverting the relation
(\ref{eq:ps-fit-full-VWI}) and (\ref{eq:ps-fit-full-AWI}) between quark
masses and pseudoscalar meson masses, and substituting $m_{\pi}^2$ and
$m_{\eta_{ss}}^2$ determined in Sec.~\ref{sec:fix}.
  
For the average up and down quark mass, we set $\kappa_{\rm
val}^{(1)}=\kappa_{\rm val}^{(2)}=\kappa_{\rm sea}$ and evaluate the
hopping parameter $\kappa_{ud}$ for these quarks by solving the equation
$\mps^2(\kappa_{ud};\kappa_{ud},\kappa_{ud}) = m_{\pi}^2$. The VWI quark
mass is then determined by $m_{ud}^{\rm VWI} =(1/\kappa_{ud} -
1/\kappa_c)/2$ where $\kappa_c$ is the critical hopping parameter where the
pseudoscalar meson mass made of sea quarks vanishes $\mps(\kappa_{\rm val}
= \kappa_{\rm sea} = \kappa_c) = 0$.

An alternative definition for the VWI quark mass, called partially quenched
VWI quark mass (VWI,PQ), has been proposed in
Ref.~\cite{GuptaLattice97}. The partially quenched (PQ) chiral limit is
defined as the point of $\kappa_{\rm val}$ where the pseudoscalar meson
mass vanishes for fixed $\kappa_{\rm sea}$, and the corresponding hopping
parameter is denoted as $\kappa_c^{\rm PQ}$. As apparent from
Fig.~\ref{fig:chiralPS}, values of $\kappa_c^{\rm PQ}$ exhibit a clear
dependence on $\kappa_{\rm sea}$ and coincide with $\kappa_c$ only in the
limit $\kappa_{\rm sea}=\kappa_c$. The proposal in
Ref.~\cite{GuptaLattice97} consists of defining the quark mass via
$m_{ud}^{\rm VWI,PQ} = (1/\kappa_{ud} - 1/\kappa_c^{\rm PQ})/2$ where for
$\kappa_c^{\rm PQ}$ the value at $\kappa_{\rm sea}=\kappa_{ud}$ is
substituted. This is equivalent to a fictitious situation where the
simulation is performed with dynamical quarks at their physical value of up
and down quarks, the spectrum of pseudoscalar mesons is measured for
several values of the valence quark and the chiral limit is defined at the
point where masses of pseudoscalar mesons vanish.

A third determination of the average up and down quark mass is obtained
using the AWI definition of quark mass. It is unambiguously determined from
Eq.~(\ref{eq:ps-fit-full-AWI}) by setting $m^{\rm AWI}_{\rm val} = m^{\rm
AWI}_{\rm sea}$ and solving for $\mps^2 = m_{\pi}^2$.

The determination of the strange quark mass is made in a similar
way. Keeping the sea quark mass fixed at the average up and down quark mass
determined above, {\it i.e.,} $\kappa_{\rm sea}=\kappa_{ud}$ in
Eq.~(\ref{eq:ps-fit-full-VWI}) and $m^{\rm AWI}_{\rm sea} = m^{\rm
AWI}_{ud}$ in Eq.~(\ref{eq:ps-fit-full-AWI}), we calculate the point of
strange quark by tuning $\kappa_{\rm val}$ or $m^{\rm AWI}_{\rm val}$ so
that $\mps^2$ equals $m_{\eta_{ss}}^2$ obtained from the spectrum analysis.

Since $m_{\eta_{ss}}^2$ depends on the physical input, the strange quark
mass also depends on this input, and we consider the two cases where the
$K$ meson mass and the $\phi$ meson mass are used as input. In an exact
parallel with the average up and down quark mass, we calculate the strange
quark mass with three definitions.

Bare quark masses are converted to renormalized quark masses in the
modified minimal subtraction ($\overline{\rm MS}$) scheme at $\mu = 1/a$ by
the use of one-loop renormalization constants and improvement coefficients,
summarized in Appendix~\ref{app:Z}. For the two definitions of VWI quark
mass this consists of a conversion of the form
\begin{equation}
m^{\rm VWI}_R = Z_m \left ( 1+b_m\frac{m^{\rm VWI}}{u_0} \right )
\frac{m^{\rm VWI}}{u_0},
\end{equation}
while the renormalized AWI quark mass is obtained with
\begin{equation}
m^{\rm AWI}_R = \frac{Z_A \left ( 1+b_A\frac{m^{\rm VWI}}{u_0} \right ) }
                     {Z_P \left ( 1+b_P\frac{m^{\rm VWI}}{u_0} \right ) }
m^{\rm AWI}.
\end{equation}
Since $(b_A-b_P)m^{\rm VWI}/u_0 = -0.0019g^2_{\overline{\rm MS}}m^{\rm
VWI}/u_0 \approx 0.0006 \ll 1$ is negligible even for the strange quark, we
have ignored this contribution. After conversion to the $\overline{\rm
MS}$ scheme we employ the three-loop beta function to run quark masses to a
common scale of $\mu =2$~GeV. Numerical results are listed in
Table~\ref{tab:quarkmass}.

\subsection{Continuum results and systematic uncertainties}

\begin{figure*}[p]
\centerline{
\epsfxsize=7cm \epsfbox{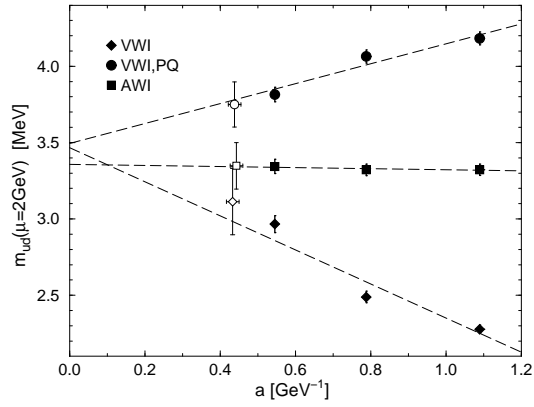}
}
\caption{Average up and down quark mass for three different definitions in
full QCD. Lines are from linear extrapolations to the continuum limit made
separately for each definition.}
\label{fig:mud_full}
\end{figure*}

\begin{figure*}[p]
\centerline{
\epsfxsize=7cm \epsfbox{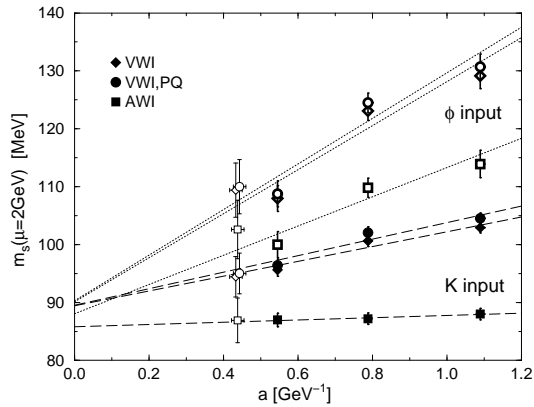}
}
\caption{Strange quark mass for three different definitions and two
different experimental inputs in
full QCD. Lines are from linear extrapolations to the continuum limit made
separately for each definition.}
\label{fig:ms_full}
\end{figure*}

Quark masses are plotted as a function of the lattice spacing in
Figs.~\ref{fig:mud_full} and \ref{fig:ms_full}. In these figures we also
show lines for continuum extrapolations performed for each definition of
quark mass separately. For extrapolation we employ fits linear in the
lattice spacing, corresponding to the leading order scaling violation. We
only include data from runs at three lattice spacings for extrapolation,
leaving the run at $\beta=2.2$ for a cross-check. Results of these
extrapolations are given in Table~\ref{tab:quarkmass}.

For $m_{ud}$ scaling violations are very small if the AWI definition is
used. The difference between the value at the coarsest lattice spacing and
the continuum value from a linear extrapolation is only 1\%. In contrast,
the two other definitions show sizable scaling violations. The partially
quenched quark mass at the coarsest lattice spacing is 20\% higher than in
the continuum limit while the VWI quark mass is lower by 34\%. Furthermore,
the VWI quark masses exhibit some curvature.

The situation is similar for the strange quark mass when the $K$ meson mass
is used as input. Scaling violations are small for the AWI quark mass,
amounting to a value 3\% higher at the coarsest lattice spacing than in the
continuum limit. For the two VWI quark masses, on the other hand, this
difference amounts to 15\%. If the $\phi$ meson mass is used as input
scaling violations are larger. In this case even the AWI quark mass is 30\%
larger at the coarsest lattice spacing than in the continuum limit and for
the two VWI quark masses the difference is as large as 45\%.

Having data at only three lattice spacings, it is difficult to explore
scaling violation for each definition of quark mass in further detail. An
important observation for linear continuum extrapolation is the fact that
the different fits to each definition converge in the continuum limit
within two-sigma of statistics (see Table~\ref{tab:quarkmass}). In
particular, VWI quark masses, where the largest scaling violations are
observed, are consistent with AWI masses, where scaling violations are
generally small. This leads us to perform a further fit, linear in the
lattice spacing and having a common continuum value. With such fits we
obtain $m_{ud}=3.44(9)$ with $\chi^2/N_{\rm DF}= 2.9$ and $m_s = 88.3(2.1)$
with $\chi^2/N_{\rm DF} = 1.3$ ($K$ input) or $m_s = 89.5(4.3)$ with
$\chi^2/N_{\rm DF} = 3.0$ ($\phi$ input). These masses lie between the ones
from individual fits and can be considered as a weighted average. We
utilize these numbers for central values of quark masses.

The errors quoted above are only statistical. Systematic errors arise from
the continuum extrapolation, the chiral extrapolation at each lattice
spacing, and from the use of one-loop renormalization factors in relating
the lattice values of quark masses to those for the continuum.

One way to examine systematic errors in the continuum extrapolation is to
include higher order terms in the combined fits. Such fits, however, are
unstable and do not lead to higher confidence levels, in particular for
$m_{ud}$. We therefore estimate uncertainties of the continuum
extrapolation from the spread of values obtained by separate fits to data
from the three definitions. Taking differences between the values from
separate fits and that from the combined fit leads to the errors quoted in
Table~\ref{tab:quarkerrors}.

We estimate the error from chiral extrapolation by changing the fit
formula. The functional form used for the determination of physical points,
and hence quark masses, is given with Eq.~(\ref{eq:vec-fit-full}). Changing
this to the alternative form of Eq.~(\ref{eq:vec-fit-full-15}) has several
effects which, combined together, lead to a decrease of the continuum value
by 2--8\% from the main analysis. This is used as estimate of the lower
error. For the upper error we add cubic terms $m^3$ to the formulae
(\ref{eq:ps-fit-full-VWI}) and (\ref{eq:ps-fit-full-AWI}) for pseudoscalar
mesons as functions of quark masses. This results into an increase of the
quark masses at each lattice spacing and also in the continuum limit.

Turning to the problem of renormalization factors, we list one-loop
corrections in Table~\ref{tab:Zfactors}. Their contribution is at most 13\%
at the strongest coupling, and hence we may expect higher loop
contributions to be smaller. Since a non-perturbative determination of the
renormalization factors is yet to be made for our improved actions, we
estimate effects of higher order corrections with a shift of the matching
scale from $\mu=1/a$ to $\mu=\pi/a$, and with use of an alternative
definition of the coupling given in Eq.~(\ref{eq:gMS_P}). The former leads
to an increase by 2\%, while the latter leads to a decrease of 5--7\%.

Finally we add the statistical and the systematic errors listed in
Table~\ref{tab:quarkerrors} in quadrature to obtain the total error. This
leads to the final values
\begin{equation}
m_{ud}^{\overline{MS}}(2 \, {\rm GeV}) = 3.44^{+0.14}_{-0.22}
\;\; {\rm MeV},
\label{eq:ud-mass}
\end{equation}
for the average up and down quark mass and
\begin{eqnarray}
m_s^{\overline{MS}}(2 \, {\rm GeV}) &=& 
     88^{+4}_{-6}\;\; {\rm MeV} \;\;\; M_K  \; {\rm input}, 
\label{eq:smassK}\\
&=&  90^{+5}_{-11}\;\; {\rm MeV} \;\;\; M_\phi \; {\rm input}.
\label{eq:smassPhi}
\end{eqnarray}
for the strange quark mass.

\subsection{Sea quark effects on light quark masses}

\begin{figure*}[p]
\centerline{
\epsfxsize=7cm \epsfbox{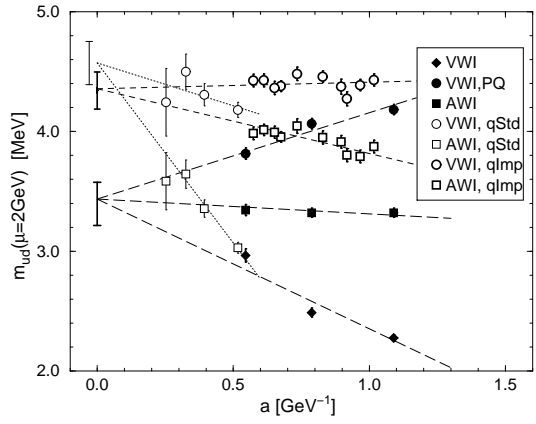}
}
\caption{Comparison of average up and down quark mass in quenched and full
QCD. Lines are from combined linear continuum extrapolations as described
in the text.}
\label{fig:mud}
\end{figure*}

\begin{figure*}[p]
\centerline{
              \epsfxsize=7cm \epsfbox{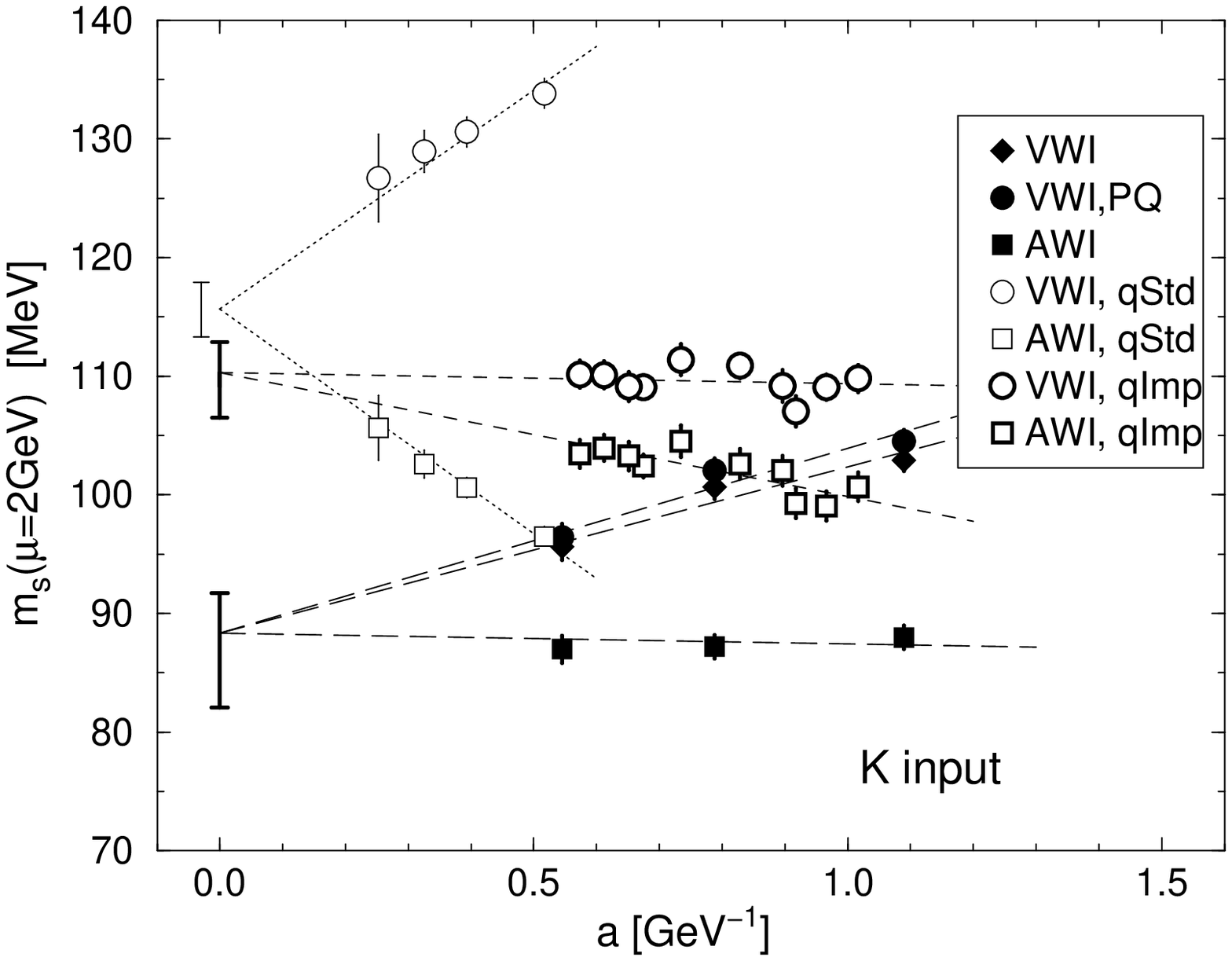}
\hspace{5mm}  \epsfxsize=7cm \epsfbox{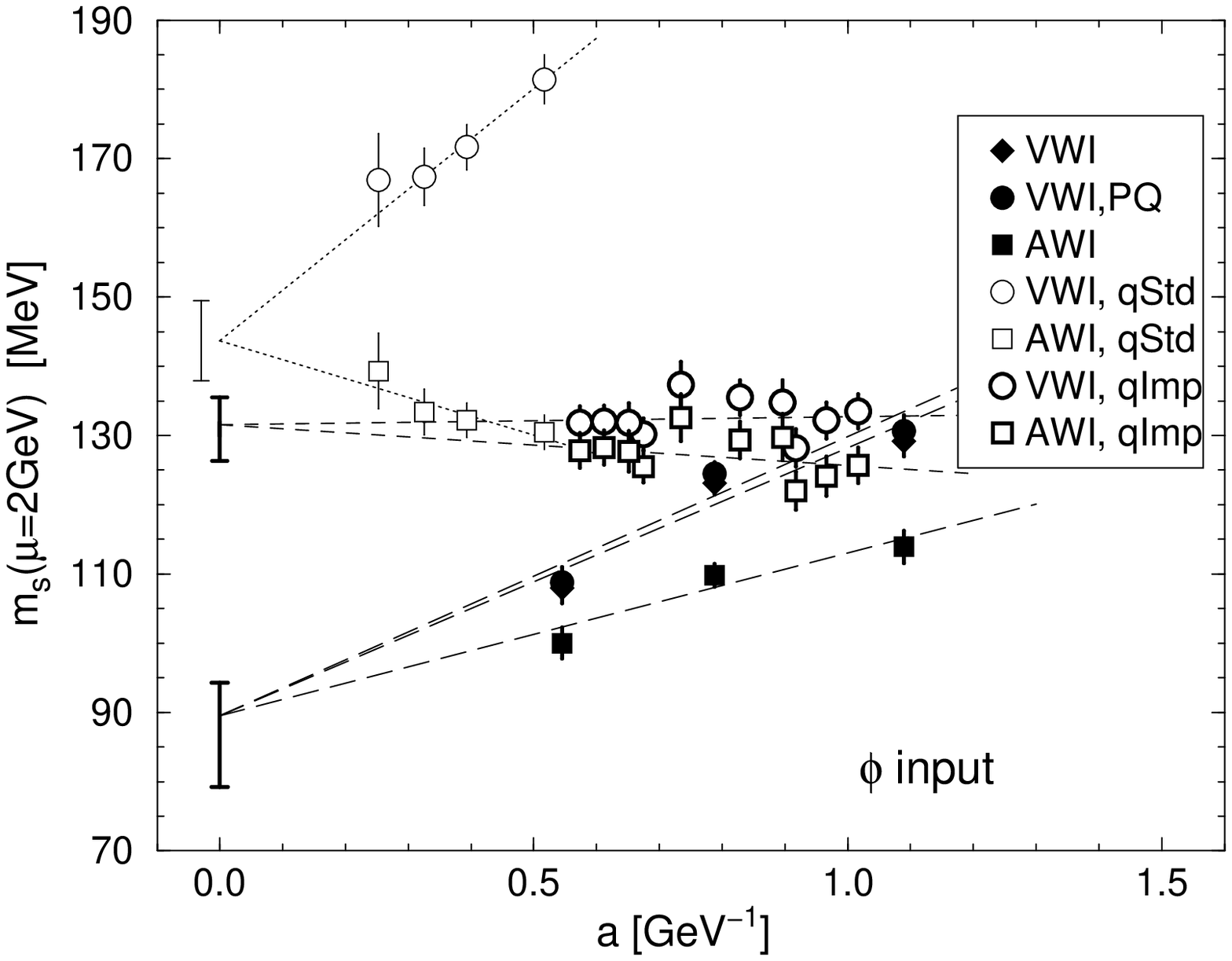}
}
\caption{Comparison of strange quark mass in quenched and full QCD using as
experimental input the $K$ meson mass (left figure) or the $\phi$ meson
mass (right figure). Lines are from combined linear continuum
extrapolations as described in the text.}
\label{fig:ms}
\end{figure*}

In Figs.~\ref{fig:mud} and \ref{fig:ms} we compare quark masses in full QCD
(filled symbols) with those in quenched QCD (open symbols). The quenched
results for improved actions (thick open symbols) are obtained from the
analysis of Sec.~\ref{sec:quenched} in parallel to those of full QCD.
There is no ambiguity in choice of the critical hopping parameter and so
there is only one definition of VWI quark mass. We also show quark masses
for the standard action reported in Ref.~\cite{quench.CPPACS} (thin open
symbols).

Long dashed lines are from the combined fit for full QCD, for which the
errors drawn in the continuum limit includes the systematic errors. The
continuum limits for quenched QCD are estimated with a combined linear
continuum extrapolation. They are listed together with quark masses in full
QCD in Table~\ref{tab:quark-continuum}.
 
Comparing the two quenched calculations of quark masses we first note that
scaling violations are visibly reduced for the improved action. This is
most noticeable for the strange quark mass where masses from improved
actions show a flat dependence against the lattice spacing $a$, while they
exhibit a sizable slope for the standard action. Nonetheless quark masses
in the continuum limit from the two calculations are in good agreement.

This confirms an inconsistency of 20--30\% in the quenched estimate of the
strange quark mass~\cite{quench.CPPACS}, depending on whether the $K$ meson
mass or the $\phi$ meson mass is used as input.

A comparison of full and quenched QCD establishes that the effect of
dynamical quarks decreases estimates of quark masses. This point was
previously argued from renormalization-group running of the gauge coupling
and quark masses in Ref.~\cite{FNALquark}. For two dynamical flavors
examined in the present work $m_{ud}$ becomes smaller by about 25\%. For
the strange quark the decrease is 20--25\% using $K$ as input, and 30--35\%
for $\phi$ as input.

In two-flavor full QCD the strange quark mass is consistent between the two
different inputs within the errors of 5--10\%. This is caused by a
different amount of decrease between quenched and full QCD. Thus the
inconsistency in the strange quark mass of quenched QCD almost disappears
in the presence of two flavors of sea quarks. This is directly related to
the finding in Sec.~\ref{sec:effects-meson} that the $K-K^*$ and the
$K-\phi$ mass splittings show a close agreement with experiment while there
is a clear discrepancy for quenched QCD.


\section{Decay constants}
\label{sec:decay}

\subsection{Pseudoscalar meson decay constants}

The pseudoscalar decay constant $f_{\rm PS}$ is defined from matrix
elements of the axial vector current through the relation
\begin{equation}
\langle 0 | A_4 | {\rm PS} \rangle = f_{\rm PS} \mps.
\end{equation}
We include the $O(a)$ improvement term in the axial vector current, and
employ one-loop renormalization constants as described in
Appendix~\ref{app:Z}. The decay constant is evaluated from the formula
\begin{equation}
f_{\rm PS} = 2\kappa u_0 Z_A \left ( 1+ b_A \frac{m}{u_0} \right ) 
\frac{C_A^s}{C_P^s}\sqrt{\frac{2C_P^l}{\mps}}.
\end{equation}
Here for $m$ we substitute the VWI,PQ quark mass, superscripts $l$ and $s$
distinguish local and smeared operators, and various amplitudes are
extracted in the following steps. The pseudoscalar mass $m_{\rm PS}$ and
the amplitude $C_P^s$ are determined from
\begin{equation}
\langle P^l(t) P^s(0) \rangle = C_P^s 
\left [ \exp(-\mps t) + \exp(-\mps (L_t-t) ) \right ].
\end{equation}
Values of $\mps$ are listed in 
Appendix~\ref{app:masses}. Keeping the mass fixed, we extract $C_P^l$ and 
$C_A^s$ from the fits
\begin{eqnarray}
\langle P^l(t) P^l(0) \rangle & = & C_P^l 
\left [ \exp(-\mps t) + \exp(-\mps (L_t-t) \right ], \\
\langle A_4^l(t) P^s(0) \rangle & = & C_A^s 
\left [ \exp(-\mps t) - \exp(-\mps (L_t-t) \right ].
\end{eqnarray}

\begin{figure*}[p]
\centerline{
              \epsfxsize=7cm \epsfbox{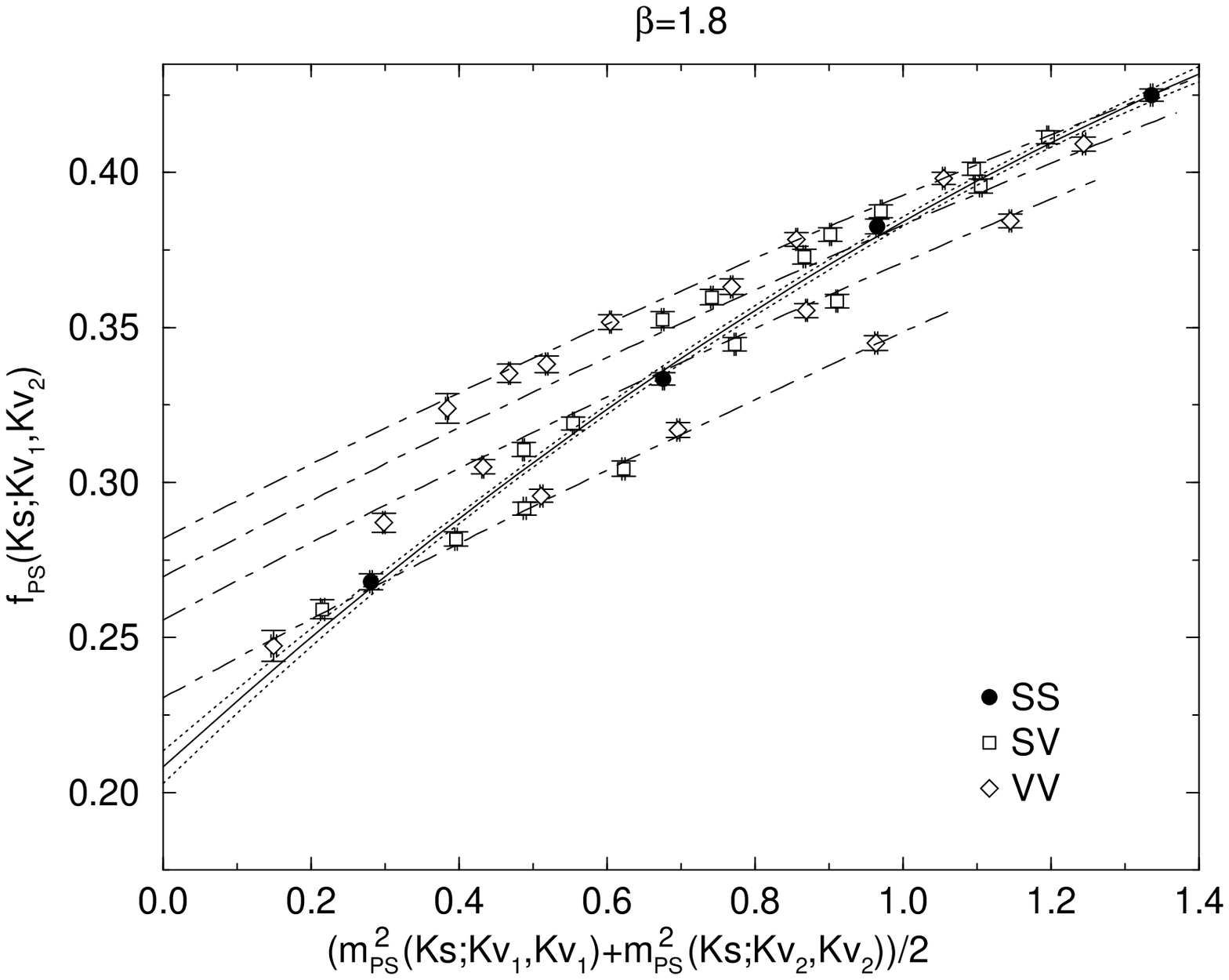}
\hspace{5mm}  \epsfxsize=7cm \epsfbox{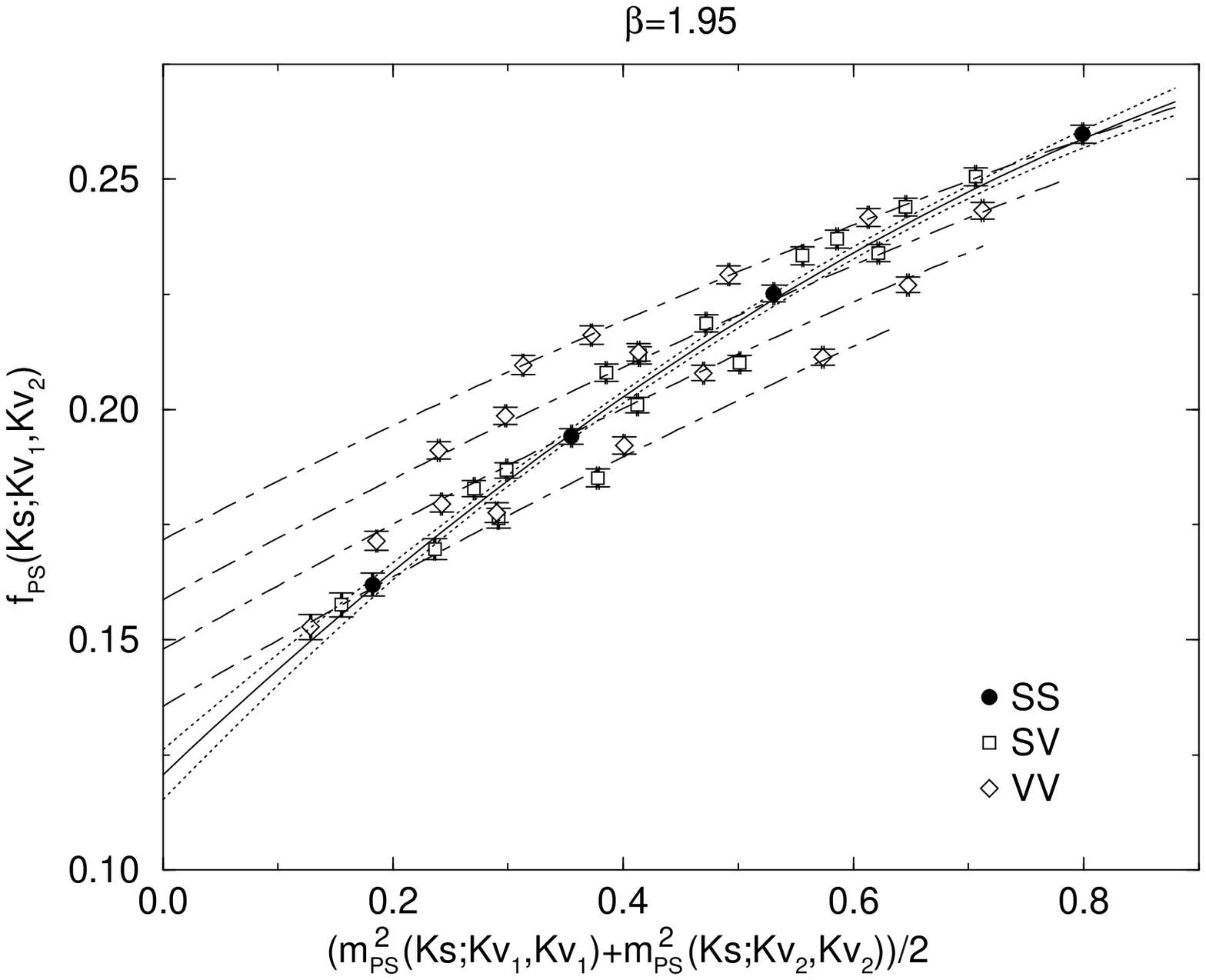}
}
\vspace{2mm}
\centerline{
              \epsfxsize=7cm \epsfbox{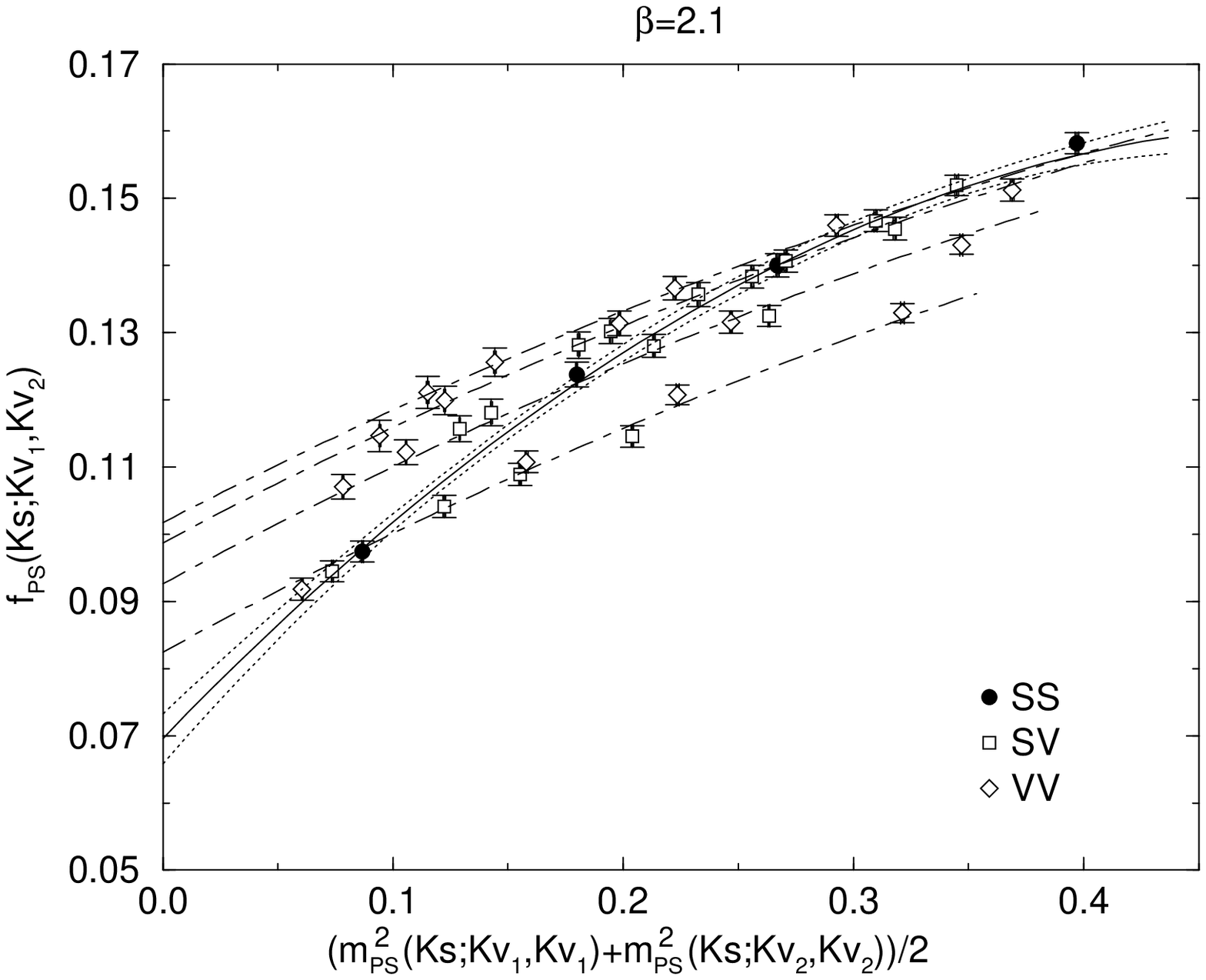}
\hspace{5mm}  \epsfxsize=7cm \epsfbox{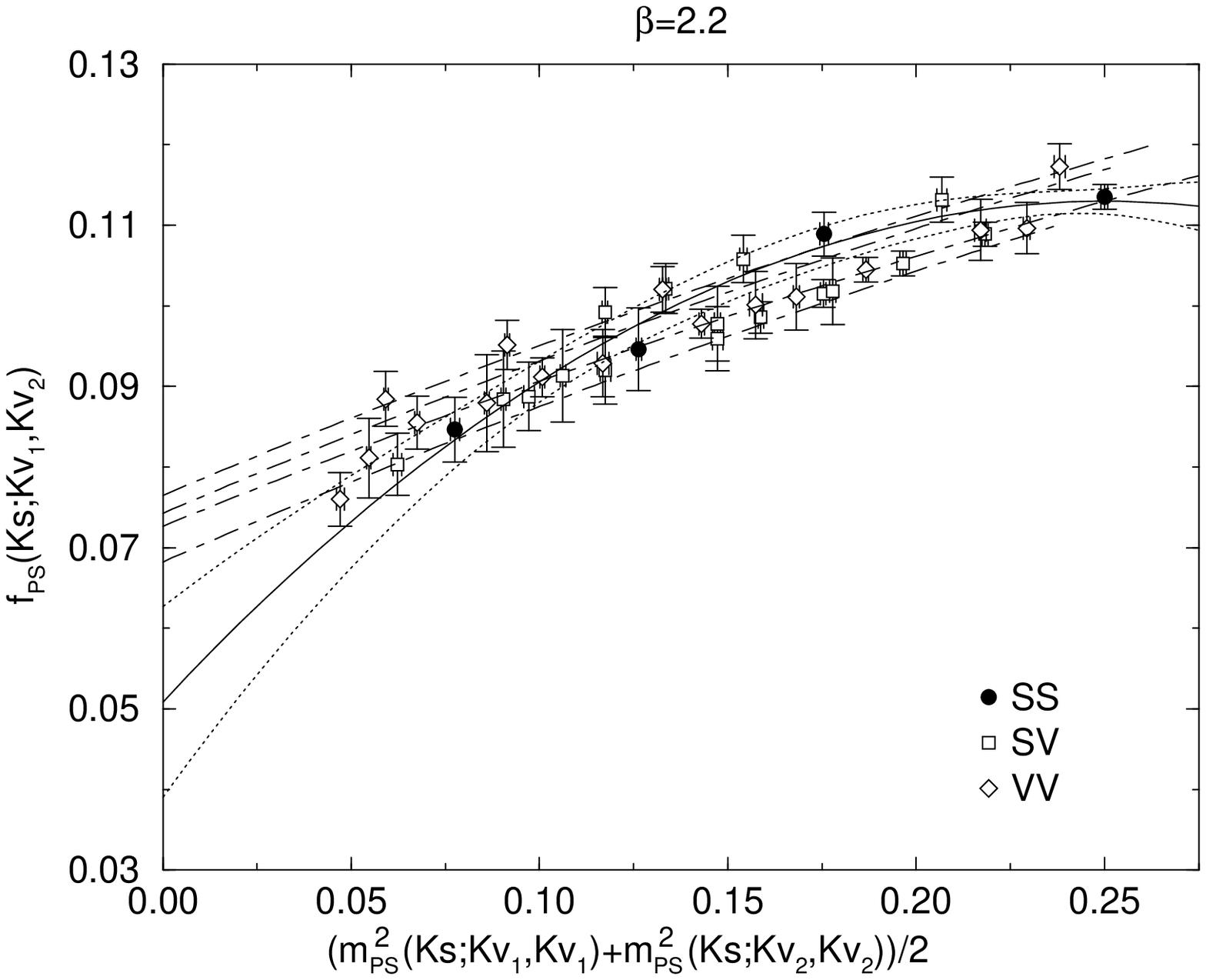}
}
\caption{Chiral extrapolations of pseudoscalar decay constants. Lines are
from fits with Eq.~(\ref{eq:decay-fit-full}).}
\label{fig:chiralDecay}
\end{figure*}

The chiral extrapolation of the decay constant is carried out in the same
way as for vector meson masses. Hence we employ a combined fit in sea and
valence quarks of the form,
\begin{equation}
f_{\rm PS} \left 
  (\kappa_{\rm sea};\kappa_{\rm val}^{(1)},\kappa_{\rm val}^{(2)} \right )
    =  A^F + B^F_s \mu_{\rm sea} + B^F_v \mu_{\rm val} 
       +  C^F_s \mu_{\rm sea}^2 + C^F_v \mu_{\rm val}^2 
       +  C^F_{sv} \mu_{\rm sea} \mu_{\rm val},    
\label{eq:decay-fit-full}
\end{equation}
where the $\mu$'s have the same meaning as in Sec.~\ref{sec:fit-vec}.
Pseudoscalar decay constants together with fits with
Eq.~(\ref{eq:decay-fit-full}) are shown in
Fig.~\ref{fig:chiralDecay}. Parameters of the fit are given in
Table~\ref{tab:decay-fit-full}.

Setting in Eq.~(\ref{eq:decay-fit-full}) $\mu_{\rm sea} = m_\pi^2$ and
$\mu_{\rm val} = m_\pi^2$ or $\mu_{\rm val} = (m_\pi^2 + m_K^2)/2$ obtained
from the spectrum analysis in Sec.~\ref{sec:fix}, $f_\pi$ and $f_K$ are
obtained in lattice units. Decay constants in physical units are finally
calculated using the lattice spacing from the rho meson mass and are listed
in Table~\ref{tab:FPhysical}.

\begin{figure*}[p]
\centerline{
\epsfxsize=8cm \epsfbox{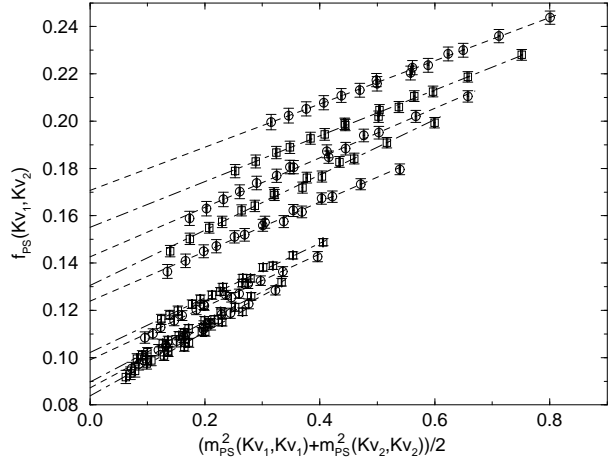}
}
\caption{Chiral extrapolations of pseudoscalar decay constants in quenched 
QCD.} 
\label{fig:chiralDecayQ}
\end{figure*}

The extraction of decay constants in quenched simulations is made similarly
to that in full QCD. For chiral extrapolation a simpler version of
Eq.~(\ref{eq:decay-fit-full}) ignoring sea quark mass dependence is used,
and the quadratic term $\mu_{\rm val}^2$ is dropped as linear fits already
yield good $\chi^2$ as illustrated in Fig.~\ref{fig:chiralDecayQ}. $f_\pi$
and $f_K$ obtained from calculations in quenched QCD are quoted in
Table~\ref{tab:FPhysical}.

\begin{figure*}[p]
\centerline{
              \epsfxsize=8cm \epsfbox{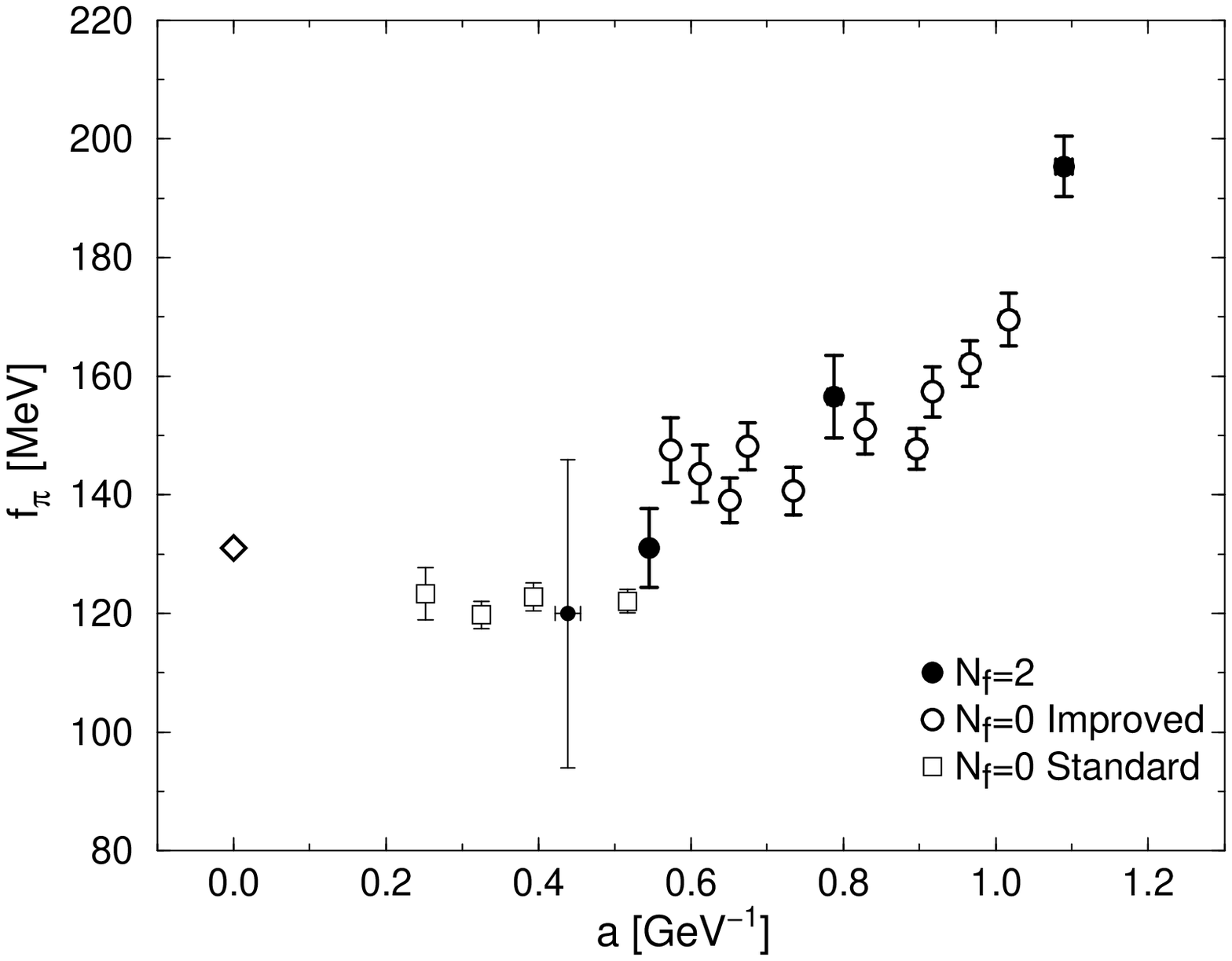}
\hspace{5mm}  \epsfxsize=8cm \epsfbox{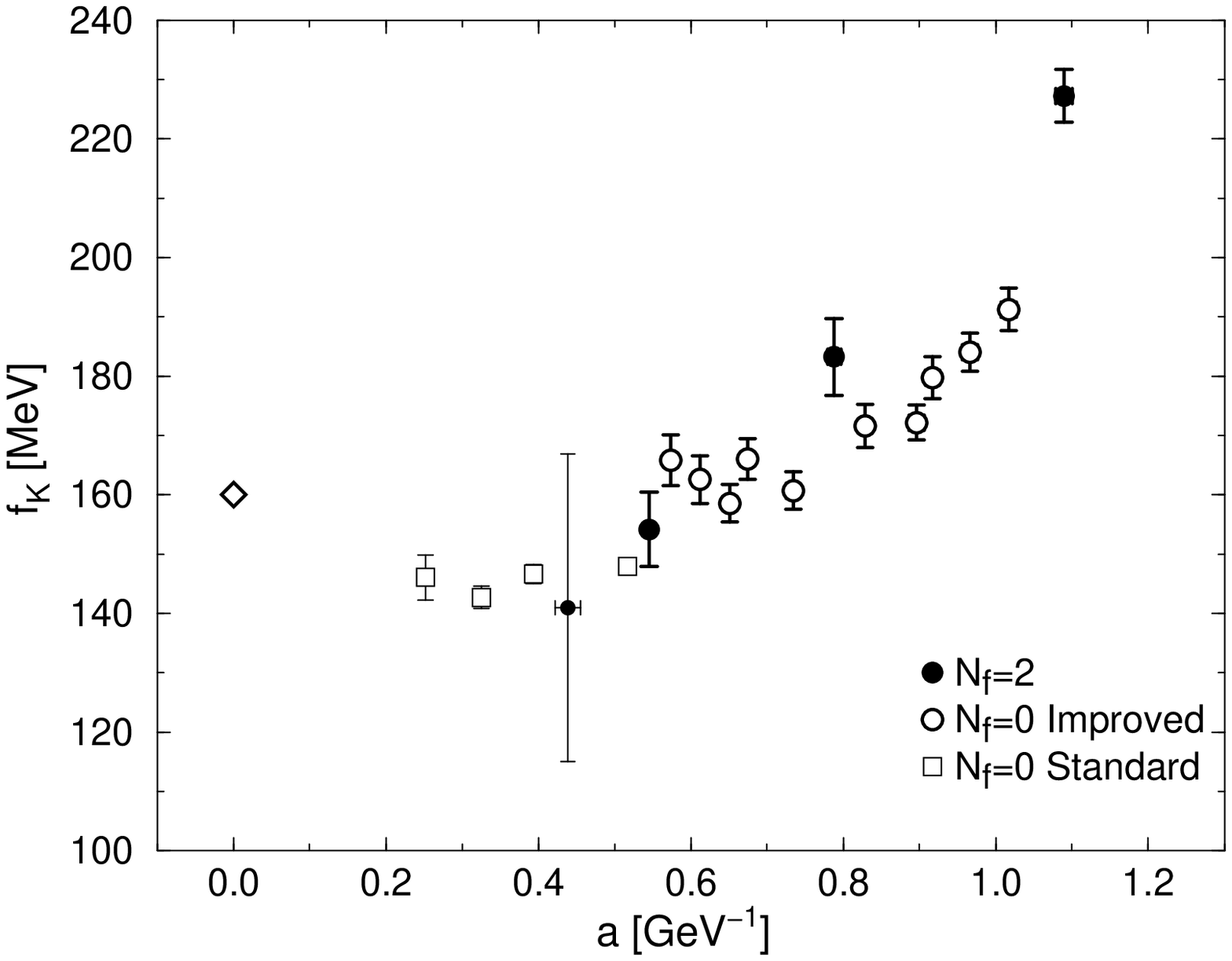}
}
\caption{Lattice spacing dependence of pseudoscalar decay constants
$f_\pi$ and $f_K$ in full QCD (filled circles) and quenched QCD with
improved actions (large open circles) or standard action (small open
squares). The strange quark mass used in the
calculation of $f_K$ is fixed with the $K$ meson mass as input.}
\label{fig:DecayFullQuench}
\end{figure*}

In Fig.~\ref{fig:DecayFullQuench} we show the lattice spacing dependence of
$f_\pi$ and $f_K$ in full and quenched QCD. For a comparison we also
include results obtained in quenched QCD with the standard
action~\cite{quench.CPPACS}. The most noticeable point is large violation
of scaling in full QCD. The values at the coarsest lattice spacing
$a=0.22$~fm are 50\% larger than that at the finest lattice spacing of
$a=0.11$~fm. Scaling violation is milder for quenched QCD, but still decay
constants at $a=0.22$~fm are 15\% larger than those at $a=0.11$~fm.

\begin{figure*}[p]
\centerline{
\epsfxsize=8cm \epsfbox{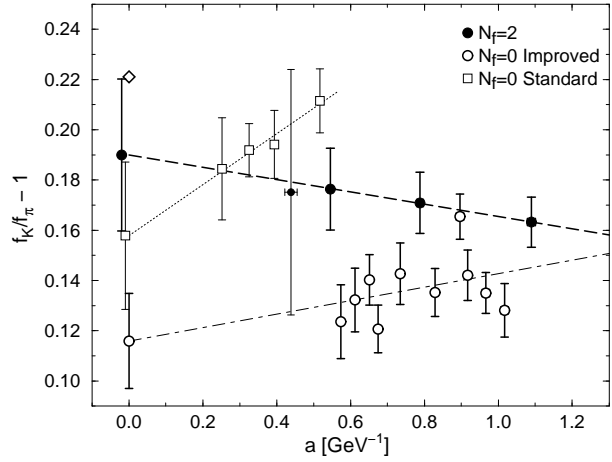}
}
\caption{Comparison of $f_K/f_\pi-1$ in full and quenched QCD. Fit lines
are linear for all data.}
\label{fig:DecayRatioFullQuench}
\end{figure*}

The origin of large scaling violation in the pseudoscalar decay constant is
not clear at present. Possible origins are contributions of higher order
corrections in the renormalization factors and $O(a)$ terms in the axial
vector currents. A suggestive hint ponting toward these origins is
provided by the ratio $f_K/f_\pi-1$ for which such corrections may largely
cancel out. As shown in Fig.~\ref{fig:DecayRatioFullQuench}, one observes
much reduced scaling violation for this quantity. Furthermore, a trend of
increase toward the experimental value as effects of sea quarks are
included is also apparent.

\subsection{Vector meson decay constants}

Vector meson decay constants are defined as
\begin{equation}
\langle 0 | V_i | {\rm V} \rangle = \epsilon_i F_{\rm V} \mv,
\end{equation}
where $\epsilon_i$ is a polarization vector and $\mv$ is the mass of the
vector meson. 

The numerical procedure employed to calculate vector meson decay constants
parallels the one for pseudoscalar decay constants. As discussed in
Sec.~\ref{sec:massmeas}, the rho correlator with smeared source is fit with
\begin{equation}
\langle V^l(t) V^s(0) \rangle = C_V^s 
\left [ \exp(-\mv t) + \exp(-\mv (L_t-t)) \right ],
\end{equation}
which determines $\mv$ and $C_V^s$. Using $\mv$ as input we make fits to
the correlator 
\begin{equation}
\langle V^l(t) V^l(0) \rangle = C_V^l 
\left [ \exp(-\mv t) + \exp(-\mv (L_t-t)) \right ],
\end{equation}
where the amplitude $C_V^l$ is the only fit parameter. Renormalized vector
meson decay constants are then obtained through 
\begin{equation}
F_{\rm V} = 2\kappa u_0 Z_V \left ( 1+ b_V \frac{m}{u_0} \right )
\sqrt{\frac{2C_V^l}{\mv}},
\end{equation}
where expressions for perturbative renormalization factors are given in
Appendix~\ref{app:Z}, and for $m$ we substitute the VWI,PQ quark mass. We
note in passing that we do not include the improvement term $c_V
\tilde{\partial}_\nu T_{n\mu\nu}$ in Eq.~(\ref{eq:VecRen}), since the
corresponding correlator has not been measured.

\begin{figure*}[p]
\centerline{
              \epsfxsize=7cm \epsfbox{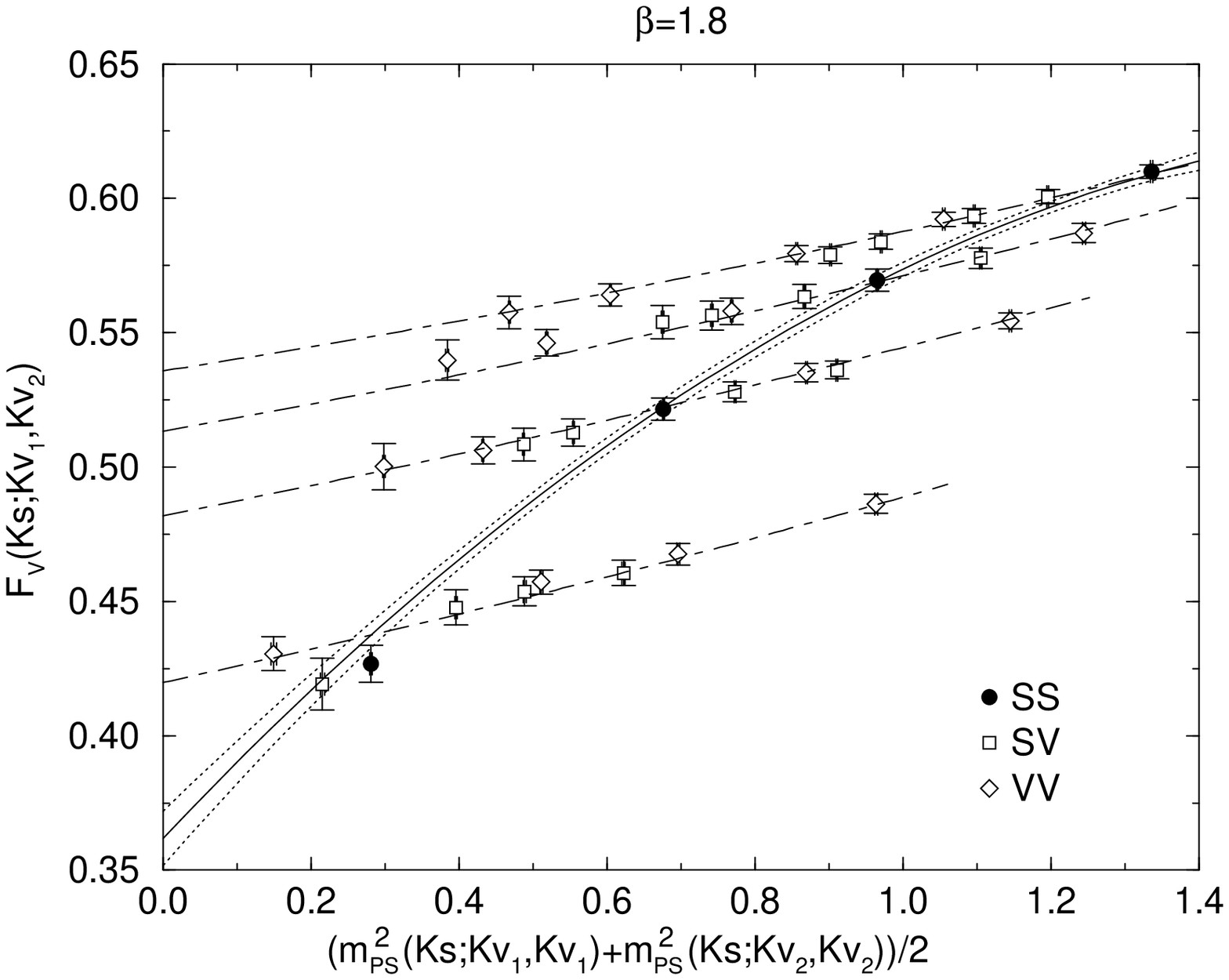}
\hspace{5mm}  \epsfxsize=7cm \epsfbox{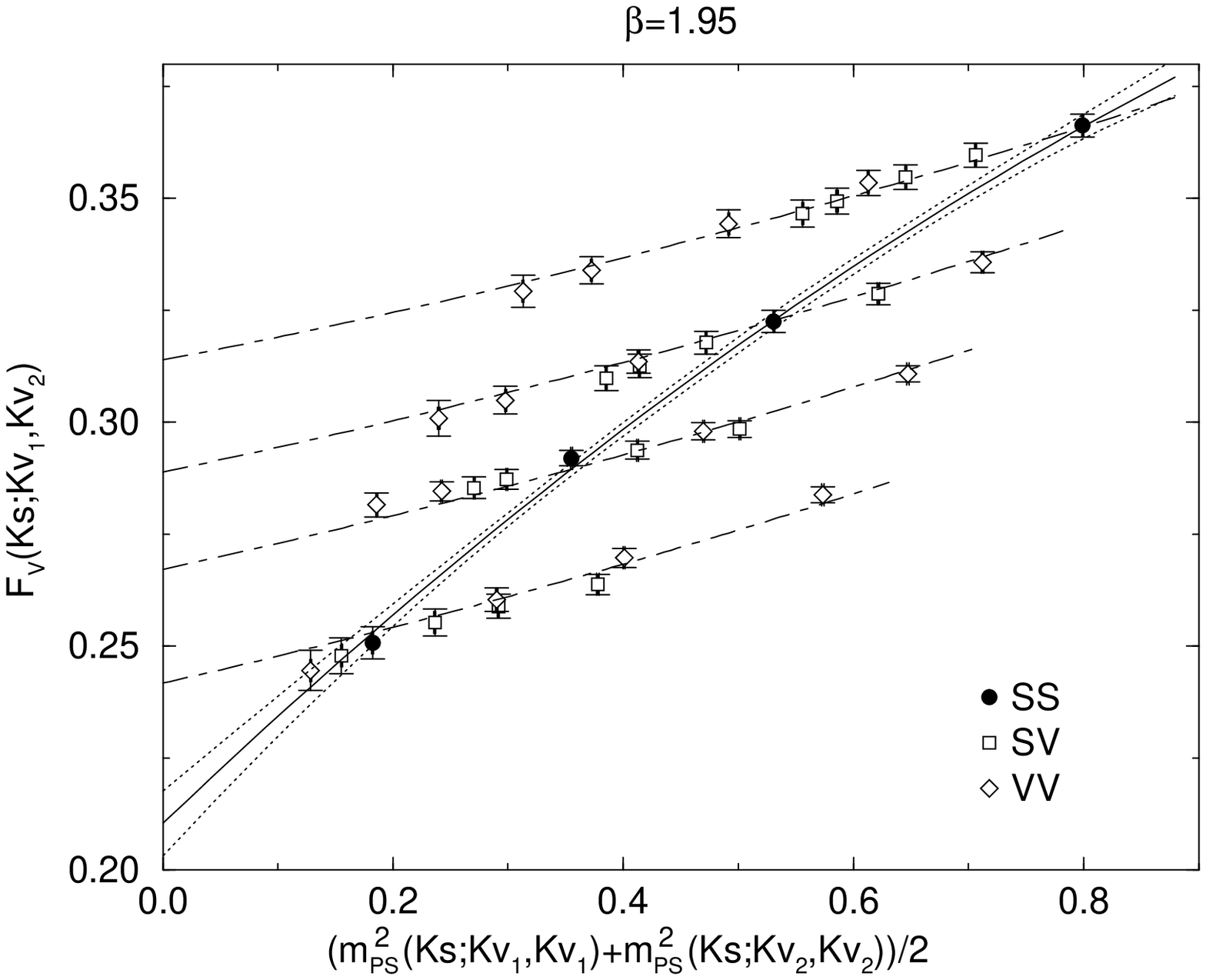}
}
\vspace{2mm}
\centerline{
              \epsfxsize=7cm \epsfbox{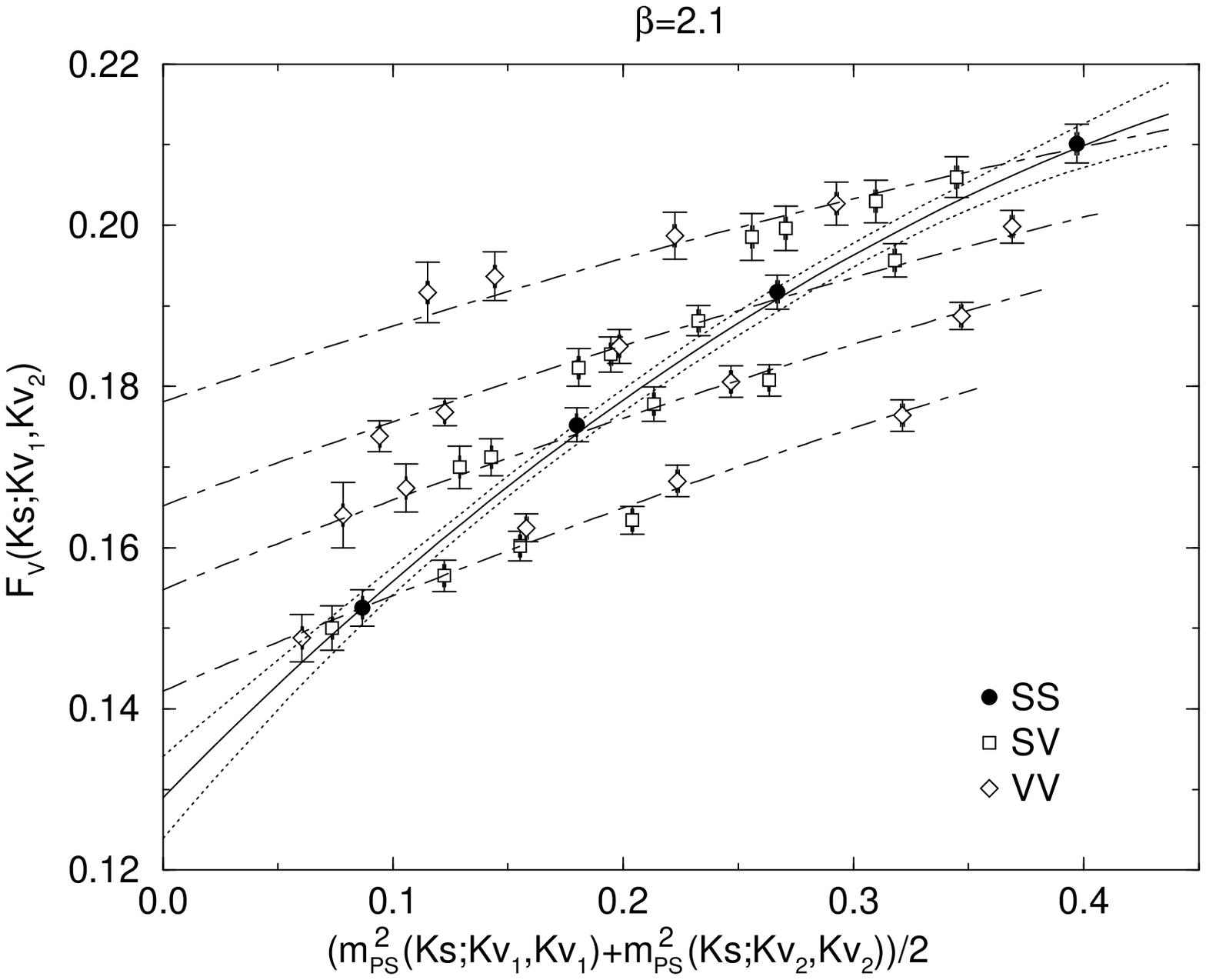}
\hspace{5mm}  \epsfxsize=7cm \epsfbox{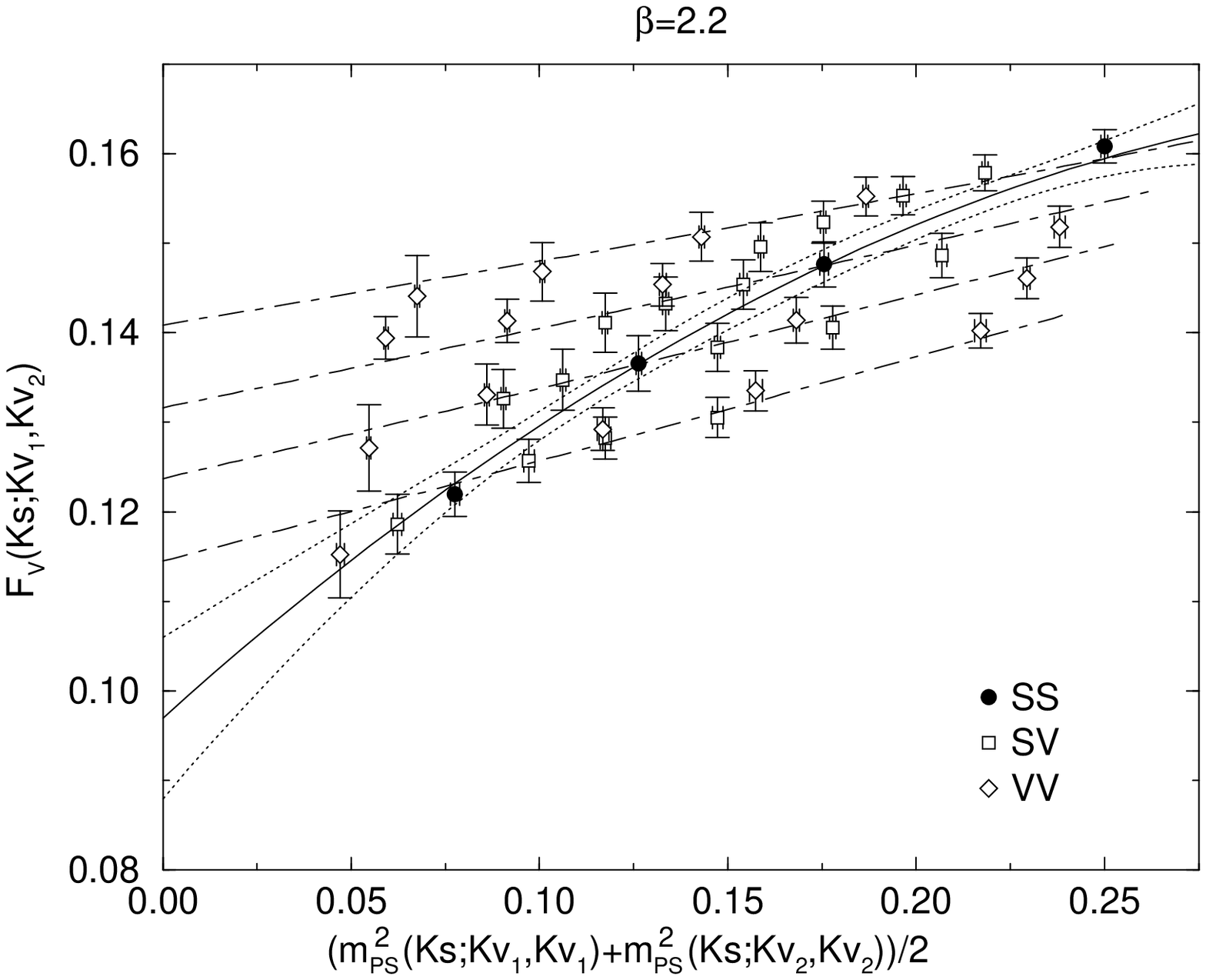}
}
\caption{Chiral extrapolations of vector meson decay constants. Lines are
from fits with Eq.~(\ref{eq:decay-fit-full}).}
\label{fig:chiralVDecay}
\end{figure*}

\begin{figure*}[p]
\centerline{
\epsfxsize=8cm \epsfbox{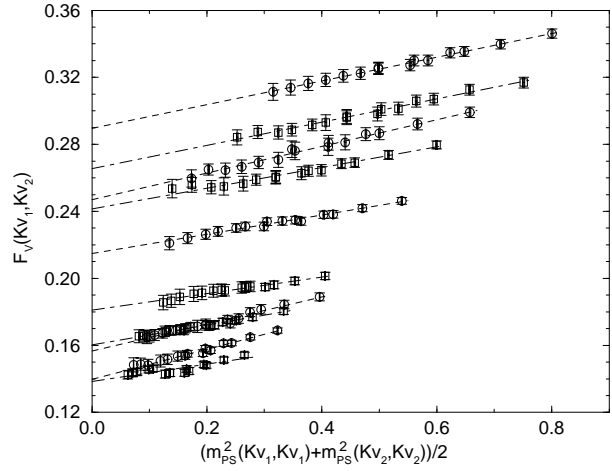}
}
\caption{Chiral extrapolation of vector meson decay constants in quenched QCD.}
\label{fig:chiralVDecayQ}
\end{figure*}

For chiral extrapolations we again employ combined quadratic fits as
defined by Eq.~(\ref{eq:decay-fit-full}). These fits describe the data
well, as shown in Fig.~\ref{fig:chiralVDecay}. Fit parameters are given in
Table~\ref{tab:decay-fit-full}. Vector meson decay constants obtained from
quenched simulations are plotted in Fig.~\ref{fig:chiralVDecayQ}. As for
pseudoscalar decay constants they are well described by linear fits. Final
values of $F_\rho$, $F_{K^*}$ and $F_\phi$ in physical units are listed in
Table~\ref{tab:FPhysical} for both full and quenched QCD.

\begin{figure*}[p]
\centerline{
              \epsfxsize=8cm \epsfbox{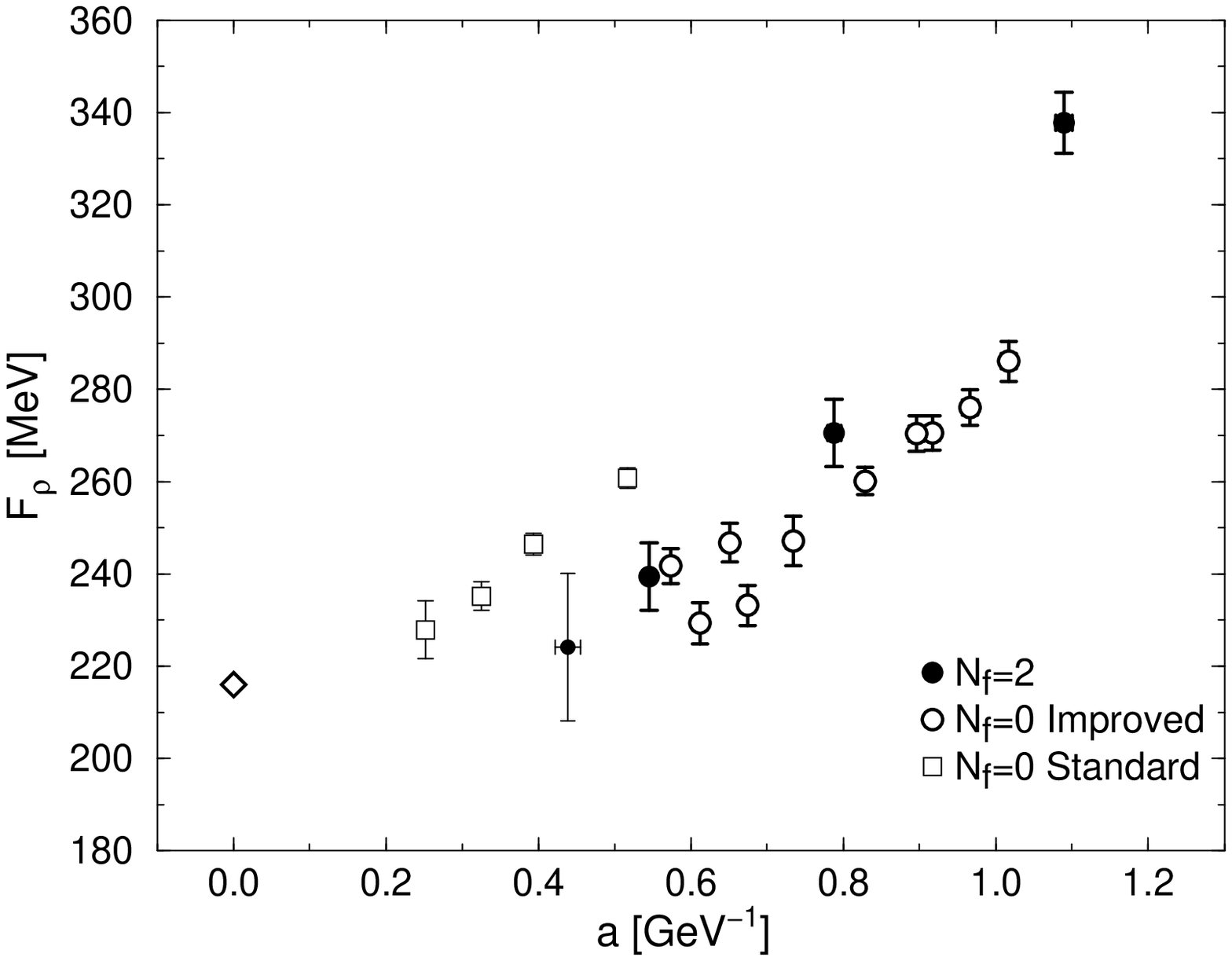}
\hspace{5mm}  \epsfxsize=8cm \epsfbox{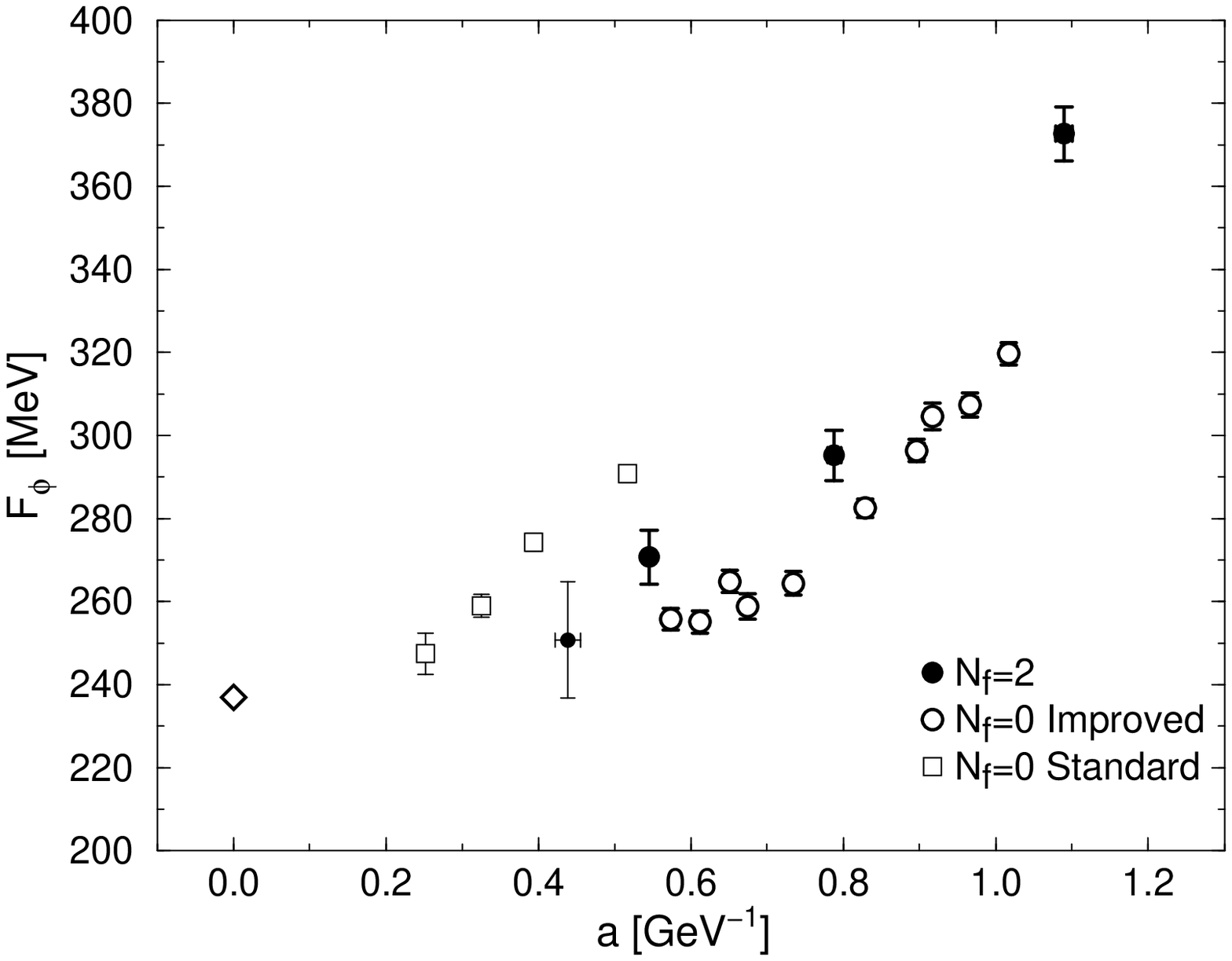}
}
\caption{Lattice spacing dependence of decay constants
 $F_\rho$ and $F_\phi$ in full and quenched QCD.} 
\label{fig:VDecayFullQuench}
\end{figure*}

The lattice spacing dependence of $F_\rho$ and $F_\phi$ in full and
quenched QCD is shown in Fig.~\ref{fig:VDecayFullQuench}. We again include
results obtained in quenched QCD with the standard
action~\cite{quench.CPPACS} for comparison. Vector meson decay constants in
full QCD exhibit scaling violations similar to those found for pseudoscalar
decay constants; e.g., $F_\rho$ is 40\% larger at $a = 0.22$~fm than at $a
= 0.11$~fm. Consequently, a continuum extrapolation poses similar
difficulties as for pseudoscalar decay constants.

\begin{figure*}[p]
\centerline{
\epsfxsize=8cm \epsfbox{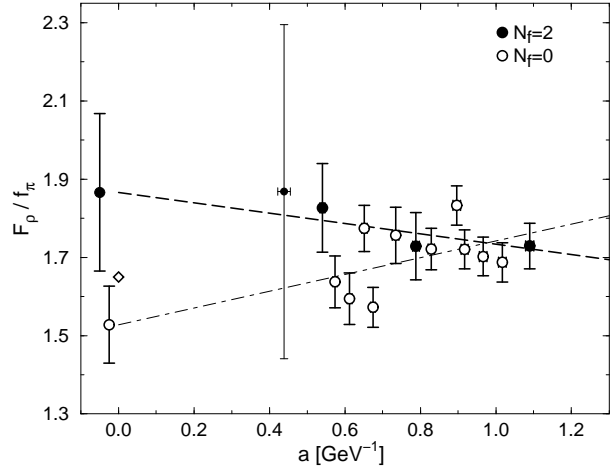}
}
\caption{Ratio of pseudoscalar and vector decay constants in full and
quenched QCD.}
\label{fig:VPSDecayRatio}
\end{figure*}

Since scaling violation is similar in vector and pseudoscalar decay
constants, one may examine the ratio of $F_\rho$ to $f_\pi$. The lattice
spacing dependence is much reduced for this quantity (see
Fig.~\ref{fig:VPSDecayRatio}), and $F_\rho/f_\pi$ is consistent with
experiment within the error of 5--10\%. In contrast to pseudoscalar decay
constants, sea quark effects are not apparent.

\subsection{Non-perturbative renormalization factors for vector currents}

For the clover quark action one can define a conserved vector current which 
reads 
\begin{equation}
V^C_i(n) = \frac{1}{2} \left \{
\overline{f}_{n+\hat{\mu}}U^{\dag}_{n,\mu}(\gamma_i+1) g_n +
\overline{f}_n U_{n,\mu}(\gamma_i-1) g_{n+\hat{\mu}}
\right \}.
\end{equation}
The non-renormalization of this current can be used to obtain a 
non-perturbative estimate of the renormalization
constant for the local current~\cite{Maiani,Lewis} according to 
the relation,
\begin{equation}
Z_V^{\rm NP} = \frac{ \langle 0 | V^C_i | {\rm V} \rangle }
                    { \langle 0 | V_i   | {\rm V} \rangle }. 
\label{eq:ZvNP}
\end{equation}

\begin{figure*}[p]
\centerline{
\epsfxsize=8cm \epsfbox{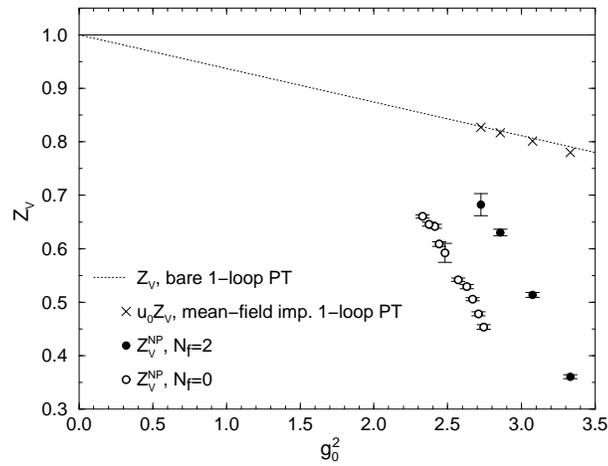}
}
\caption{Perturbative and non-perturbative $Z$-factors for vector current
at zero quark mass.} 
\label{fig:g-ZvNP}
\end{figure*}

The non-perturbative renormalization factors obtained from
Eq.~(\ref{eq:ZvNP}) and extrapolated to zero quark mass are plotted as a
function of the gauge coupling constant in Fig.~\ref{fig:g-ZvNP}. In the
same figure we also plot mean-field improved one-loop perturbative
renormalization factors as calculated in
Appendix~\ref{app:Z}. Non-perturbative values of $Z_V$ are significantly
smaller than those obtained from perturbation theory. This may be partly
due to corrections of $O(a)$ which are necessarily included in $Z_V$
calculated from Eq.~(\ref{eq:ZvNP})~\cite{Maiani,Lewis,Martinelli}.

\begin{figure*}[p]
\centerline{
\epsfxsize=8cm \epsfbox{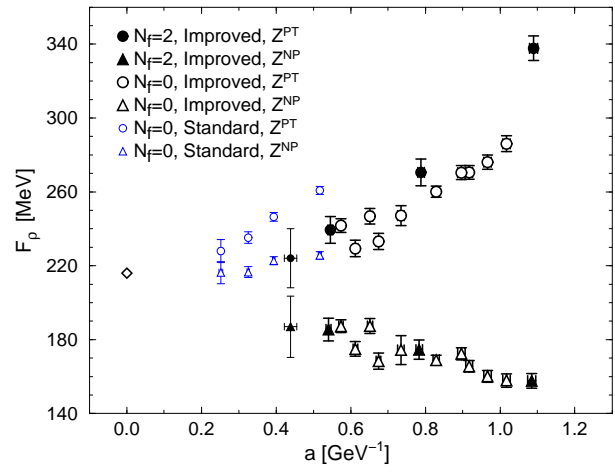}
}
\caption{Comparison of $F_\rho$ in full and quenched QCD with perturbative
(circles) and non-perturbative $Z$ factors (triangles).}
\label{fig:VDecayPNP}
\end{figure*}

In Fig.~\ref{fig:VDecayPNP} we compare $F_\rho$ determined with either
perturbative or non-perturbative renormalization factors. We observe that
decay constants calculated with $Z_V^{\rm NP}$ exhibit a much flatter
behavior as a function of the lattice spacing. We take this as an
encouraging indication that a further study with non-perturbative
renormalization factors will help moderate an apparently large scaling
violation in the pseudoscalar and vector decay constants.


\section{Conclusions}
\label{sec:conclusions}

In this article we have presented a simulation of lattice QCD fully
incorporating the dynamical effects of up and down quarks. A salient
feature of our work, going beyond previous two-flavor dynamical
simulations, is an attempt toward continuum extrapolation through
generation of data at three values of lattice spacings within a single set
of simulations. In order to deal with the large computational requirement
that ensue in such an attempt, we have used improved quark and gluon
actions. This has allowed us to work with lattice spacings in the range
$a\approx 0.22 - 0.11$~fm, which is twice as coarse as the range $a\simlt
0.1$~fm needed for the standard plaquette gluon and Wilson quark actions.
Still, this work would have been difficult without the CP-PACS computer
with a peak speed of 614 GFLOPS. With a typical sustained efficiency for
configuration generation of 30--40\%, the total CPU time spent for the
present work equals 415 days of saturated use of the CP-PACS, of which 318
days were for configuration generation and 84 days for measurements.

A major physics issue we addressed with our simulation was the origin of a
systematic discrepancy of the quenched spectrum from
experiment~\cite{quench.CPPACS}. Our new quenched simulation employing the
same improved actions as for full QCD has quantitatively confirmed the
results of Ref.~\cite{quench.CPPACS} for both mesons and baryons.

For mesons, masses in two-flavor full QCD become much closer to experiment
than those in quenched QCD. Using the $K$ meson mass to fix the strange
quark mass, the difference between quenched QCD and experiment of
$2.6^{+0.3}_{-0.9}$\% for the $K^*$ meson mass and of $4.1^{+0.5}_{-1.6}$\%
for the $\phi$ meson mass is reduced to $0.7^{+1.1}_{-1.7}$\% and
$1.3^{+1.8}_{-2.5}$\% in full QCD. When the $\phi$ meson mass is used as
input, the difference in the $K^*$ meson mass is less than 1\% for both
quenched and full QCD, while the deviation from experiment for the $K$
meson mass is reduced from $8.5^{+1.6}_{-3.8}$\% in quenched QCD to
$1.3^{+5.3}_{-5.3}$\% in full QCD. Similarly the $J$ parameter takes a
value $J=0.440^{+0.061}_{-0.031}$ in two-flavor full QCD, which is much
closer to the experimental value $J\approx 0.48$ compared to
$J=0.375^{+0.039}_{-0.009}$ in quenched QCD. We take these results as
evidence of sea quark effects in the meson spectrum.

A common point in reaching this conclusion is the importance of continuum
extrapolation. Differences between quenched and full QCD meson masses are
less obvious at finite lattice spacings but the slope of the continuum
extrapolation is different between them. Unexpectedly, the scaling
violation for full QCD is apparently larger than for quenched QCD with the
same improved actions. A possible origin of this feature is the common
choice of $c_{SW}$ we made for the two cases while the correct $c_{SW}$
necessary to remove $O(a)$ scaling violations need not be the same.

Full QCD baryon masses exhibit the pattern that the difference from
experiment increases with decreasing strange quark content. While masses of
$\Xi$ and $\Omega$ are in agreement with experiment, the nucleon mass
differs most from experiment among the octet, being larger by 10\%, and the
$\Delta$ among the decuplet by 13\%. This pattern of disagreement suggests
that finite-size effects sizably distort light baryon masses for an
$La\approx 2.5$~fm spatial size employed in our study. We leave detailed
finite-size analyses in full QCD for future investigations, however.

The sea quark effects in the meson sector have an interesting consequence
that the light quark masses decrease by about 25\% in two-flavor full QCD
compared to quenched QCD. An inconsistency of 20--30\% in the strange quark
mass for quenched QCD, depending on the particle used as input, disappears
in full QCD within the errors of 5--10\%.

In contrast to the encouraging results above, meson decay constants exhibit
large scaling violations which obstruct a continuum extrapolation. We have
found this trend to be common through light pseudoscalar and vector decay
constants of this work as well as in heavy-light decay
constants~\cite{CPPACS-fB-clover,CPPACS-fB-NRQCD}. Possibly this problem
arises from two-loop and higher order corrections in the renormalization
factors not included in our analyses. An indication for this explanation is 
given by a much flatter behavior of vector meson decay constants when using 
a non-perturbative renormalization factor derived from a conserved vector
current. 

While we consider that the present work has brought sizable progress in our
effort toward fully realistic simulations of QCD, it is also clear that a
number of gaps have to be filled in future studies. One of them is an
examination of finite-size effects, particularly for baryons. Another is
the exploration of lighter values of sea quark masses below
$\mps/\mv\approx 0.6$ for better control of the chiral extrapolation, and
generation of data at more points in the lattice spacing for a better
control of continuum extrapolations. Important in the latter context will
be the use of non-perturbative improvement coefficients and renormalization
factors. Finally, the inclusion of a dynamical strange quark will be
necessary to remove the last uncontrolled approximation.


\acknowledgements

This work is supported in part by Grants-in-Aid of the Ministry of
Education (Nos. 09304029, 10640246, 10640248, 10740107, 11640250, 11640294,
11740162, 12014202, 12304011, 12640253, 12740133, 13640260). AAK and TM are
supported by the JSPS Research for the Future Program (No. JSPS-RFTF
97P01102). GB, SE, TK, KN and HPS are JSPS Research Fellows.


\appendix

\section{Hadron masses}
\label{app:masses}

In Tables~\ref{tab:meson}--\ref{tab:bary_nondeg} we set out the hadron
masses measured in full QCD simulations. We list fitting ranges,
$\chi^2/N_{\rm DF}$ and masses in lattice units for all values of $\beta$
and all combinations of $\kappa_{\rm sea}$ and $\kappa_{\rm val}^{(i)}$. We
quote errors determined with the jack-knife method with a bin size of 10
configurations or 50 HMC trajectories.

\section{Jack-knife analysis for full QCD simulations}
\label{app:jack}

In quenched simulations masses of hadrons with different quark content are
obtained from the same gauge configurations and are therefore
correlated. Often the quality of data does not allow a correlated chiral
extrapolation and it is usual practice to resort to uncorrelated fits. By
using the jack-knife method errors of fit parameters can still be correctly
determined.

At first sight the situation seems simpler for full QCD with valence quarks
equal to sea quarks. Separate runs have to be made for different sea quark
masses, and are manifestly uncorrelated. Errors on parameters of chiral
fits can be correctly calculated from an uncorrelated $\chi^2$
fit. Nevertheless, the jack-knife method is extremely useful even in this
case. Since the fit parameters are often highly correlated, the
determination of the error of derived quantities can not be made with naive
error propagation. The jack-knife method takes such correlations into
account correctly. Moreover, in the setup of two-flavor QCD entire sets of
hadron masses with different valence quark content are measured on the same
configurations created with a given sea hopping parameter. Combined fits
according to the method of Sec.~\ref{sec:chiral} have correlations between
some of the data, and therefore one is in a similar situation to quenched
QCD.

A difference from quenched QCD is that there are as many sets of gauge
configurations as sea quarks in the simulation. They are mutually
independent and can differ in number between runs with various sea
quarks. A generalization is implemented in the following way. First, hadron
masses are determined with the usual jack-knife method. This yields mass
estimates $m_{\rm H}^{(J)i}(\kappa^k_{\rm sea})$ for each jack-knife
ensemble obtained by omitting the gauge configuration number $i$ from the
run with sea hopping parameter number $k$. Mean values and variances are
defined by
\begin{eqnarray}
 m_{\rm H}(\kappa^k_{\rm sea}) & = &
 \frac{1}{N_k} \sum_{i=1}^{N_k}
 m_{\rm H}^{(J)i}(\kappa^k_{\rm sea}), \\
 \left ( \Delta m_{\rm H}(\kappa^k_{\rm sea}) \right )^2 & = &
 \frac{N_k - 1}{N_k} \sum_{i=1}^{N_k}
 \left ( m_{\rm H}^{(J)i}(\kappa^k_{\rm sea}) 
       - m_{\rm H}(\kappa^k_{\rm sea}) 
 \right )^2.
\label{eq:jack}
\end{eqnarray}
Chiral fits are then carried out by replacing mean values $m_{\rm
H}(\kappa^k_{\rm sea})$ with jack-knife estimates $m_{\rm
H}^{(J)i}(\kappa^k_{\rm sea})$ for the sea hopping parameter number $k$
while keeping masses at all other sea hopping parameters at their mean
value. This procedure gives error estimates $(\Delta P)_k$ as above where
$P$ stands for a fit parameter or a quantity derived from fit
parameters. Since runs at different sea quarks are uncorrelated, the total
error $\Delta P$ is determined by quadratic addition $(\Delta P)^2 = \sum_k
((\Delta P)_k)^2$. Errors quoted throughout this paper are determined with
this method.

\section{Renormalization factors and improvement coefficients}
\label{app:Z}

In this appendix we summarize renormalization factors and improvement
coefficients used in the calculation of matrix elements and quark
masses. Perturbative calculations to one loop have been carried out in
Refs.~\cite{aoki,taniguchi}.

For the coupling constant we adopt a mean-field improved
value~\cite{TadImp} in the modified minimal subtraction ($\overline{\rm
MS}$) scheme obtained in the following way. We start with the one-loop
perturbative relation between the bare and $\overline{\rm MS}$ couplings
for the RG improved gauge action and the $O(a)$-improved Wilson quark
action~\cite{taniguchi},
\begin{equation}
  \frac{1}{g^2_{\overline{\rm MS}}(\mu)} = 
  \frac{1}{g^2} + 0.1000 + 0.03149 N_f +
  \frac{11-\frac{2}{3}N_f}{8\pi^2}\log(\mu a).
\end{equation}
The formula is reorganized so that $1/g^2$ becomes the coefficient in front
of $F_{\mu\nu}^2$ in the continuum limit after the mean field
approximation. Using the one-loop expressions~\cite{RGIA} $P = 1 - 0.1402
g^2$ and $R=1 - 0.2689 g^2$ for the expectation value of the plaquette
$P=\langle W^{1\times 1}\rangle$ and the $1\times 2$ rectangle $R=\langle
W^{1\times 2}\rangle$, we obtain the relation
\begin{equation}
  \frac{1}{g^2_{\overline{\rm MS}}(\mu)}
    = \frac{c_0 P + 8 c_1 R}{g^2} - 0.1006 + 0.03149 N_f +
      \frac{11-\frac{2}{3}N_f}{8\pi^2}\log(\mu a).
\label{eq:gMS_PR}
\end{equation}
Tadpole-improvement is realized by using nonperturbatively measured values
of $P$ and $R$. For full QCD we use values extrapolated to the chiral
limit of the sea quark. Numerical values of $P$ and $R$ used in the
calculation are given in Tables~\ref{tab:wloop} and \ref{tab:Qwloop}.

As an alternative we define the tadpole improved coupling constant with the
usual procedure which only uses the plaquette $P$,
\begin{equation}
  \frac{1}{\tilde{g}^2_{\overline{\rm MS}}(\mu)} = 
  \frac{P}{g^2} + 0.2402 + 0.03149 N_f +
      \frac{11-\frac{2}{3}N_f}{8\pi^2}\log(\mu a).
\label{eq:gMS_P}
\end{equation}

The VWI quark mass is renormalized with
\begin{equation}
m_R = Z_m \left ( 1+b_m\frac{m}{u_0} \right ) \frac{m}{u_0},
\end{equation}
where
\begin{equation}
Z_m = 1 + g^2_{\overline{\rm MS}}(\mu) \left ( 
         0.0400-\frac{1}{4\pi^2}\log(\mu a)^2 \right ),
\end{equation}
and
\begin{equation}
b_m = -\frac{1}{2} - 0.0323 g^2_{\overline{\rm MS}}(\mu).
\end{equation}
For $u_0$
\begin{equation}
u_0 = P^{1/4} = \left (1- \frac{0.8412}{\beta} \right )^{1/4}
\end{equation}
is used.

The local pseudoscalar density $P_n=\overline{\psi}_n\gamma_5\psi_n$ 
is renormalized with
\begin{equation}
P_n^R = 2\kappa u_0 Z_P \left ( 1+ b_P \frac{m}{u_0} \right ) P_n,
\end{equation}
where
\begin{equation}
Z_P = 1 + g^2_{\overline{\rm MS}}(\mu) \left (
          -0.0523+\frac{1}{4\pi^2}\log(\mu a)^2 \right ),
\end{equation}
and
\begin{equation}
b_P = 1 + 0.0397 g^2_{\overline{\rm MS}}(\mu).
\end{equation}

The renormalized axial vector current $A_\mu^R$, improved to $O(g^2a)$, is
obtained through
\begin{equation}
\label{eq:AxVecRen}
A_{n\mu}^R = 2\kappa u_0 Z_A \left ( 1+ b_A \frac{m}{u_0} \right ) 
              \left ( A_{n\mu} + c_A \tilde{\partial}_\mu P_n   \right ),
\end{equation}
where $A_{n\mu}=\overline{\psi}_ni\gamma_{\mu}\gamma_5\psi_n$ is the bare
local current and $\tilde{\partial}_\mu$ the symmetric lattice
derivative. Perturbative expressions for the renormalization factor and the
improvement coefficients are
\begin{eqnarray}
Z_A & = & 1 - 0.0215 g^2_{\overline{\rm MS}}(\mu), \\
b_A & = & 1 + 0.0378 g^2_{\overline{\rm MS}}(\mu), \\
c_A & = & -0.0038 g^2_{\overline{\rm MS}}(\mu).
\end{eqnarray}

Similarly, the renormalized vector current $V_\mu^R$ is obtained from the
bare local vector current $V_{n\mu}=\overline{\psi}_n\gamma_{\mu}\psi_n$
and $T_{n\mu\nu}=\overline{\psi}_ni\sigma_{\mu\nu}\psi_n$ through
\begin{equation}
\label{eq:VecRen}
V_{n\mu}^R = 2\kappa u_0 Z_V \left ( 1+ b_V \frac{m}{u_0} \right ) 
              \left ( V_{n\mu} + c_V \tilde{\partial}_\nu T_{n\mu\nu}   
              \right ).
\end{equation}
Here the perturbative results are
\begin{eqnarray}
Z_V & = & 1 - 0.0277 g^2_{\overline{\rm MS}}(\mu), \\
b_V & = & 1 + 0.0382 g^2_{\overline{\rm MS}}(\mu), \\
c_V & = & -0.0097 g^2_{\overline{\rm MS}}(\mu).
\end{eqnarray}

Numerical values for coupling constants, $Z$ factors and improvement
coefficients are listed in Table~\ref{tab:Zfactors}.

\newpage


\begin{table}
\caption{Overview of simulations.
The scale $a$ is fixed by $\protect M_\rho=768.4$~MeV from
fit to vector mesons with Eq.~(\ref{eq:vec-fit-full}).} 
\label{tab:overview}
\begin{tabular}{ccccccccc}
$\beta$ & $L^3\times T$ & $c_{\rm SW}$ & $a$ [fm] & $La$ [fm] 
& \multicolumn{4}{c}{$\mps/\mv$ for sea quarks : $N_{\rm Traj}$} \\
\hline 
1.80 & $12^3{\times}24$ & $1.60$ & 0.2150(22) & 2.580(26) 
& $0.807(1) \!:\! 6250$ & $0.753(1) \!:\! 5000$  & $0.694(2) \!:\! 7000$ & 
  $0.547(4) \!:\! 5250$ \STRUT \\ 
1.95 & $16^3{\times}32$ & $1.53$ & 0.1555(17) & 2.489(27) 
& $0.804(1) \!:\! 7000$ & $0.752(1) \!:\! 7000$ & $0.690(1) \!:\! 7000$ & 
  $0.582(3) \!:\! 5000$ \\ 
2.10 & $24^3{\times}48$ & $1.47$ & 0.1076(13) & 2.583(31) 
& $0.806(1) \!:\! 4000$ & $0.755(2) \!:\! 4000$ & $0.691(3) \!:\! 4000$ & 
  $0.576(3) \!:\! 4000$ \\ 
2.20 & $24^3{\times}48$ & $1.44$ & 0.0865(33) & 2.076(79) 
& $0.799(3) \!:\! 2000$ & $0.753(4) \!:\! 2000$ & $0.705(5) \!:\! 2000$ & 
  $0.632(7) \!:\! 2000$ \\ 
\end{tabular}
\end{table}

\begin{table}
\caption{Run parameters. The employed molecular dynamics (MD) integration
schemes are introduced in Sec.~\ref{sec:generation}. $N_{\rm inv}$ is the
sum of iterations for inversions of $D^{\dagger}$ and $D$ in the evaluation
of the fermionic force during HMC. $N_{\rm Meas}$ is the number of hadron
propagator measurements. In brackets of $N_{\rm Meas}$ the numbers of
removed configurations are also given.}
\label{tab:param}
\begin{tabular}{cccccr@{.}lccr@{.}lccr@{(}l}
$\beta$ & Size & $c_{\rm SW}$ & $\kappa$ & 
MD &
\multicolumn{2}{c}{$\Delta\tau$} & 
Accept. & Stop & 
\multicolumn{2}{c}{$N_{\rm inv}$}  & 
Hour/Traj. &
$N_{\rm Traj}$ &
\multicolumn{2}{c}{$N_{\rm Meas}$} \\
\hline
1.80& $12^3{\times}24$ \STRUT & 1.60 
    & 0.1409& c) & 0&033  & 0.781& $10^{-10}$&  64&8& 0.10& 6250& 1238&12)\\
    & on 64 PU   & 
    & 0.1430& c) & 0&025  & 0.807& $10^{-10}$&  87&2& 0.15& 5000&  990&10)\\ 
 & && 0.1445& c) & 0&0167 & 0.840& $10^{-10}$& 119&5& 0.26& 3500&  690&10)\\
 & &&       & a) & 0&0065 & 0.809& $10^{-10}$& 120&4& 0.25& 3500&  692&8) \\
 & && 0.1464& a) & 0&0033 & 0.764& $10^{-10}$& 263&6& 0.92& 4280&  839&17)\\
 & &&       & b) & 0&0066 & 0.714& $10^{-10}$& 256&9& 0.90&  970&  194&0) \\
\hline
1.95& $16^3{\times}32$ \STRUT & 1.53 
    & 0.1375& c) & 0&03125& 0.732& $10^{-11}$&  95&1& 0.10& 7000& 1400&0)\\
    & on 256 PU  & 
    & 0.1390& c) & 0&025  & 0.755& $10^{-11}$& 133&3& 0.15& 7000& 1395&5)\\ 
 & && 0.1400& c) & 0&0185 & 0.761& $10^{-11}$& 187&4& 0.25& 7000& 1397&3)\\
 & && 0.1410& c) & 0&008  & 0.820& $10^{-11}$& 331&8& 0.83& 5000& 1000&0)\\
\hline
2.10& $24^3{\times}48$ \STRUT & 1.47 
    & 0.1357& b) & 0&02   & 0.759& $10^{-12}$& 151&3& 0.35& 4000&  798&2)\\ 
    & on 512 PU  & 
    & 0.1367& b) & 0&016  & 0.792& $10^{-12}$& 208&7& 0.57& 4000&  800&0)\\
 & && 0.1374& b) & 0&0143 & 0.788& $10^{-12}$& 289&3& 0.82& 4000&  798&2)\\
 & && 0.1382& b) & 0&0075 & 0.781& $10^{-12}$& 544&7& 2.72& 4000&  800&0)\\
\hline
2.20& $24^3{\times}48$ \STRUT & 1.44 
    & 0.1351& b) & 0&02   & 0.758& $10^{-12}$& 192&0& 0.42& 2000&  400&0)\\ 
    & on 512 PU  & 
    & 0.1358& b) & 0&016  & 0.826& $10^{-12}$& 254&9& 0.67& 2000&  400&0)\\
 & && 0.1363& b) & 0&0143 & 0.837& $10^{-12}$& 336&8& 0.94& 2000&  400&0)\\
 & && 0.1368& b) & 0&01   & 0.859& $10^{-12}$& 505&6& 1.90& 2000&  400&0)\\
\end{tabular}
\end{table}

\begin{table}
\caption{Parameters $A$ and $B$ used for the smearing of
quark sources.}
\label{tab:smearing}
\begin{tabular}{ccccc}
             &
$\beta=1.80$ &
$\beta=1.95$ &
$\beta=2.10$ &
$\beta=2.20$ \\
\hline
$A$   & 1.25 & 1.25 & 1.05 & 1.02  \\
$B$   & 0.50 & 0.50 & 0.17 & 0.125 \\
\end{tabular}
\end{table}

\begin{table}
\caption{ Fit ranges for extraction of potential data,
Eq.~(\ref{eq:wilfit}), and ranges of $R_{\rm min}$ and $R_{\rm max}$ used
in potential fit, Eq.~(\ref{eq:potfit}). At $\beta=2.2$, we use the same
fit ranges as those taken at $\beta=2.1$.}
\label{tab:pot-range}
\begin{tabular}{llll}
&
$\beta=1.80$ & 
$\beta=1.95$ & 
$\beta=2.10$ \\
\hline
$t$           & [2,4]                   & [3,6] 
              & [4,8]                   \\
$R_{\rm min}$ & [$\sqrt{2}$,$\sqrt{2}$] & [$\sqrt{2}$, $\sqrt{6}$]
              & [$\sqrt{5}$,3]          \\
$R_{\rm max}$ & [$2\sqrt{3}$,4]         & [$3\sqrt{5}$,8]
              & [9,$6\sqrt{5}$]         \\
\end{tabular}
\end{table}

\begin{table}
\caption{String tension $\sigma$ and Sommer scale $r_0$ at simulated
sea quark masses and in the chiral limit of the sea quark. $\kappa_1$ is the
hopping parameter corresponding to the heaviest sea quark,
$\kappa_4$ to the lightest. $\sigma^{\chi}$ and $r_0^{\chi}$ in the
chiral limit $\mps=0$ are obtained from extrapolations using
Eqs.~(\ref{eq:stringfit}) and (\ref{eq:R0fit}).}
\label{tab:sig}
\begin{tabular}{ccccccccc}
$\kappa$    & 
\multicolumn{2}{c}{$\beta=1.80$} & 
\multicolumn{2}{c}{$\beta=1.95$} & 
\multicolumn{2}{c}{$\beta=2.10$} & 
\multicolumn{2}{c}{$\beta=2.20$}  \\
            &
$\sigma$ & $r_0$ &
$\sigma$ & $r_0$ &
$\sigma$ & $r_0$ &
$\sigma$ & $r_0$ \\
\hline
$\kappa_1$  & 0.4115(96)  & 1.716(35) & 0.2078(22)  & 2.497(54)  & 
              0.08949(99) & 3.843(16) & 0.05485(17) & 4.913(21)  \\
$\kappa_2$  & 0.389(12)   & 1.799(13) & 0.1859(29)  & 2.651(42)  & 
              0.07823(90) & 4.072(15) & 0.05107(26) & 5.073(19)  \\
$\kappa_3$  & 0.3595(68)  & 1.897(30) & 0.1633(23)  & 2.821(29)  & 
              0.07195(73) & 4.236(14) & 0.04760(31) & 5.237(22)  \\
$\kappa_4$  & 0.3067(60)  & 2.064(38) & 0.1436(25)  & 3.014(33)  & 
              0.06340(51) & 4.485(12) & 0.04474(23) & 5.410(21) \\
\hline
$\mps=0$ & 0.2858(72)     & 2.175(51) & 0.1295(25)  & 3.210(52)  &
              0.05720(63) & 4.695(18) & 0.04072(29) & 5.656(33)  \\
\end{tabular}
\end{table}

\begin{table}
\caption{Estimate of integrated autocorrelation times for the gauge action
$S_g$, for the number of iterations for the inversion of the
Dirac matrix $N_{\rm inv}$, and for the effective pion mass $m_{\pi,{\rm
eff}}$ at the onset of a plateau.}
\label{tab:autocor}
\begin{tabular}{ccccc}
$\beta$ &  $\kappa$ & $\tau_{S_g}$ [$\times 1$ HMC Traj.] &  
$\tau_{N_{\rm inv}}$ [$\times 1$ HMC Traj.]  &
$\tau_{m_{\pi,{\rm eff}}}$ [$\times 5$ HMC Traj.] \\
\hline
1.80 & 0.1409 & 4.2(4)   & 3.4(3)    &  0.7(1)  \\
     & 0.1430 & 5.5(1.0) & 4.6(7)    &  0.7(1)  \\ 
     & 0.1445 & 6.9(9)   & 5.7(8)    &  0.7(2)  \\
     & 0.1464 & 9.2(1.6) & 7.8(9)    &  0.9(2)  \\
\hline
1.95 & 0.1375 & 5.6(6)   &  5.8(7)   &  0.9(2) \\
     & 0.1390 & 7.6(9)   &  8.2(1.6) &  1.4(2) \\ 
     & 0.1400 & 9.0(2.0) & 10.1(2.0) &  1.4(2) \\
     & 0.1410 & 7.8(1.4) &  9.3(2.0) &  1.0(2) \\
\hline
2.10 & 0.1357 & 3.8(5)   &  6.7(1.5) & 1.5(3) \\ 
     & 0.1367 & 4.0(7)   &  9.4(2.4) & 1.5(3) \\
     & 0.1374 & 3.1(4)   &  8.1(2.1) & 1.0(1) \\
     & 0.1382 & 5.4(1.0) & 11.0(2.1) & 1.7(4) \\
\hline
2.20 & 0.1351 & 2.7(5)   & 5.0(1.5)  &  1.5(2) \\ 
     & 0.1358 & 1.8(3)   & 4.4(1.2)  &  3.9(1.6) \\
     & 0.1363 & 2.1(3)   & 3.2(8)    &  2.5(1.0) \\
     & 0.1368 & 1.9(3)   & 4.3(1.2)  &  2.5(7) \\
\end{tabular}
\end{table}

\begin{table}
\caption{Parameters of chiral fits to pseudoscalar meson masses as a
function of $1/\kappa$ with Eq.~(\ref{eq:ps-fit-full-VWI}) (first four
rows) or as a function of the AWI quark mass with
Eq.~(\ref{eq:ps-fit-full-AWI}) (last four rows).}  
\label{tab:ps-fit-full}
\begin{tabular}{r@{.}lcr@{.}lr@{.}lr@{.}lr@{.}lr@{.}lr@{.}lr@{.}l}
\multicolumn{2}{c}{$\beta$} & 
$\chi^2/N_{\rm DF}$ & 
\multicolumn{2}{c}{$\kappa_c$} &
\multicolumn{2}{c}{$b_s$}      &
\multicolumn{2}{c}{$b_v$}      &
\multicolumn{2}{c}{$c_s$}      &
\multicolumn{2}{c}{$c_v$}      &
\multicolumn{2}{c}{$c_{sv}$}   &
\multicolumn{2}{c}{$c_{vv}$}   \\
\hline
1&80 &  $116/29$  & 0&147635(16)  & 4&562(72) & 5&400(19)  & $-11$&51(38) &
        $-3$&064(81) & 2&45(10)  & 1&646(42) \\
1&95 &  $26.6/29$ & 0&142065(13)  & 2&655(69) & 4&169(12)  & $-4$&64(48)  &
        $-0$&846(72) & 4&379(99) & 1&333(42) \\
2&10 &  $17.4/29$ & 0&138984(13)  & 0&924(55) & 3&206(13)  & $-1$&40(49)  &
        0&96(12)     & 4&38(17)  & 1&121(91) \\
2&20 &  $15.0/29$ & 0&137675(52)  & 0&55(19) & 2&685(36)   & $-1$&9(1.9)  &
        1&79(38)     & 4&52(44)  & 1&04(27) \\
\hline
\multicolumn{2}{c}{$\beta$} & 
$\chi^2/N_{\rm DF}$ & 
\multicolumn{2}{c}{ }          &
\multicolumn{2}{c}{ }          &
\multicolumn{2}{c}{$b'_v$}     &
\multicolumn{2}{c}{ }          &
\multicolumn{2}{c}{$c'_v$}     &
\multicolumn{2}{c}{$c'_{sv}$}  &
\multicolumn{2}{c}{ }\STRUT       \\
\hline
1&80 &  $75/33$   & \multicolumn{2}{c}{ } & \multicolumn{2}{c}{ } &
        5&777(25) & \multicolumn{2}{c}{ } & 
        $-1$&335(90) & 1&99(15) & \multicolumn{2}{c}{ } \\
1&95 &  $75/33$   & \multicolumn{2}{c}{ } & \multicolumn{2}{c}{ } &
        4&393(28) & \multicolumn{2}{c}{ } & 
        $-0$&33(11) & 3&81(24) & \multicolumn{2}{c}{ } \\
2&10 &  $57/33$   & \multicolumn{2}{c}{ } & \multicolumn{2}{c}{ } &
        3&188(25) & \multicolumn{2}{c}{ } & 
        1&02(14)  & 3&63(35) & \multicolumn{2}{c}{ } \\
2&20 &  $32/33$   & \multicolumn{2}{c}{ } & \multicolumn{2}{c}{ } &
        2&641(61) & \multicolumn{2}{c}{ } & 
        1&37(37)  & 3&37(86) & \multicolumn{2}{c}{ } \\
\end{tabular}
\end{table}

\begin{table}
\caption{Parameters of chiral fits to vector meson masses with
Eq.~(\ref{eq:vec-fit-full}) (first four rows) or
Eq.~(\ref{eq:vec-fit-full-15}) (last four rows).} 
\label{tab:vec-fit-full}
\begin{tabular}{r@{.}lcr@{.}lr@{.}lr@{.}lr@{.}lr@{.}lr@{.}l}
\multicolumn{2}{c}{$\beta$} & 
$\chi^2/N_{\rm DF}$ & 
\multicolumn{2}{c}{$A^V$} & 
\multicolumn{2}{c}{$B^V_s$} & 
\multicolumn{2}{c}{$B^V_v$} & 
\multicolumn{2}{c}{$C^V_s$} &  
\multicolumn{2}{c}{$C^V_v$} &  
\multicolumn{2}{c}{$C^V_{sv}$} \\
\hline
1&80 &  $25.5/30$ & 0&8241(85) & 0&206(15) & 
        0&4066(96) & $-0$&0517(87) & $-0$&0193(48) & $-0$&0471(66) \\
1&95 &  $32.8/30$ & 0&5963(66) & 0&258(23) & 
        0&567(11)  & $-0$&072(21)  & $-0$&0443(91) & $-0$&112(12) \\
2&10 &  $43.0/30$ & 0&4124(51) & 0&327(34) & 
        0&907(24)  & $-0$&143(70)  & $-0$&265(40)  & $-0$&368(46) \\
2&20 &  $6.0/30$  & 0&332(13)  & 0&467(14) & 
        1&080(57)  & $-0$&40(40)   & $-0$&35(14)   & $-0$&76(23)  \\
\hline
\multicolumn{2}{c}{$\beta$} & 
$\chi^2/N_{\rm DF}$ & 
\multicolumn{2}{c}{$A^V$} & 
\multicolumn{2}{c}{$B^V_s$} & 
\multicolumn{2}{c}{$B^V_v$} & 
\multicolumn{2}{c}{$D^V_s$} &  
\multicolumn{2}{c}{$D^V_v$} &  
\multicolumn{2}{c}{$D^V_{sv}$}\STRUT \\
\hline
1&80 &  $22.0/30$ & 0&802(10)  & 0&319(28) & 
        0&480(18) & $-0$&120(20) & $-0$&078(11) & $-0$&086(12) \\
1&95 &  $25.3/30$ & 0&5812(80) & 0&376(42) & 
        0&663(19) & $-0$&131(37) & $-0$&135(16) & $-0$&155(17) \\
2&10 &  $37.4/30$ & 0&4003(63) & 0&478(64) & 
        1&112(44) & $-0$&192(88) & $-0$&459(52) & $-0$&350(45) \\
2&20 &  $4.7/30$  & 0&320(16)  & 0&70(26)  & 
        1&32(11)  & $-0$&41(42)  & $-0$&59(15)  & $-0$&60(18)  \\
\end{tabular}
\end{table}

\begin{table}
\caption{Parameters of chiral fits to decuplet baryon masses with
Eq.~(\ref{eq:dec-fit-full}).} 
\label{tab:dec-fit-full}
\begin{tabular}{r@{.}lcr@{.}lr@{.}lr@{.}lr@{.}lr@{.}lr@{.}l}
\multicolumn{2}{c}{$\beta$} & 
$\chi^2/N_{\rm DF}$ & 
\multicolumn{2}{c}{$A^D$} & 
\multicolumn{2}{c}{$B^D_s$} & 
\multicolumn{2}{c}{$B^D_v$} & 
\multicolumn{2}{c}{$C^D_s$} &  
\multicolumn{2}{c}{$C^D_v$} &  
\multicolumn{2}{c}{$C^D_{sv}$} \\
\hline
1&80 &  $13.5/46$  & 1&360(24) & 0&461(55) & 
        0&647(36)  & $-0$&116(33) & $-0$&036(22) & $-0$&090(25) \\
1&95 &  $2.12/46$  & 1&036(17) & 0&384(65) & 
        0&816(26)  & $-0$&038(67) & $-0$&034(25) & $-0$&193(38) \\
2&10 &  $7.82/46$  & 0&704(17) & 0&67(12)  & 
        1&202(67)  & $-0$&52(23)  & $-0$&11(11)  & $-0$&48(13)  \\
2&20 &  $15.9/46$  & 0&527(28) & 1&20(34) & 
        1&64(14)   & $-1$&9(1.0)  & $-0$&75(34)  & $-0$&89(43)  \\
\end{tabular}
\end{table}

\begin{table}
\caption{Parameters of chiral fits to octet baryon masses with
Eq.~(\ref{eq:ol-fit-full}).} 
\label{tab:ol-fit-full}
\begin{tabular}{r@{.}lcr@{.}lr@{.}lr@{.}lr@{.}lr@{.}lr@{.}l}
\multicolumn{2}{c}{$\beta$} & 
$\chi^2/N_{\rm DF}$ & 
\multicolumn{2}{c}{$A^O$}       & 
\multicolumn{2}{c}{$B_s^O$}       & 
\multicolumn{2}{c}{$F_v^O$}       & 
\multicolumn{2}{c}{$D_v^O$}       &  
\multicolumn{2}{c}{$C^O_s$}     &  
\multicolumn{2}{c}{$C^O_{vv}$}  \\
\multicolumn{2}{c}{  }          & 
                                & 
\multicolumn{2}{c}{$C_v^O$}       & 
\multicolumn{2}{c}{$C_v^\Sigma$}  & 
\multicolumn{2}{c}{$C_v^\Lambda$} & 
\multicolumn{2}{c}{$C_{sv}^O$}       &  
\multicolumn{2}{c}{$C_{sv}^\Sigma$}  &  
\multicolumn{2}{c}{$C_{sv}^\Lambda$} \\
\hline
1&80 & $28/72$ & 1&080(18) & 0&303(39) & 0&2945(96) & $-0$&0685(96)
               & $-0$&056(23) & $-0$&0437(67) \\
\multicolumn{2}{c}{  } &  
               & $-0$&0507(69) & 0&0027(41) & 0&0401(59) 
               & $-0$&0411(82) & 0&0225(70) & 0&0118(65) \\
1&95 & $17/72$ & 0&804(11) & 0&219(44) & 0&3799(69) & $-0$&0959(62)
               & 0&009(45) & $-0$&0855(77) \\
\multicolumn{2}{c}{  } &  
               & $-0$&1163(78) & 0&0090(54) & 0&0952(57) 
               & $-0$&055(13)  & 0&0334(73) & 0&0039(76) \\ 
2&10 & $59/72$ & 0&5418(83) & 0&376(64) & 0&576(13) & $-0$&131(13)
               & $-0$&17(13) & $-0$&251(29) \\
\multicolumn{2}{c}{  } &  
               & $-0$&389(29) & 0&051(24) & 0&289(22)
               & $-0$&189(43) & 0&107(29) & 0&006(27) \\
2&20 & $12/72$ & 0&432(19) & 0&65(24) & 0&674(42) & $-0$&170(36)
               & $-1$&06(73) & $-0$&46(14) \\
\multicolumn{2}{c}{  } & 
               & $-0$&61(11) & 0&041(92) & 0&519(80) 
               & $-0$&32(17) & 0&20(14) & 0&03(12) \\               
\end{tabular}
\end{table}

\begin{table}
\caption{Lattice spacings and hopping parameters $\kappa_{ud}$
and $\kappa_s$.}
\label{tab:parameters}
\begin{tabular}{ccccc}
$\beta$ & $a$ [fm] & $\kappa_{ud}$ & $\kappa_{s}$ ($M_K$ input) 
& $\kappa_{s}$ ($M_\phi$ input) \\
\hline
1.80 & 0.2150(22) & 0.147540(16) & 0.143147(91) & 0.14192(16) \\
1.95 & 0.1555(17) & 0.141998(12) & 0.139279(59) & 0.138633(79) \\
2.10 & 0.1076(13) & 0.138933(12) & 0.137324(41) & 0.137105(61) \\
2.20 & 0.0865(33) & 0.137634(50) & 0.13642(11)  & 0.13622(11) \\
\end{tabular}
\end{table}

\begin{table}
\caption{Meson and baryon masses at finite lattice
spacings and in the continuum limit. Values in the continuum limit are
obtained by a fit linear in the lattice spacing to data at
$\beta=$~1.8, 1.95 and 2.1. All masses are in GeV units.}
\label{tab:physical}
\begin{tabular}{lr@{.}lr@{.}lr@{.}lr@{.}lr@{.}lr@{.}l}
channel     &  \multicolumn{2}{c}{experiment}  &
               \multicolumn{2}{c}{$\beta=1.8$}  &  
               \multicolumn{2}{c}{$\beta=1.95$} &  
               \multicolumn{2}{c}{$\beta=2.1$}  &  
               \multicolumn{2}{c}{$\beta=2.2$}  &  
               \multicolumn{2}{c}{Continuum} \\
\hline
$N$        &  0&9396 &
   1&016(16)  & 1&040(13)  & 1&016(17)  & 1&007(43)  & 1&034(36)\\
$\Delta$   &  1&232 &
   1&270(23)  & 1&332(18)  & 1&310(30)  & 1&225(56)  & 1&392(58)\\
\hline                     
\multicolumn{12}{c}{$M_K$ input}\\                     
\hline                     
$\eta_{ss}$&  \multicolumn{2}{c}{--}    &
   0&69154(21)& 0&69578(19)& 0&69769(29)& 0&69838(91)& 
\multicolumn{2}{c}{--}       \\
$K^*$      &  0&8961 &
   0&8685(16) & 0&8708(13) & 0&8813(19) & 0&8774(46) & 0&8902(38)\\
$\phi$     &  1&0194 &
   0&9660(27) & 0&9710(23) & 0&9895(33) & 0&9832(84) & 1&0066(67)\\ 
$\Lambda$  &  1&1157 &
   1&149(15)  & 1&165(12)  & 1&147(16)  & 1&132(38)  & 1&158(33)\\
\end{tabular}
\end{table}

\begin{table}
\caption{Parameters of quenched QCD simulations. Coupling constants $\beta$ 
are chosen, so that measured values of $\sigma$ correspond to the ones in
full QCD given in Table~\ref{tab:sig}.} 
\label{tab:quench}
\begin{tabular}{ccccccc}
$\beta$ &  Size & $c_{\rm SW}$ & $a$ [fm]   & $La$ [fm] 
 & $\sigma$ & $r_0$ \\
\hline
2.187 & $16^3{\times}32$ & 1.439 & 0.2004(20) & 3.206(31) 
 & 0.2157(32) & 2.494(35) \\
2.214 & $16^3{\times}32$ & 1.431 & 0.1903(19) & 3.045(31) 
 & 0.1949(25) & 2.621(43) \\    
2.247 & $16^3{\times}32$ & 1.422 & 0.1807(18) & 2.891(29) 
 & 0.1713(18) & 2.801(28) \\
2.281 & $16^3{\times}32$ & 1.412 & 0.1765(20) & 2.824(32) 
 & 0.1487(17) & 3.001(36) \\
2.334 & $16^3{\times}32$ & 1.398 & 0.1632(16) & 2.611(26) 
 & 0.1241(14) & 3.289(23) \\
2.416 & $24^3{\times}48$ & 1.378 & 0.1446(18) & 3.471(42) 
 & 0.0921(10) & 3.824(13) \\
2.456 & $24^3{\times}48$ & 1.370 & 0.1328(13) & 3.188(30) 
 & 0.0800(16) & 4.080(16) \\ 
2.487 & $24^3{\times}48$ & 1.363 & 0.1284(14) & 3.081(34) 
 & 0.0725(11) & 4.286(15) \\
2.528 & $24^3{\times}48$ & 1.355 & 0.1206(13) & 2.895(30) 
 & 0.0637(11) & 4.570(21) \\ 
2.575 & $24^3{\times}48$ & 1.345 & 0.1130(11) & 2.713(27) 
 & 0.0561(7)  & 4.887(16) \\
\end{tabular}
\end{table}

\begin{table}
\caption{Parameters of chiral fits to meson masses in quenched QCD with
Eqs.~(\ref{eq:ps-fit-quench-VWI}), (\ref{eq:ps-fit-quench-AWI}) and
(\ref{eq:vec-fit-quench}).}
\label{tab:meson-fit-quench}
\begin{tabular}{r@{.}lr@{.}lr@{.}lr@{.}lr@{.}lr@{.}lr@{.}lr@{.}lr@{.}l}
\multicolumn{2}{c}{$\beta$} & 
\multicolumn{2}{c}{$\kappa_c$} &
\multicolumn{2}{c}{$b_v$}      &
\multicolumn{2}{c}{$c_v$}      &
\multicolumn{2}{c}{$c_{vv}$}   &
\multicolumn{2}{c}{$b'_v$}     &
\multicolumn{2}{c}{$c'_v$}     &
\multicolumn{2}{c}{$A^V$}     &
\multicolumn{2}{c}{$B_v^V$}   \\
\hline
2&187 & 0&141666(12) & 4&660(25) & $-1$&59(13) & 1&631(97) &
        4&684(48)    & 0&66(32) &
        0&7735(77)   & 0&4229(87) \\
2&214 & 0&140999(15) & 4&496(28) & $-1$&42(13) & 1&612(84) &
        4&582(44)    & 0&20(26) &
        0&7349(74)   & 0&447(10)  \\
2&247 & 0&140239(19) & 4&408(40) & $-1$&68(24) & 1&63(14)  &
        4&370(44)    & 0&25(28) &
        0&6975(71)   & 0&480(11)  \\ 
2&281 & 0&139587(15) & 4&211(37) & $-1$&24(25) & 1&50(14)  &
        4&162(36)    & 0&38(27) &
        0&6816(77)   & 0&473(12)  \\ 
2&334 & 0&138728(13) & 3&849(27) & $-0$&44(18) & 1&37(12)  &
        3&854(42)    & 0&72(29) &
        0&6302(63)   & 0&513(11)  \\ 
2&416 & 0&137633(7)  & 3&434(17) & 0&32(20)    & 1&21(14)  &
        3&388(34)    & 1&22(30) &
        0&5586(69)   & 0&571(16)  \\ 
2&456 & 0&137179(6)  & 3&258(15) & 0&70(14)    & 1&05(11)  &
        3&220(26)    & 1&06(29) &
        0&5128(49)   & 0&647(13)  \\ 
2&487 & 0&136852(7)  & 3&168(20) & 0&51(16)    & 1&07(11)  &
        3&094(22)    & 1&20(24) &
        0&4956(56)   & 0&662(16)  \\ 
2&528 & 0&136493(7)  & 2&951(23) & 1&33(16)    & 0&81(12)  &
        2&918(20)    & 1&05(21) &
        0&4656(49)   & 0&708(15)  \\ 
2&575 & 0&136116(8)  & 2&781(18) & 1&60(18)    & 0&69(14)  &
        2&776(28)    & 0&81(30) &
        0&4364(44)   & 0&757(16)  \\ 
\end{tabular}
\end{table}

\begin{table}
\caption{Meson and baryon masses at finite lattice
spacings and in the continuum limit in quenched QCD.
All masses are in GeV units.}
\label{tab:physicalNf0}
\begin{tabular}{lr@{.}lr@{.}lr@{.}lr@{.}lr@{.}lr@{.}lr@{.}lr@{.}lr@{.}lr@{.}l}
\multicolumn{21}{c}{$M_K$ input}\\                     
\hline
$\beta$      & \multicolumn{2}{c}{$m_{K^*}$} &
               \multicolumn{2}{c}{$m_{\phi}$} &
               \multicolumn{2}{c}{$m_{N}$} &
               \multicolumn{2}{c}{$m_{\Sigma}$} &
               \multicolumn{2}{c}{$m_{\Lambda}$} &
               \multicolumn{2}{c}{$m_{\Xi}$} &
               \multicolumn{2}{c}{$m_{\Delta}$} &
               \multicolumn{2}{c}{$m_{\Sigma^*}$} &
               \multicolumn{2}{c}{$m_{\Xi^*}$} &
               \multicolumn{2}{c}{$m_{\Omega}$} \\
\hline   
2.187 & 0&8690(11) & 0&9695(23) & 
        0&991(18) & 1&173(14) & 1&126(14) & 1&277(10) &
        1&324(18) & 1&420(15) & 1&516(12) & 1&612(10) \\ 
2.214 & 0&8694(14) & 0&9704(28) & 
        0&987(17) & 1&169(13) & 1&133(12) & 1&2893(90) &
        1&339(19) & 1&434(16) & 1&529(14) & 1&625(11) \\ 
2.247 & 0&8710(15) & 0&9736(31) & 
        1&014(15) & 1&189(11) & 1&135(12) & 1&2802(89) &
        1&334(16) & 1&430(14) & 1&526(11) & 1&623(10) \\ 
2.281 & 0&8675(16) & 0&9665(32) & 
        0&958(19) & 1&152(13) & 1&092(14) & 1&248(11) &
        1&274(19) & 1&379(16) & 1&483(13) & 1&587(11) \\ 
2.334 & 0&8684(13) & 0&9684(25) & 
        0&953(17) & 1&143(13) & 1&096(12) & 1&2561(88) &
        1&305(15) & 1&403(14) & 1&501(12) & 1&599(10) \\ 
2.416 & 0&8673(15) & 0&9662(31) & 
        0&943(17) & 1&133(12) & 1&076(13) & 1&2322(99) &
        1&283(20) & 1&384(17) & 1&486(14) & 1&587(12) \\
\end{tabular}
\end{table}

\begin{table}
\caption{Renormalized quark masses (in MeV) in the $\overline{\rm MS}$-scheme
at $\mu=2$~GeV at finite lattice spacings in full and quenched QCD. Values 
in the continuum limit obtained with separate linear fits to each
definition are also listed. For full QCD data at $\beta=$~2.2 were not
included in these fits.}
\label{tab:quarkmass}
\begin{tabular}{cr@{.}lr@{.}lr@{.}lr@{.}lr@{.}lr@{.}lr@{.}lr@{.}lr@{.}l}
$\beta$   & 
\multicolumn{2}{c}{$m_{ud}^{\rm VWI}$}     & 
\multicolumn{2}{c}{$m_{ud}^{\rm VWI,PQ}$}      &  
\multicolumn{2}{c}{$m_{ud}^{\rm AWI}$}         & 
\multicolumn{2}{c}{$m_s^{\rm VWI}$}($K$)     & 
\multicolumn{2}{c}{$m_s^{\rm VWI,PQ}$}($K$)      &  
\multicolumn{2}{c}{$m_s^{\rm AWI}$}($K$)         & 
\multicolumn{2}{c}{$m_s^{\rm VWI}$}($\phi$)  & 
\multicolumn{2}{c}{$m_s^{\rm VWI,PQ}$}($\phi$)   &  
\multicolumn{2}{c}{$m_s^{\rm AWI}$}($\phi$)  \\
\hline
\multicolumn{19}{c}{$N_f =2$ Full QCD} \\
\hline
1.80 &   2&277(27) &   4&183(42) &   3&322(37)  
     & 102&92(92)  & 104&54(93)  &  88&0(1.0)  
     & 129&1(2.2)  & 130&7(2.2)  & 113&9(2.4)  \\
1.95 &   2&489(38) &   4&064(43) &   3&321(38)
     & 100&65(98)  & 102&08(99)  &  87&2(1.0)
     & 123&1(1.7)  & 124&5(1.7)  & 109&8(1.7)  \\
2.10 &   2&966(55) &   3&816(47) &   3&344(46)
     &  95&6(1.1)  &  96&4(1.1)  &  87&0(1.2)
     & 108&0(2.2)  & 108&8(2.2)  & 100&0(2.2) \\
2.20 &   3&11(22)  &   3&75(15)  &   3&35(15)
     &  94&4(3.5)  &  95&0(3.5)  &  86&9(3.9)
     & 109&4(4.7)  & 110&0(4.7)  & 102&6(5.0) \\
$a\to 0$& 3&47(10) &   3&50(10)  &  3&36(9)
     & 89&4(2.3)   &  89&5(2.3)  & 85&8(2.4)
     & 90&1(4.9)   &  90&3(4.9)  & 88&1(4.9) \\
$\chi^2/N_{\rm DF}$ 
     & 10&8        &   2&4       &  0&07     
     & 2&1         &   2&7       &  0&03     
     & 6&0         &   6&5       &  2&4   \\
\end{tabular}
\end{table}

\begin{table}
\caption{Breakdown of contributions to total error of full QCD quark masses
in the continuum limit.}
\label{tab:quarkerrors}
\begin{tabular}{ccccc}
                       & Stat.    & Cont. Ext. & Chiral   & Z factor  \\
\hline           
$m_{ud}$               & $+2.6\%$ & $+1.7\%$   & $+1.2\%$ & $+2.3\%$  \\
                       & $-2.6\%$ & $-2.3\%$   & $-2.3\%$ & $-5.0\%$  \\
$m_s$ ($M_K$ input)    & $+2.4\%$ & $+1.4\%$   & $+1.6\%$ & $+2.2\%$  \\
                       & $-2.4\%$ & $-2.8\%$   & $-2.2\%$ & $-5.6\%$  \\
$m_s$ ($M_\phi$ input) & $+4.8\%$ & $+0.9\%$   & $+1.5\%$ & $+1.7\%$  \\
                       & $-4.8\%$ & $-1.6\%$   & $-7.6\%$ & $-6.9\%$  \\
\end{tabular}
\end{table}

\begin{table}
\caption{Continuum limit quark masses in the $\overline{\rm
MS}$ scheme at $\mu=2$~GeV (in MeV).}
\label{tab:quark-continuum}
\begin{tabular}{ccccc}
        & action & $m_{ud}$  & \multicolumn{2}{c}{$m_s$} \\
        &        &           & $M_K$ input & $M_\phi$ input \\
\hline
$N_f=2$\STRUT & impr.  & $3.44^{+0.14}_{-0.22}$  & 
            $88^{+4}_{-6}$  &  $90^{+5}_{-11}$ \\[0.5ex]
$N_f=0$ & impr.  & $4.36^{+0.14}_{-0.17}$  &
                   $110^{+3}_{-4}$  &  $132^{+4}_{-6}$ \\[0.5ex]
$N_f=0$ & stand.\cite{quench.CPPACS} & 4.57(18)                &
                   $116(3)$        &  $144(6)$ \\[0.5ex]
\end{tabular}
\end{table}

\begin{table}
\caption{Parameters of chiral fits to pseudoscalar decay constants (upper
part) and vector decay constants (lower part) with
Eq.~(\ref{eq:decay-fit-full}).}  
\label{tab:decay-fit-full}
\begin{tabular}{r@{.}lcr@{.}lr@{.}lr@{.}lr@{.}lr@{.}lr@{.}l}
\multicolumn{2}{c}{$\beta$} & 
$\chi^2/N_{\rm DF}$ & 
\multicolumn{2}{c}{$A^F$} & 
\multicolumn{2}{c}{$B^F_s$} & 
\multicolumn{2}{c}{$B^F_v$} & 
\multicolumn{2}{c}{$C^F_s$} &  
\multicolumn{2}{c}{$C^F_v$} &  
\multicolumn{2}{c}{$C^F_{sv}$} \\
\hline
1&80 &  $47.1/30$  & 0&2082(52) & 0&086(13) & 
        0&1309(61) & $-0$&0228(84) & $-0$&0110(32) & $-0$&0069(41) \\
1&95 &  $14.6/30$  & 0&1207(54) & 0&087(22) & 
        0&1502(56) & $-0$&029(21)  & $-0$&0253(44) & $-0$&0266(68) \\
2&10 &  $14.8/30$  & 0&0696(37) & 0&168(35)  & 
        0&189(12)  & $-0$&218(73)  & $-0$&101(23)  & $-0$&029(30)  \\
2&20 &  $10.6/30$  & 0&051(12)  & 0&29(15) & 
        0&211(32)  & $-0$&79(44)   & $-0$&123(92)  & $-0$&08(15)  \\
\hline
1&80 &  $11.0/30$  & 0&362(10)  & 0&226(24) & 
        0&065(13)  & $-0$&072(15)  & $0$&009(7)    & $-0$&016(10) \\
1&95 &  $17.3/30$  & 0&2105(73) & 0&184(28) & 
        0&061(12)  & $-0$&068(27)  & $0$&020(11)   & $-0$&015(15) \\
2&10 &  $8.0/30$   & 0&1290(51) & 0&160(41)  & 
        0&130(22)  & $-0$&092(90)  & $-0$&049(39)  & $-0$&080(52)  \\
2&20 &  $9.7/30$   & 0&0970(90) & 0&25(10) & 
        0&128(45)  & $-0$&29(33)   & 0&02(12)  & $-0$&24(16)  \\
\end{tabular}
\end{table}

\begin{table}
\caption{Decay constants at finite lattice
spacings in full and quenched QCD. All decay constants are in GeV units.}
\label{tab:FPhysical}
\begin{tabular}{r@{.}lr@{.}lr@{.}lr@{.}lr@{.}lr@{.}lr@{.}lr@{.}lr@{.}l}
\multicolumn{2}{c}{$\beta$} & 
\multicolumn{2}{c}{$f_{\pi}$} &
\multicolumn{2}{c}{$f_K$ ($K$)} &
\multicolumn{2}{c}{$f_K$ ($\phi$)} &
\multicolumn{2}{c}{$F_{\rho}$} &
\multicolumn{2}{c}{$F_{K^*}$ ($K$)} &
\multicolumn{2}{c}{$F_{K^*}$ ($\phi$)} &
\multicolumn{2}{c}{$F_{\phi}$ ($K$)} &
\multicolumn{2}{c}{$F_{\phi}$ ($\phi$)}  \\
\hline
\multicolumn{18}{c}{$N_f =2$ Full QCD} \\
\hline
1&80 &  0&1954(51) & 0&2273(45) & 0&2359(43) 
     &  0&3378(66) & 0&3546(61) & 0&3595(61) & 0&3726(65) & 0&3833(65)  \\
1&95 &  0&1565(70) & 0&1832(65) & 0&1896(64)            
     &  0&2705(73) & 0&2823(64) & 0&2854(63) & 0&2952(60) & 0&3019(58)  \\
2&10 &  0&1311(66) & 0&1542(63) & 0&1573(63)            
     &  0&2394(73) & 0&2555(65) & 0&2577(64) & 0&2708(65) & 0&2749(65)  \\
2&20 &  0&120(26)  & 0&141(26)  & 0&145(26)             
     &  0&224(16)  & 0&237(14)  & 0&2397(14) & 0&251(14)  & 0&2555(14)  \\
\hline
\multicolumn{18}{c}{$N_f =0$ Quenched QCD} \\
\hline
2&187 & 0&1695(44) & 0&1912(36) & 0&1966(33)        
      & 0&2861(44) & 0&3029(32) & 0&3070(30) & 0&3197(26) & 0&3280(25)  \\
\end{tabular}
\end{table}

\begin{table}
\caption{Meson masses and AWI quark masses.}
\label{tab:meson}
\begin{tabular}{ccr@{.}lcr@{.}lr@{.}lcr@{.}lr@{.}lr@{.}l}
$\kappa_{\rm val}^{(1)}$ &$\kappa_{\rm val}^{(2)}$ 
&\multicolumn{2}{c}{$\mps$} &$[t_{\rm min},t_{\rm max}]$
&\multicolumn{2}{c}{$\chi^2/N_{\rm DF}$}  
&\multicolumn{2}{c}{$\mv$} &$[t_{\rm min},t_{\rm max}]$
&\multicolumn{2}{c}{$\chi^2/N_{\rm DF}$}
&\multicolumn{2}{c}{$\mps/\mv$}  
&\multicolumn{2}{c}{$m_q^{\rm AWI}$}\\[0.5ex]  
\hline
\multicolumn{16}{c}{$\beta=1.80$, $\kappa_{\rm sea} = 0.1409$}\\
\hline
0.1409 & 0.1409 & 1&15601(61) & [5,12] &  1&1(9)  
                & 1&4330(13)  & [5,11] &  1&4(1.2)
                & 0&80669(71) 
                & 0&22483(43)        \\
0.1409 & 0.1430 & 1&09336(63) & [5,12] &  1&2(9)
                & 1&3908(15)  & [5,11] &  1&4(1.1)
                & 0&78616(80) 
                & 0&20056(40)        \\ 
0.1409 & 0.1445 & 1&04660(64) & [5,12] &  1&3(9)
                & 1&3603(16)  & [5,11] &  1&3(1.1) 
                & 0&76938(88) 
                & 0&18328(39)        \\ 
0.1409 & 0.1464 & 0&98441(68) & [5,12] &  1&5(1.0)
                & 1&3217(18)  & [5,11] &  1&0(1.0) 
                & 0&7448(10) 
                & 0&16141(38)        \\ 
0.1409 & 0.1474 & 0&94996(71) & [5,12] &  1&6(1.1)
                & 1&3016(20)  & [5,11] &  0&7(8) 
                & 0&7298(11) 
                & 0&14984(38)        \\ 
0.1430 & 0.1430 & 1&02733(65) & [5,12] &  1&3(1.0)
                & 1&3479(16)  & [5,11] &  1&3(1.1) 
                & 0&76219(91) 
                & 0&17649(38)        \\
\end{tabular}
\end{table}

\begin{table}
\caption{Baryon masses with degenerate quark combinations.}
\label{tab:bary_deg}
\begin{tabular}{cr@{.}lcr@{.}lr@{.}lcr@{.}l}
$\kappa_{\rm val}$ 
&\multicolumn{2}{c}{$m_{\rm N}$} &$[t_{\rm min},t_{\rm max}]$
&\multicolumn{2}{c}{$\chi^2/N_{\rm DF}$}  
&\multicolumn{2}{c}{$m_{\rm \Delta}$} &$[t_{\rm min},t_{\rm max}]$
&\multicolumn{2}{c}{$\chi^2/N_{\rm DF}$}\\  
\hline
\multicolumn{11}{c}{$\beta=1.80$, $\kappa_{\rm sea} = 0.1409$}\\
\hline
0.1409 & 2&2596(49) & [6,12] &  0&9(8) 
       & 2&4151(71) & [6,11] &  2&7(1.8) \\  
0.1430 & 2&1085(62) & [6,12] &  0&5(6) 
       & 2&2853(94) & [6,11] &  1&4(1.3) \\  
0.1445 & 1&9950(82) & [6,12] &  0&3(5) 
       & 2&193(13)  & [6,11] &  0&6(8) \\  
0.1464 & 1&8355(78) & [5,11] &  0&3(5) 
       & 2&061(13)  & [5,9]  &  0&3(6) \\  
0.1474 & 1&746(11)  & [5,11] &  0&4(6) 
       & 1&999(19)  & [5,9]  &  0&3(6) \\  
\hline
\multicolumn{11}{c}{$\beta=1.80$, $\kappa_{\rm sea} = 0.1430$}\\
\end{tabular}
\end{table}

\begin{table}
\caption{Baryon masses with non-degenerate quark combinations.}
\label{tab:bary_nondeg}
\begin{tabular}{cccr@{.}lcr@{.}lr@{.}lcr@{.}lr@{.}lcr@{.}l}
$\kappa_{\rm val}^{(1)}$ &$\kappa_{\rm val}^{(2)}$ &$\kappa_{\rm val}^{(3)}$ 
&\multicolumn{2}{c}{$m_{\rm \Sigma}$} &$[t_{\rm min},t_{\rm max}]$
&\multicolumn{2}{c}{$\chi^2/N_{\rm DF}$} 
&\multicolumn{2}{c}{$m_{\rm \Lambda}$} &$[t_{\rm min},t_{\rm max}]$
&\multicolumn{2}{c}{$\chi^2/N_{\rm DF}$} 
&\multicolumn{2}{c}{$m_{\rm \Delta}$} &$[t_{\rm min},t_{\rm max}]$
&\multicolumn{2}{c}{$\chi^2/N_{\rm DF}$} \\[0.5ex]
\hline
\multicolumn{18}{c}{$\beta=1.80$, $\kappa_{\rm sea} = 0.1409$}\\
\hline
0.1430 & 0.1430 & 0.1409  & 2&1634(57)  & [6,12] &  0&6(7)
                          & 2&1557(57)  & [6,12] &  0&6(7)  
                          & 2&3280(84)  & [6,11] &  1&8(1.4) \\  
0.1445 & 0.1445 & 0.1409  & 2&0924(68)  & [6,12] &  0&4(6)  
                          & 2&0778(68)  & [6,12] &  0&5(6)  
                          & 2&267(10)   & [6,11] &  1&1(1.1) \\  
0.1464 & 0.1464 & 0.1409  & 1&9981(64)  & [5,11] &  0&4(6)  
                          & 1&9750(94)  & [6,11] &  0&3(6)  
                          & 2&188(14)   & [6,11] &  0&4(6) \\  
0.1474 & 0.1474 & 0.1409  & 1&9501(80)  & [5,11] &  0&3(5)  
                          & 1&919(14)   & [6,11] &  0&2(5)  
                          & 2&148(19)   & [6,11] &  0&3(5) \\  
0.1409 & 0.1409 & 0.1430  & 2&2064(52)  & [6,12] &  0&7(7)  
                          & 2&2135(53)  & [6,12] &  0&7(8)  
                          & 2&3700(76)  & [6,11] &  2&2(1.6) \\  
0.1409 & 0.1409 & 0.1445  & 2&1670(56)  & [6,12] &  0&6(7)  
                          & 2&1798(57)  & [6,12] &  0&6(7)  
                          & 2&3392(83)  & [6,11] &  1&9(1.5) \\
\end{tabular}
\end{table}

\begin{table}
\caption{$1\times 1$ and $1 \times 2$ Wilson loops in full QCD at each
simulated sea quark mass and extrapolated to the chiral limit.}
\label{tab:wloop}
\begin{tabular}{cccc}
$\beta$ &  $\kappa$ & $\langle W^{1\times 1} \rangle$ & 
$\langle W^{1\times 2} \rangle$ \\
\hline
1.80 & 0.1409 & 0.490527(30) & 0.232159(35) \\
     & 0.1430 & 0.495049(39) & 0.237880(53) \\ 
     & 0.1445 & 0.499361(37) & 0.243370(49) \\
     & 0.1464 & 0.507204(57) & 0.253308(78) \\
     & $\mps=0$ & 0.51471(34)  & 0.26274(45)        \\
\hline
1.95 & 0.1375 & 0.553355(20) & 0.305089(27) \\
     & 0.1390 & 0.556667(21) & 0.309890(31) \\ 
     & 0.1400 & 0.559143(21) & 0.313473(34) \\
     & 0.1410 & 0.561884(27) & 0.317457(36) \\
     & $\mps=0$ & 0.56518(20)  & 0.32228(31)        \\
\hline
2.10 & 0.1357 & 0.5980283(76) & 0.362139(12) \\ 
     & 0.1367 & 0.5992023(76) & 0.363979(12) \\
     & 0.1374 & 0.6000552(67) & 0.365297(10) \\
     & 0.1382 & 0.6010819(84) & 0.366883(13) \\
     & $\mps=0$ & 0.602197(64)  & 0.36862(10)       \\
\hline
2.20 & 0.1351 & 0.620027(10) & 0.390976(16) \\ 
     & 0.1358 & 0.620616(7)  & 0.391911(12) \\
     & 0.1363 & 0.621035(8)  & 0.392570(11) \\
     & 0.1368 & 0.621490(8)  & 0.393289(12) \\
     & $\mps=0$ & 0.62233(22)  & 0.39465(26)        \\
\end{tabular}
\end{table}

\begin{table}
\caption{$1\times 1$ and $1 \times 2$ Wilson loops in quenched QCD.}
\label{tab:Qwloop}
\begin{tabular}{ccc}
$\beta$ & $\langle W^{1\times 1} \rangle$ & $\langle W^{1\times 2} \rangle$ \\
\hline
2.187 & 0.5921968(62) & 0.3438930(92)  \\
2.214 & 0.5991994(56) & 0.3533512(86)  \\
2.247 & 0.6072343(59) & 0.3642759(92)  \\
2.281 & 0.6149775(51) & 0.3748875(80)  \\
2.334 & 0.6261248(48) & 0.3902635(80)  \\
2.416 & 0.6415604(20) & 0.4117512(34)  \\
2.456 & 0.6484512(17) & 0.4214117(29)  \\
2.487 & 0.6535537(21) & 0.4286002(36)  \\
2.528 & 0.6600072(20) & 0.4377232(34)  \\ 
2.575 & 0.6670422(18) & 0.4477145(25)  \\
\end{tabular}
\end{table}

\begin{table}
\caption{Numerical values for coupling constants, $Z$ factors and
improvement coefficients in full QCD. $Z$ factors and
improvement coefficients are evaluated using $g^2_{\overline{\rm
MS}}(1/a)$. We also quote run factors used for running quark masses from
$\mu=1/a$ to $\mu=2$~GeV with the three-loop beta function.}
\label{tab:Zfactors}
\begin{tabular}{cccccccccccccccc}
$\beta$ &
$g^2_{\overline{\rm MS}}(1/a)$ &
$\tilde{g}^2_{\overline{\rm MS}}(1/a)$ &
$g^2_{\overline{\rm MS}}(\pi/a)$ &
$\tilde{g}^2_{\overline{\rm MS}}(\pi/a)$ & 
$Z_m$ & 
$b_m$ &
$Z_P$ &
$b_P$ &
$Z_A$ &
$Z_A/Z_P$ &
$b_A$ &
$c_A$ &
$Z_V$ &
$b_V$ &
run factor\\[0.5ex]
\hline
1.80 &  3.155 & 2.185 & 2.188 & 1.673 & 1.126 & $-0.602$ & 0.835 & 1.125 & 
0.932 & 1.116 & 1.119 & $-0.0120$ & 0.913 & 1.121 & 0.8662 \\     
1.95 &  2.816 & 2.054 & 2.019 & 1.595 & 1.113 & $-0.591$ & 0.853 & 1.112 & 
0.939 & 1.102 & 1.106 & $-0.0107$ & 0.922 & 1.108 & 0.9239 \\    
2.10 &  2.567 & 1.946 & 1.888 & 1.529 & 1.103 & $-0.583$ & 0.866 & 1.102 & 
0.945 & 1.091 & 1.097 & $-0.0098$ & 0.929 & 1.098 & 0.9885 \\     
2.20 &  2.429 & 1.882 & 1.812 & 1.489 & 1.097 & $-0.578$ & 0.873 & 1.096 & 
0.948 & 1.086 & 1.092 & $-0.0092$ & 0.933 & 1.093 & 1.0219 \\
\end{tabular}
\end{table}

\newpage


\newpage

\newpage

\newpage

\newpage

\newpage

\newpage

\newpage

\newpage

\newpage

\newpage

\newpage

\newpage

\newpage

\newpage

\newpage

\end{document}